\documentclass[11pt,a4paper]{article}
\pdfoutput=1
\usepackage{jheppub}

\newcommand{\be}{\begin{equation}}
\newcommand{\ee}{\end{equation}}
\newcommand{\bea}{\begin{eqnarray}}
\newcommand{\eea}{\end{eqnarray}}

\newcommand{\eqn}[1]{(\ref{#1})}

\newcommand{\mt}[1]{\textrm{\tiny #1}}
\def\nc {N_\mt{c}}

\newcommand{\sac}{\, , \qquad}

\newcommand{\uh}{u_\mt{H}}

\newcommand{\gym}{g_\mt{YM}}

\newcommand{\cf}{{\cal F}}
\newcommand{\cb}{{\cal B}}
\newcommand{\ch}{{\cal H}}
\newcommand{\cn}{{\cal N}}
\newcommand{\cl}{{\cal L}}

\newcommand{\vp}{\varphi}
\newcommand{\umax}{u_{\mt{max}}}

\newcommand{\Lmax}{L_{\mt{max}}}
\newcommand{\adiss}{a_\mt{diss}}
\newcommand{\Tdiss}{T_\mt{diss}}
\newcommand{\vtrans}{v_\mt{trans}}
\newcommand{\vlim}{v_\mt{lim}}
\newcommand{\size}{0.6}
\newcommand{\sizetwo}{0.6}
\newcommand{\dist}{3mm}

\title{Quarkonium dissociation by anisotropy}
\author[a]{Mariano Chernicoff,}
\author[b]{Daniel Fern\'andez,}
\author[b,c]{David Mateos,}
\author[d,e]{and Diego Trancanelli} 
\affiliation[a]{Department of Applied Mathematics and Theoretical Physics, Centre for Mathematical Sciences,
Wilberforce Road, Cambridge, CB3 0WA, UK}
\affiliation[b]{Departament de F\'\i sica Fonamental \&  Institut de 
Ci\`encies del Cosmos (ICC), Universitat de Barcelona (UB), Mart\'{\i}  i 
Franqu\`es 1, E-08028 Barcelona, Spain} 
\affiliation[c]{Instituci\'o Catalana de Recerca i Estudis Avan\c cats (ICREA),
Passeig Llu\'\i s Companys 23, E-08010, Barcelona, Spain} 
\affiliation[d]{Instituto de F\'\i sica, Universidade de S{\~a}o Paulo, 05314-970 S{\~a}o Paulo, Brazil}
\affiliation[e]{Department of Physics, University of Wisconsin, Madison, WI
53706, USA}  

\date{\today}

\abstract{We compute the screening length for quarkonium mesons moving through an anisotropic, strongly coupled ${\cal N}=4$ super Yang-Mills plasma by means of its gravity dual. We present the results for arbitrary velocities and orientations of the mesons, as well as for arbitrary values of the anisotropy. The anisotropic screening length  can be larger or smaller than the isotropic one, and this depends on whether the comparison is made at equal temperatures or at equal entropy densities. For generic motion we find that:  (i) mesons dissociate above a certain critical value of the anisotropy, even at zero temperature; (ii) there is a limiting velocity for mesons in the plasma, even at zero temperature; (iii) in the ultra-relativistic limit  the screening length scales as $(1-v^2)^{\epsilon}$ with $\epsilon =1/2$, in contrast with the isotropic result $\epsilon =1/4$. 
}  
  

\emailAdd{M.Chernicoff@damtp.cam.ac.uk}
\emailAdd{daniel@ffn.ub.edu} 
\emailAdd{dmateos@icrea.cat} 
\emailAdd{dtrancan@fma.if.usp.br} 
 
\begin{document}

\begin{flushright}
DAMTP-2012-58 \\
ICCUB-12-319 \\
MAD-TH-12-03
\end{flushright}

\maketitle
\setlength{\parskip}{8pt}


\section{Introduction}
\label{intro}

A remarkable conclusion from the experiments at the Relativistic Heavy Ion Collider (RHIC)  \cite{rhic,rhic2} and at the Large Hadron Collider (LHC) \cite{lhc} is that the quark-gluon plasma (QGP) does not behave as a weakly coupled gas of quarks and gluons, but rather as a strongly coupled fluid \cite{fluid,fluid2}. This places limitations on the applicability of perturbative methods. The lattice formulation of Quantum Chromodynamics (QCD) is also of limited utility, since for example it is not well suited for studying real-time phenomena. This has provided a strong motivation for understanding the dynamics of strongly coupled non-Abelian plasmas through the gauge/string duality \cite{duality,duality2,duality3} (see \cite{review} for a recent review of applications to the QGP). In general, a necessary requirement for the string description to be tractable is that the plasma be infinitely strongly coupled, $\lambda=\gym^2 \nc \to \infty$. Of course, the real-world QGP is not infinitely strongly coupled, and its dynamics involves a complex combination of both weak and strong coupling physics that depend on the possibly multiple scales that characterize the process of interest. The motivation for studying string models is that they provide examples in which explicit calculations can be performed from first principles at strong coupling, in particular in the real-time domain. The hope is then that, by understanding the weak and the strong coupling limits, one may be able to bracket the dynamics of the real-world QGP, which lies somewhere in between. 

During the initial stage after the collision the plasma is far from equilibrium, and after a certain time a hydrodynamic description becomes applicable. If one thinks of hydrodynamics as a gradient expansion around a locally isotropic system, it is somewhat surprising that the hydrodynamic description actually becomes applicable when the  longitudinal and transverse pressures are still significantly different. This can be explicitly seen, for example, in holographic descriptions \cite{Chesler:2009cy,Chesler:2010bi,Heller:2011ju,Heller:2012je} in which gravity provides a  valid description all the way from the far-from-equilibrium phase to the locally isotropic phase, across the intermediate hydrodynamic-but-still-anisotropic phase. Thus, during most of the time that viscous hydrodynamics is applied, the plasma created in a heavy ion collision is anisotropic, with the level of anisotropy in fact increasing as one approaches the edge of the system. The fact that the range of time and space over which the QGP is anisotropic is larger than traditionally assumed has  provided additional motivation for the study of anisotropic plasmas.

In this paper we will investigate the effect of an intrinsic anisotropy on the screening length between a quark-antiquark pair in a strongly coupled plasma. As we will review below, the plasma is static because it is held in anisotropic equilibrium by an external force \cite{prl,jhep}. We will discuss all the caveats in more detail below, but we emphasize from the beginning that there are several reasons why, in terms of potential extrapolations to the real-world QGP, our results must be interpreted with  caution. First, the sources of anisotropy in the QGP created in a heavy ion collision and in our system are different. In the QGP the anisotropy is dynamical in the sense that it is due to the initial distribution of particles in momentum space, which will evolve in time and eventually become isotropic. In contrast, in our case the anisotropy is due to an external source that keeps the system in an equilibrium anisotropic state that will not evolve in time. Nevertheless, we hope that our system might provide a good toy model for processes whose characteristic time scale is sufficiently shorter than the time scale controlling the evolution of a dynamical plasma.  

The second caveat concerns the fact that, even in an static situation,  different external sources can be chosen to hold the plasma in equilibrium, so one may wonder to what extent the results depend on this choice. We will provide a partial answer to this question in Sec.~\ref{discussion}, where we will explain that our qualitative results, for example the ultrarelativistic limit, do not depend on the details of our solution but only on a few general features. Nevertheless, it would still be very interesting to compute the same observables in other strongly coupled, static, anisotropic plasmas. Only then a general picture would emerge that would allow one, for example, to understand which observables are robust, in the sense that they are truly insensitive to the way in which the plasma is held in anisotropic equilibrium, and which ones are model-dependent. Obviously it is the first type of observables that have a better chance of being relevant for the real-world QGP. Our paper should be regarded as a first step in this general program. 

We will consider the screening length in the case in which the quark-antiquark pair is at rest in the plasma as well as the case in which it is moving through the plasma. 
For this purpose we will examine a string with both endpoints on the boundary of an asymptotically AdS spacetime  \cite{prl,jhep} that is dual to an anisotropic ${\cal N}=4$ super Yang-Mills plasma. The gravity solution possesses an anisotropic horizon, it is completely regular on and outside the horizon, and it is solidly embedded in type IIB string theory. For these reasons it provides an ideal toy model in which questions about  anisotropic effects at strong coupling can be addressed from first principles. For the particular case of a quark-antiquark pair at rest, the screening length has also been computed \cite{Rebhan:2012bw} in a different model \cite{sing} of a strongly coupled, anisotropic plasma. The results exhibit some differences with respect to those presented here. While this may indicate some model dependence of the screening length, it is important to note that the solution of \cite{sing} possesses a naked singularity. Although this is a rather benign singularity, its presence introduces a certain amount of ambiguity in the calculations,  which can only be performed by prescribing  somewhat ad hoc boundary conditions at the singularity. In any case, this discussion is another indication that it would be interesting to compute the screening length in a larger class of models in order to ascertain which of its features  are model-independent.  

To avoid any possible confusion, we clarify from the beginning that the quarks and antiquarks that we will consider are infinitely massive, i.e.~the bound states that we will consider are the analogue  of heavy quarkonium mesons in QCD. Thus, the reader should always have the word `quarkonium' in mind despite the fact that we will often refer to these states simply as `mesons', `heavy mesons', `quark-antiquark bound states', `dipoles', etc. This is specially relevant in the ultra-relativistic limit of the screening length, to which we will pay particular attention since it can be determined analytically. We emphasize that our results correspond to sending the quark and antiquark masses to infinity first, and then sending $v\to 1$. In particular, this means that in any future attempt to connect our results to the phenomenology of the QGP, this connection can only be made to the phenomenology of heavy quarkonium moving through the plasma.

The screening length for quarkonium mesons at rest in the anisotropic plasma of \cite{prl,jhep} has been previously studied in \cite{giataganas,Rebhan:2012bw}. Our Sec.~\ref{static}  has some overlap with these references and, wherever they overlap, our results agree with theirs. Other physical properties of the anisotropic plasma that have been calculated include its shear viscosity \cite{rebhan_viscosity,mamo}, the drag force on a heavy quark \cite{cfmt,giataganas}, the jet quenching parameter \cite{giataganas,jet,Rebhan:2012bw}, and the energy lost by a rotating quark \cite{fadafan}. The phase diagram of the zero-coupling version of the model considered in \cite{prl,jhep} has been studied in \cite{rebhan3}.  Dissociation of baryons in the isotropic ${\cal N}=4$ plasma has been analyzed in \cite{krishnan}.


\section{Gravity solution}
\label{grav}
\begin{figure}[tb]
\begin{center}
\begin{tabular}{cc}
\includegraphics[scale=0.82]{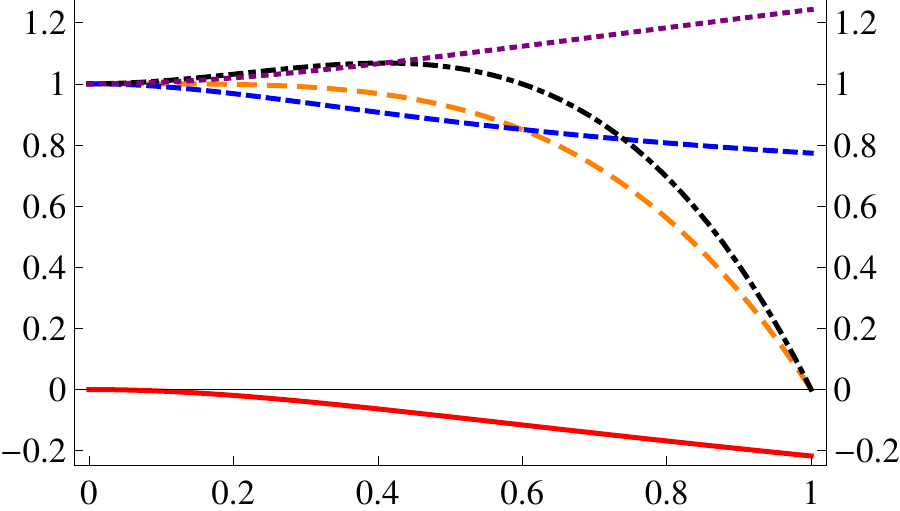}
\put(-109,-10){\small $u/\uh$}
\put(-109,-14){$$}
\put(-150,17){$\phi$}
\put(-70,118){$\ch$}
\put(-30,88){$\cb$}
\put(-40,65){$\cf$}
\put(-82,68){$\cf \cb$}
&
\includegraphics[scale=0.78]{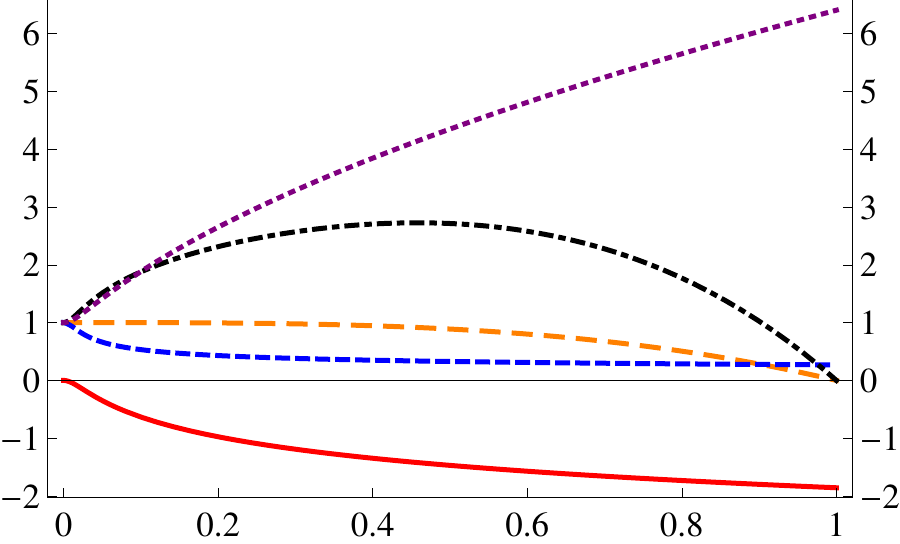}
\put(-109,-10){\small $u/\uh$}
\put(-109,-14){$$}
\put(-55,17){$\phi$}
\put(-70,111){$\ch$}
\put(-100,52){$\cf \cb$}
\put(-100,32){$\cb$}
\put(-53,67){$\cf$}
\end{tabular}
\caption{\small Metric functions for $a/T\simeq 4.4$ (left) and $a/T\simeq 86$ (right).
\label{plots}
}
 \end{center} 
 \end{figure}
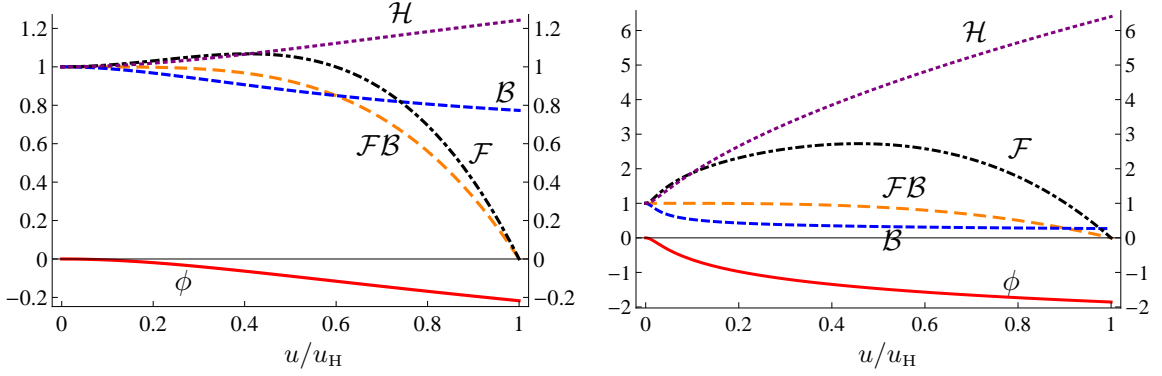

The type IIB supergravity solution of \cite{prl,jhep} in the string frame takes the form
\bea
&&\hskip -.35cm 
ds^2 =  \frac{L^2}{u^2}
\left( -\cf \cb\, dt^2+dx^2+dy^2+ \ch dz^2 +\frac{ du^2}{\cf}\right) +
L^2 e^{\frac{1}{2}\phi} d\Omega_5^2, 
\,\,\,\,\,\,\label{sol1} \\
&& \hskip -.35cm \chi = az \sac \phi=\phi(u) \,,
\label{sol2}
\eea
where $\chi$ and $\phi$ are the axion and the dilaton, respectively, and $(t,x,y,z)$ are the gauge theory coordinates. Since there is rotational invariance in the $xy$-directions, we will refer to these as the transverse directions, and to $z$ as the longitudinal direction. $\cf, \cb$ and $\ch$ are functions of the holographic radial coordinate $u$ that were determined numerically in \cite{prl,jhep}. Their form for two values of $a/T$ is plotted in Fig.~\ref{plots}. 
The horizon lies at $u=\uh$, where $\cf=0$, and the boundary at $u=0$, where $\cf=\cb=\ch=1$ and $\phi=0$. The metric near the boundary asymptotes to $AdS_5 \times S^5$. Note that the axion is linear in the $z$-coordinate. The proportionality constant $a$ has dimensions of mass and is a measure of the anisotropy. The axion profile is dual in the gauge theory to a position-dependent theta parameter of the form $\theta \propto z$. This acts as an isotropy-breaking external source that forces the system into an anisotropic equilibrium state. 

\begin{figure}[tb]
\begin{center}
\includegraphics[scale=0.71]{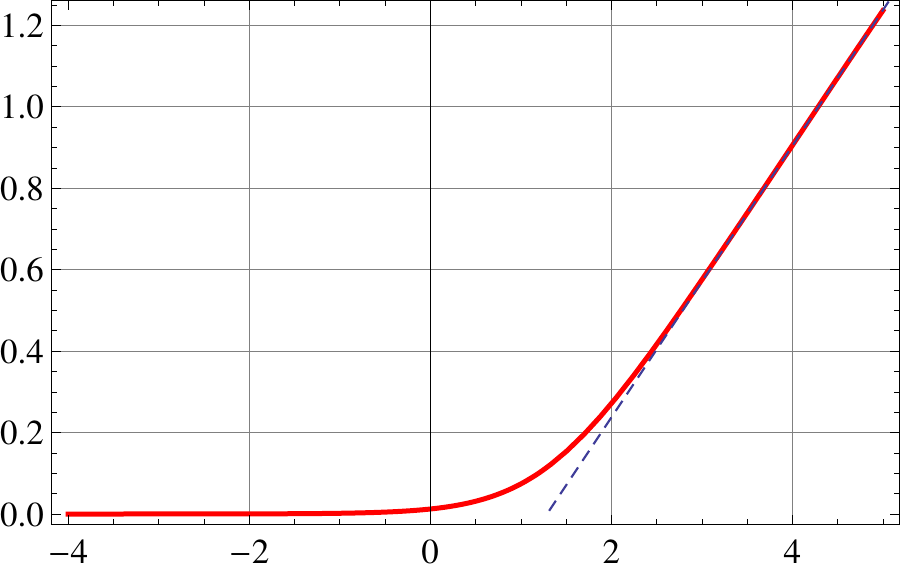}
\put(-109,-10){\small $\log (a/T)$}
\put(-208,40){\rotatebox{90}
{$\log (s/ s_\mt{iso})$}}
\caption{\small Log-log plot of the entropy density per unit 3-volume in the $xyz$-directions as a function of 
$a/T$, with $s_\mt{iso}$ defined as in eqn.~\eqn{siso}. The dashed blue line is a straight line with slope $1/3$. 
\label{scalings}
}
 \end{center} 
 \end{figure}

If $a=0$ then the solution reduces to the isotropic black D3-brane solution dual to the isotropic $\cn=4$ theory at finite temperature. In this case
\be
\cb=\ch=1 \sac \chi=\phi=0 \sac \cf = 1-\frac{u^4}{\uh^4}
\sac \uh = \frac{1}{\pi T} 
\label{iso}
\ee
and the entropy density takes the form 
\be
s_\mt{iso}= \frac{\pi^2}{2} \nc^2 T^3 \,. 
\label{siso}
\ee
Fig.~\ref{scalings} shows the entropy density per unit 3-volume in the $xyz$-directions of the anisotropic plasma as a function of the dimensionless ratio $a/T$, normalized to the entropy density of the isotropic plasma at the same temperature. At small $a/T$ the entropy density scales as in the isotropic case, whereas at large $a/T$ it scales as \cite{ALT,prl,jhep}
\be
s = c_\mt{ent} \nc^2 a^{1/3} T^{8/3} \,, \qquad\qquad [a/T \gg 1]
\label{larges}
\ee
where $c_\mt{ent}$ is a constant that can be determined numerically. The transition between the two asymptotic behaviors of the entropy density takes place at $a/T \simeq 3.7$. 

For later use  we list here the near-boundary behavior of the different functions that determine the solution \eqn{sol2}: 
\bea
{\cal F}&=&1+\frac{11}{24}a^2u^2+\left({\cal F}_4+\frac{7}{12}a^4\log u\right)u^4+O(u^6)\,,\cr
{\cal B}&=&1-\frac{11}{24}a^2u^2+\left({\cal B}_4-\frac{7}{12}a^4\log u\right)u^4+O(u^6)\,,\cr
{\cal H}&=&1+\frac{1}{4}a^2u^2-\left(\frac{2}{7}{\cal B}_4-\frac{5}{4032}a^4-\frac{1}{6}a^4\log u\right)u^4+O(u^6)\,.
\label{expansion}
\eea
The coefficients ${\cal F}_4$ and ${\cal B}_4$ depend on $a$ and $T$ and are known analytically in the limits of low, and high temperature and numerically for intermediate regimes \cite{jhep}.

A feature of the solution \eqn{sol2} that played an important role in the analysis of \cite{prl,jhep} is the presence of a conformal anomaly. Its origin lies in the fact that diffeomorphism invariance in the radial direction $u$ gets broken in the process of renormalization of the on-shell supergravity action. In the gauge theory this means that scale invariance is broken by the renormalization process. One manifestation of the anomaly is the fact that, unlike the entropy density, other thermodynamic quantities do not depend solely on the ratio $a/T$ but on $a$ and $T$ separately. Fortunately, this will not be the case for the screening length, as we will see below.

\begin{figure}[tb]
\begin{center}
\includegraphics[scale=0.80]{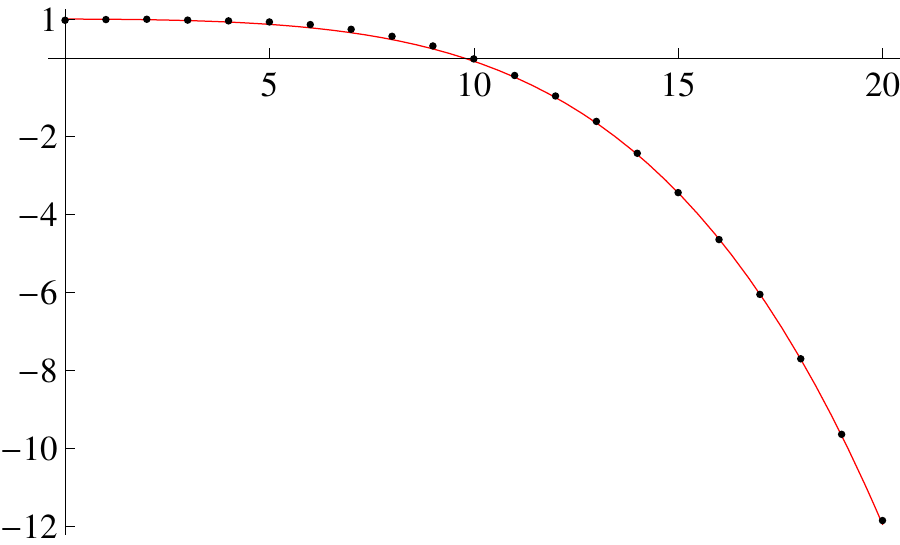}
 \begin{picture}(0,0)
   \put(-20,90){$a/T$}
      \put(-230,50){\rotatebox{0}{$\alpha$}}
 \end{picture}
\caption{\small  Ratio \eqn{ratio} as a function of $a/T$. The blue dots are the actual values of the ratio, and the red curve is the fit \eqn{fit}. 
\label{fitplot}
}
 \end{center} 
 \end{figure}
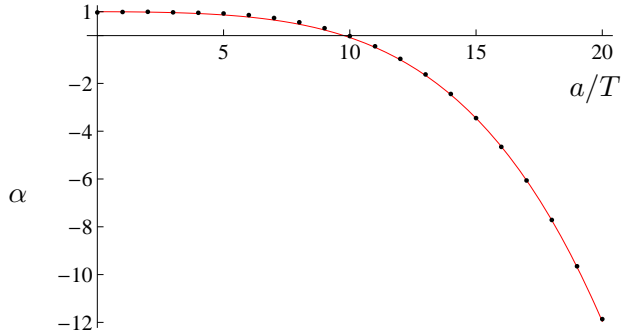

To facilitate a (rough) comparison of the anisotropy in our system to that in other anisotropic plasmas it is useful to consider the ratio
\be
\alpha = \frac{4 E + P_\perp -P_\mt{L}}{3 T s} \,,
\label{ratio}
\ee
where $E$  is the energy density and $P_\perp, P_\mt{L}$ are the transverse and longitudinal pressures, respectively. In addition to being dimensionless, this ratio has the virtue that it does not depend on $a$ and $T$ separately, but only on the combination $a/T$. For the isotropic ${\cal N}=4$ super Yang-Mills plasma $\alpha=1$, whereas for $0< a/T \lesssim 20$ the ratio is well approximated by the expression
\be
\alpha \simeq 1 - 0.0036 \left( \frac{a}{T}\right)^2 - 0.000072 \left( \frac{a}{T}\right)^4
\,, \label{fit}
\ee
as shown in Fig.~\ref{fitplot}.

At various points we will refer to the limit $T= 0$ of the anisotropic plasma. The zero-temperature version of the solution \eqn{sol2} was found in \cite{ALT}. In this case the string-frame metric exhibits a naked curvature singularity deep in the infra-red, and the Einstein-frame metric exhibits infinite tidal forces \cite{tidal,tidal2}. However, we emphasize that, for any finite temperature, the singularity is hidden behind the horizon and the solution is completely regular on and outside the horizon, exhibiting no pathologies of any type. Thus we will think of the $T=0$ results as those obtained by taking the limit $T\to 0$ of the finite-temperature results. Moreover, regulating the infra-red geometry in this or any other way is actually unnecessary for most of the physics of quarkonium dissociation. The reason is that, as we will see, in the limit in which $a/T$ becomes large the penetration depth into the AdS bulk of the string that is dual to the quarkonium meson becomes very small. As a result, the dissociation is entirely controlled by the metric near the boundary, which is insensitive to the infra-red behavior described above. 


\section{Preliminaries}
\label{preli}
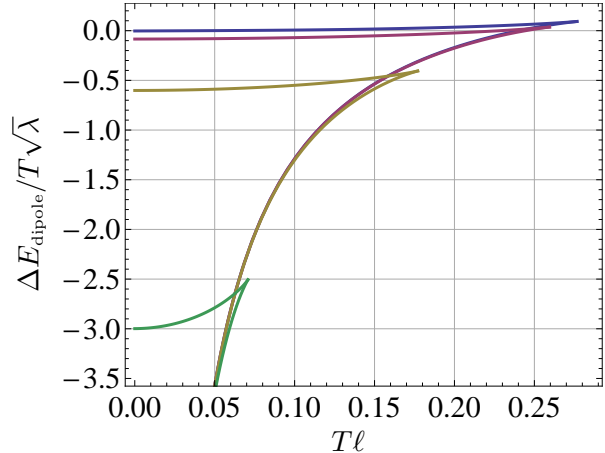
\begin{figure}[t!]
\begin{center}
\includegraphics[scale=0.7]{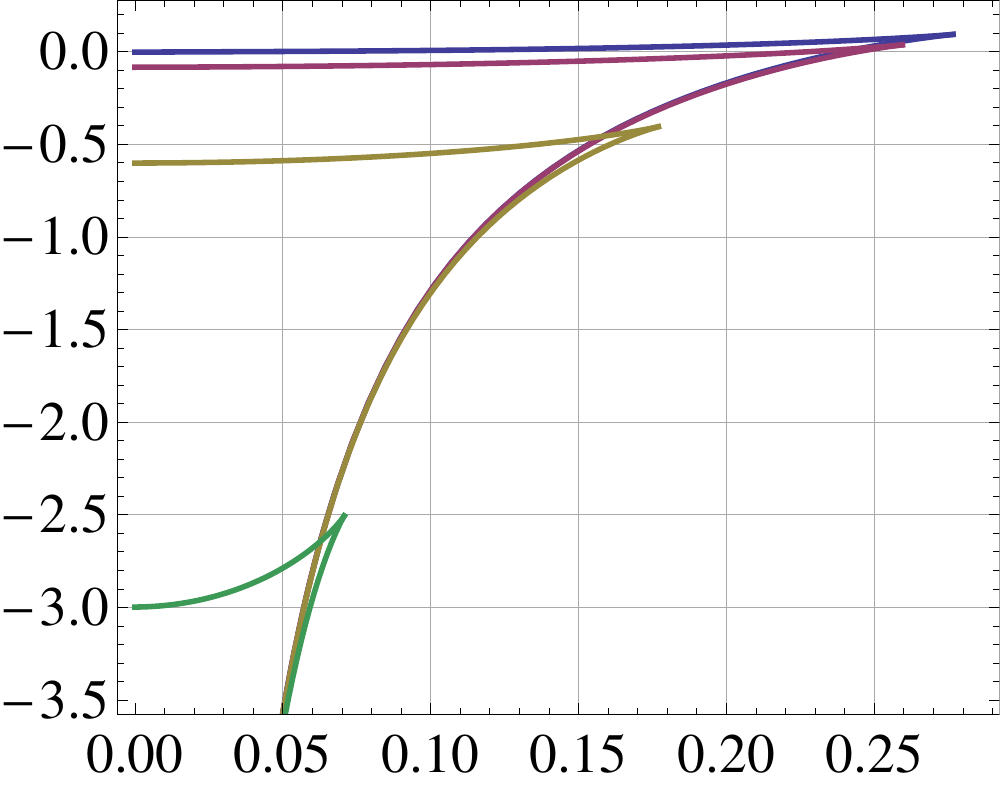}
 \begin{picture}(0,0)
   \put(-225,50 ){\rotatebox{90}
   {$\Delta E_\mt{dipole}/T \sqrt{\lambda}$}}
    \put(-105,-10){$T \ell $}
 \end{picture}
\caption{Energy difference, as defined in \eqn{E}, between a bound and an unbound quark-antiquark pair moving through the isotropic plasma \eqn{iso} with velocities (from the rightmost curve to the leftmost curve) $v=0,0.35,0.85,0.996$. The dipole is oriented orthogonally to its velocity. For $v<\vtrans$ one has $L_s < \Lmax$, whereas for $v>\vtrans$ one finds $L=\Lmax$, where $\vtrans \simeq 0.45$ is the transition velocity between the two behaviors. At $v=0$ the screening length and the maximum separation are $L_\mt{s}\simeq 0.24/ T$ and $L_\mt{max}\simeq 0.27/ T$, respectively. 
\label{isopic}}
\end{center}
\end{figure}

\begin{figure}[tb]
\begin{center}
\begin{tabular}{cc}
\includegraphics[scale=0.70]{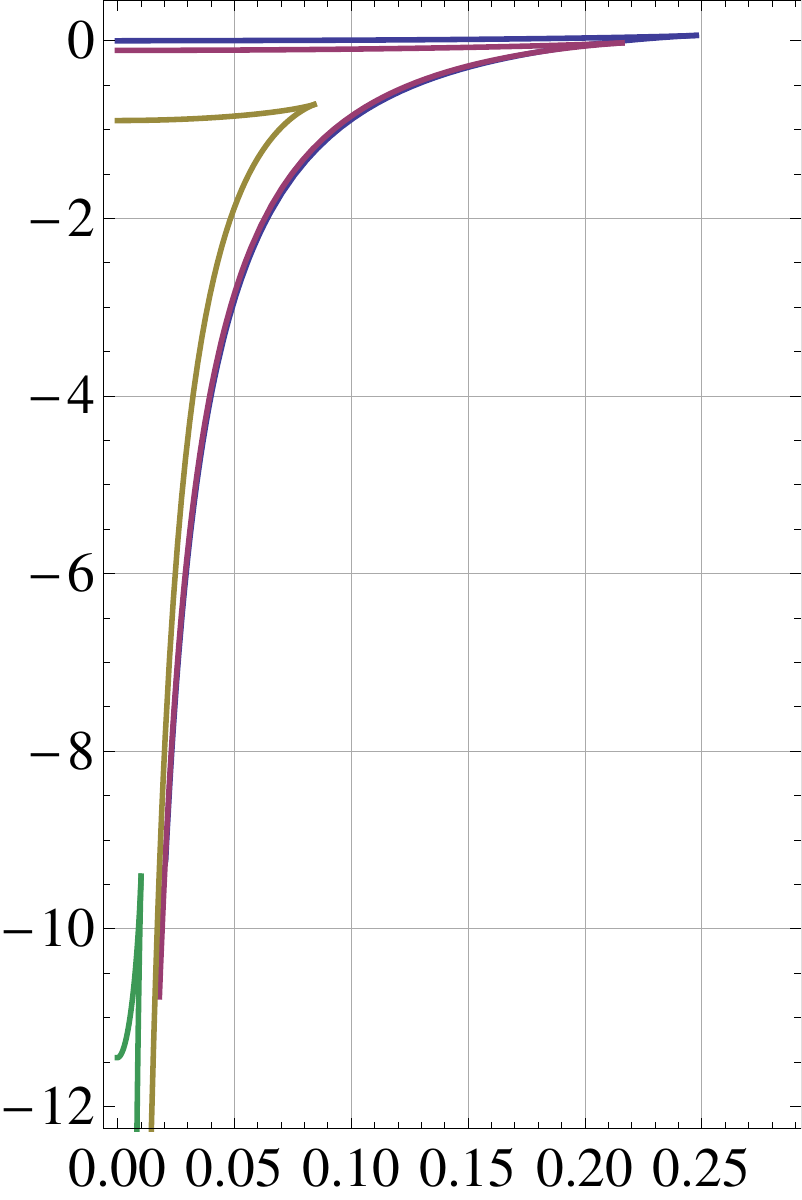}
   \put(-175,90 ){\rotatebox{90}
   {$\Delta E_\mt{dipole}/T \sqrt{\lambda}$}}
    \put(-78,-10){$T \ell $}
\qquad\qquad
\includegraphics[scale=0.70]{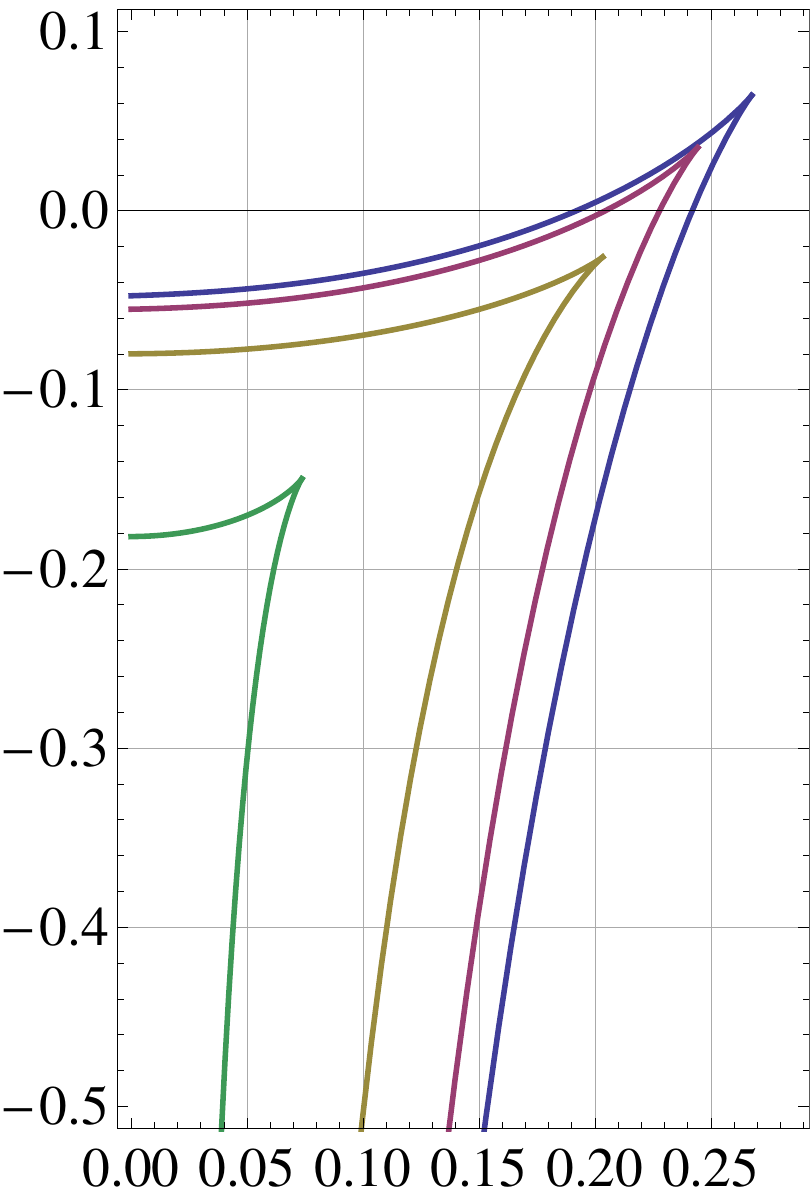}
   \put(-179,90 ){\rotatebox{90}
   {$\Delta E_\mt{dipole}/T \sqrt{\lambda}$}}
    \put(-78,-10){$T \ell $}
\end{tabular}
\caption{\small 
Energy difference in an anisotropic plasma, as defined in \eqn{E}, between a bound and an unbound quark-antiquark pair oriented along the transverse direction $x$ and moving along the anisotropic  direction $z$. All the curves on the left correspond to $a/T=12.2$ and different velocities (from the rightmost curve to the leftmost curve) $v=0,0.35,0.85,0.996$. All the curves on the right correspond to the same velocity $v=0.25$ and different anisotropies (from the rightmost curve to the leftmost curve) $a/T=0,6.5,43,744$. For these anisotropies the corresponding transition velocities are respectively given by $\vtrans = 0.45, 0.29, 0.19, 0.11$. 
\label{potentials22}
}
 \end{center} 
 \end{figure}

In this paper we define the screening length $L_s$ as the separation between a quark and an antiquark such that for $\ell < L_s$ ($\ell > L_s$) it is energetically favorable for the quark-antiquark pair to be bound (unbound) \cite{static1,static2}. Obviously this satisfies $L_s \leq L_\mt{max}$, where $L_\mt{max}$ is the maximum separation $L_\mt{max}$ for which a bound quark-antiquark solution exists. We will determine $L_s$ by comparing the action $S(\ell)$ of the bound pair, which is a function of the quark-antiquark separation $\ell$, to the action $S_\mt{unbound}$ of the unbound system, i.e.~by computing:
\be
\Delta S (\ell) = S(\ell) - S_\mt{unbound} \,.
\label{criterion}
\ee
The screening length is the maximum value of $\ell$ for which $\Delta S$ is positive (since we will work in Lorentzian signature). This may correspond to the value of $\ell$ at which $\Delta S$ crosses zero, in which case $L_s < \Lmax$, or the maximum value of $\ell$ for which a bound state exists, in which case $L_s = \Lmax$.  In the Euclidean version of our calculations, this criterion corresponds to determining which configuration has the lowest free energy, which is therefore the configuration that is thermodynamically preferred. As shown in Fig.~\ref{isopic}, for a meson moving through the isotropic plasma \eqn{iso} one has $L_s < L_\mt{max}$ for $v<\vtrans$, whereas for $v>\vtrans$ one finds that $L_s = \Lmax$, where $\vtrans \simeq 0.45$ is the transition velocity between the two behaviors \cite{liu1,liu2,chernicoff1}.
These qualitative features extend to the anisotropic case, as we have illustrated in Fig.~\ref{potentials22}. The transition velocity decreases with the anisotropy, so for large $a/T$ one has $L_s=L_\mt{max}$ except for very low velocities. Similarly, if the ultra-relativistic limit $v\to 1$ is taken at fixed $a$ and $T$, then obviously $v  > \vtrans$ and again   $L_s=L_\mt{max}$. 

All our calculations will be done in the rest frame of the quark-antiquark pair, to which we will refer as the dipole rest frame. Since any observable can be easily translated between this frame and the plasma rest frame, we will speak interchangeably of `mesons in a plasma wind' and of `mesons in motion in the plasma'. We emphasize however that all the  physical quantities that we will present, e.g.~the screening length, are computed in the dipole rest frame. 

The actions are scalar quantities, so $\Delta S_\mt{dipole}=\Delta S_\mt{plasma}$. Moreover, in the dipole rest frame we have 
\be
\Delta S_\mt{dipole} = - {\cal T} \Delta E_\mt{dipole} \,,
\label{E}
\ee
since the dipole is static in its own rest frame. In this expression $E_\mt{dipole}$ is the energy (as opposed to the free energy) of the configuration and ${\cal T}=\int dt$ is the length of the integration region in time. Thus we see that our criterion, which is based on comparing the actions, can also be thought  of as a comparison between the energies of the bound and the unbound configurations in the dipole rest frame. 

We will see that the ultraviolet divergences in the string action associated to integrating all the way to the boundary of AdS cancel out in the difference \eqn{criterion}, and neither the bound nor the unbound actions possess infrared divergences associated to integrating all the way down to the horizon. This can be verified explicitly and it also follows from their relation to the energy in the rest frame of the dipole: While the energy of the unbound string pair possesses an infrared  logarithmic divergence in the plasma rest frame \cite{yaffe}, no such divergence is present in the dipole rest frame (see e.g.~the discussion in \cite{chernicoff1}). 


\section{Static dipole in an anisotropic plasma}
\label{static}
In an anisotropic plasma the screening length depends on the relative orientation between the dipole and the anisotropic direction $z$. Given the rotational symmetry in the $xy$-plane we  assume without loss of generality that the dipole lies in the $xz$-plane, at an angle $\theta$ with the $z$-axis. We thus choose the static gauge $t=\tau, \sigma=u$ and specify the string embedding as
\be
x \to \sin \theta \, x(u) \sac 
z \to  \cos \theta \, z(u)  \,.
\ee
The string action takes the form
\be
S = -\frac{L^2}{2\pi\alpha'} \, 2 \, 
\int dt \int_0^{\umax} du \frac{1}{u^2} 
\sqrt{\cb \left( 1+\cf\ch \cos^2 \theta \, z'^2 
+ \cf \sin^2 \theta \, x'^2\right) }
\,,
\ee
where the 2 comes from the two branches of the string and  $\umax$ will be determined below. The conserved momenta associated to translation invariance in the $x,z$ directions are given by
\bea
\Pi_x &=& \frac{1}{\sin \theta} 
\frac{\partial \cal L}{\partial x'} = 
\frac{\cb\cf \sin\theta \, x'}
{u^2 \sqrt{\cb \left( 1+\cf\ch \cos^2 \theta \, z'^2 
+ \cf \sin^2 \theta \, x'^2\right) }} \,, \\
\Pi_z &=& \frac{1}{\cos \theta} 
\frac{\partial \cal L}{\partial z'} = 
\frac{\cb\cf\ch \cos\theta \, z'}
{u^2 \sqrt{\cb \left( 1+\cf\ch \cos^2 \theta \, z'^2 
+ \cf \sin^2 \theta \, x'^2\right) }} \,.
\eea
Inverting these relations we find
\be
x' = \frac{\sqrt{\ch} \csc \theta \, u^2 \, \Pi_x}
{\sqrt{\cf} \sqrt{\cb \cf \ch -u^4 
\left(\Pi_z^2 + \ch \, \Pi_x^2 \right)} } \sac
z' = \frac{\sec \theta\,  u^2 \, \Pi_z}
{\sqrt{\cf \ch} \sqrt{\cb \cf \ch -u^4 
\left(\Pi_z^2 + \ch \, \Pi_x^2 \right)} } \,.
\ee
Substituting back in the action we arrive at
\be
S = -\frac{L^2}{2\pi\alpha'} \, 2 \, 
\int dt \int_0^{\umax} du \frac{1}{u^2} 
\frac{\cb \sqrt{\cf \ch}}
{\sqrt{\cb \cf \ch -u^4 
\left(\Pi_z^2 + \ch \, \Pi_x^2 \right)}}
\,.
\label{onshell1}
\ee

For a U-shaped string describing a bound quark-antiquark pair the turning point $\umax$ is determined in terms of the momenta by the condition that $x'(\umax)=z'(\umax)\to \infty$. This happens if $\umax=\uh$, in which case $\cf (\umax)=0$, or if 
\be
\left. \cb \cf \ch -u^4 \left(\Pi_z^2 + \ch \, \Pi_x^2 \right)
\right|_{\umax} =0 \,.
\label{together}
\ee
The first possibility is not physically relevant because the second possibility is always realized first, meaning that the string turns around at $\umax < \uh$, before reaching the horizon. The only exception is the case
$\Pi_x=\Pi_z=0$, but this corresponds to $x'=z'=0$, namely to an unbound pair of strings that descend from the boundary straight down to the horizon. 

The momenta are determined by the boundary conditions that require the string endpoints to lie a distance $\ell$ apart from each other: 
\be
\frac{\ell}{2} =  \int_0^{\umax} du \, x' = 
 \int_0^{\umax} du \, z' \,.
\ee
These two equations, together with \eqn{together}, can be solved numerically to express the momenta and $\umax$ in terms of $\ell$. In this way the on-shell action \eqn{onshell1} for a bound pair becomes a function of $\ell$ alone. In order to determine $L_s$ we subtract from this action the action of a static, unbound quark-antiquark pair, which is described by two straight strings hanging down from the boundary to the horizon. The action of this unbound pair is equal to \eqn{onshell1} with the momenta set to zero and the range of integration extended down to the horizon:
\be
S_\mt{unbound} =  -\frac{L^2}{2\pi\alpha'} \, 2 \, 
\int dt \int_0^{\uh} du \frac{\sqrt{\cb}}{u^2} 
\,.
\label{unbound}
\ee
We  obtain the screening length by numerically determining the value of $\ell$ at which the difference \mbox{$S(\ell)-S_\mt{unbound}$} crosses zero, since in the static case we always have $L_s<L_\mt{max}$. The result for this difference  as a function of $\ell$ in the isotropic plasma  \cite{static1,static2} described by eqn.~\eqn{iso} is plotted in Fig.~\ref{isopic}, from which we see that the screening length is
\be
L_\mt{iso} (T) \simeq \frac{0.24}{T} 
\qquad\qquad \mbox{[static dipole]}\,.
\label{staticT}
\ee
The scaling with the temperature is  expected on dimensional grounds. In the isotropic case the temperature and the entropy density are related simply through \eqn{siso}, so this result can be recast as
\be
L_\mt{iso} (s) \simeq 0.24 \left( \frac{\pi^2 \nc^2}{2s} \right)^{1/3}
\qquad\qquad \mbox{[static dipole]}\,,
\label{staticS}
\ee
which will be useful later.

\begin{figure}[tb]
\begin{center}
\begin{tabular}{cc}
\includegraphics[scale=0.75]{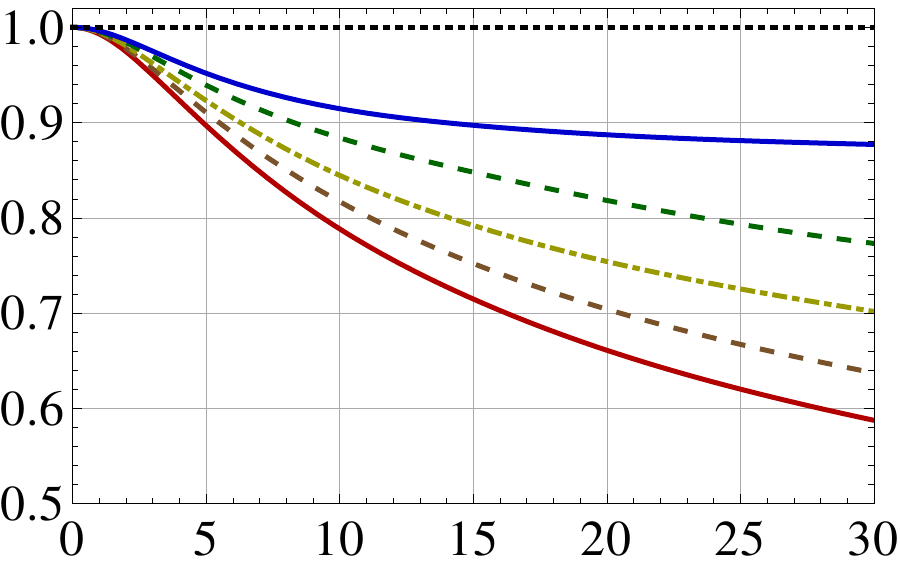}
\put(-109,-10){\small $a/T$}
\put(-215,40){\rotatebox{90}{$L_s/L_{\text{iso}}(T)$}}
\qquad
\includegraphics[scale=0.75]{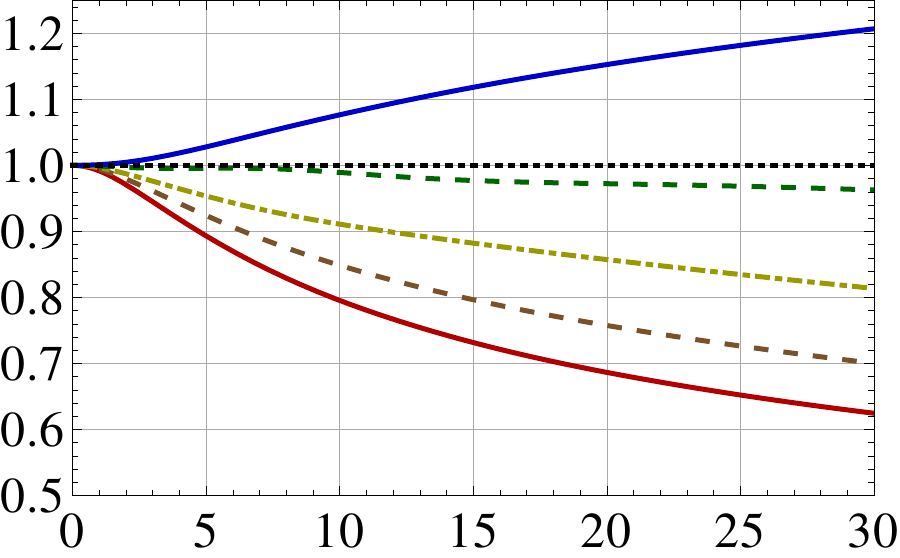}
\put(-109,-10){\small $a\nc^{2/3}/s^{1/3}$}
\put(-210,40){\rotatebox{90}{$L_s/L_{\text{iso}}(s)$}}
\end{tabular}
\caption{\small Screening length as a function of the anisotropy for a static quark-antiquark dipole lying at an angle with the $z$-direction (from top to bottom on the right-hand side of the plot) $\theta=\pi/2, \pi/3, \pi/4, \pi/6, 0$. The screening length is plotted in the appropriate units to facilitate comparison with the isotropic result for a plasma at the same temperature (left), or at the same entropy density (right). The isotropic result is given in eqs.~(\ref{staticT}) and (\ref{staticS}).
\label{staticscreening}
}
 \end{center} 
 \end{figure}

\begin{figure}[tb]
\begin{center}
\begin{tabular}{cc}
\includegraphics[scale=0.81]{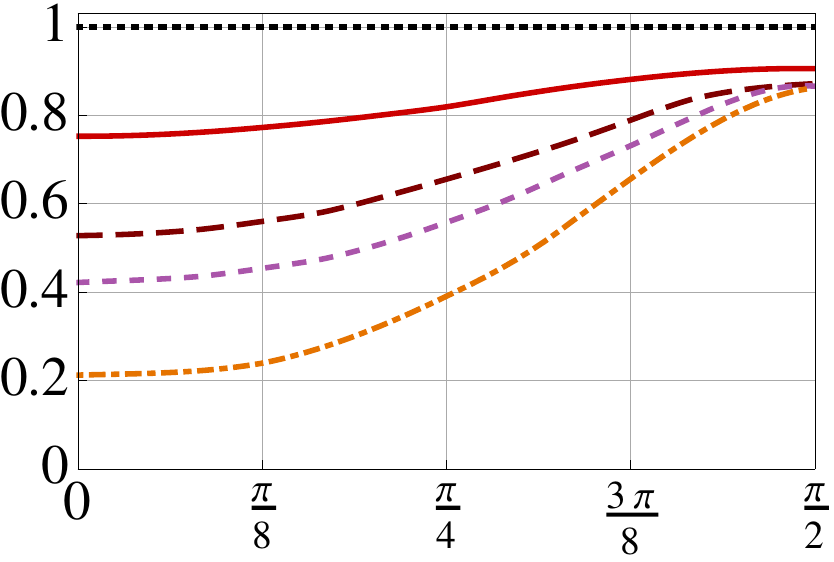}
\put(-100,-10){\small $\theta$}
\put(-215,40){\rotatebox{90}{$L_s/L_{\text{iso}}(T)$}}
\qquad
\includegraphics[scale=0.81]{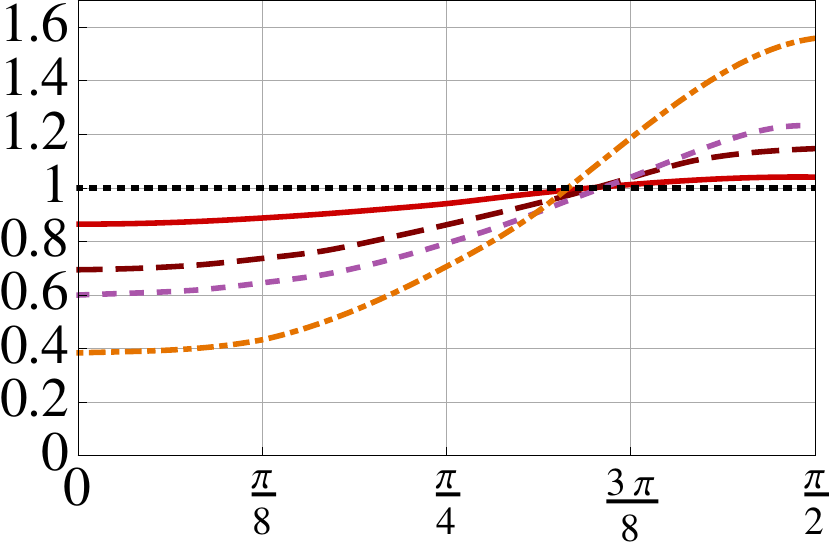}
\put(-100,-10){\small $\theta$}
\put(-210,40){\rotatebox{90}{$L_s/L_{\text{iso}}(s)$}}
\end{tabular}
\caption{Screening length for a quark-antiquark dipole lying at an angle $\theta$ with the $z$-direction for anisotropies $a/T=12.2$ (red, solid), 42.6 (maroon, coarsely dashed), 86 (violet, dashed), 744 (orange, dot-dashed). The corresponding values in units of the entropy density are (in the same order) $a\nc^{2/3}/s^{1/3}=6.2, 19, 35, 242$. The screening length is plotted in the appropriate units to facilitate comparison with the isotropic result for a plasma at the same temperature (left), or at the same entropy density (right). The isotropic result is given in eqs. (\ref{staticT}) and (\ref{staticS}).
\label{staticscreening1}
}
 \end{center} 
 \end{figure}

The results in the anisotropic case are plotted in 
Figs.~\ref{staticscreening} and \ref{staticscreening1}. 
 Fig.~\ref{staticscreening} shows the screening length, for several orientations of the dipole, as a function of the anisotropy measured in units of the temperature (left) and the entropy density (right). The reason for working with both normalizations is that we wish to compare the screening length in the anisotropic plasma to that in the isotropic plasma, and this can be done at least in two different ways: the two plasmas can be taken to have the same temperatures but different entropy densities, or the same entropy densities but different temperatures. Fig.~\ref{staticscreening1} shows the screening length as a function of the dipole orientation  for several values of the anisotropy. 
 
 We see from  Fig.~\ref{staticscreening}(left) that $L_s$ decreases monotonically as $a$ increases, for any dipole orientation, if the temperature is kept fixed. We also see from Fig.~\ref{staticscreening1}(left) that this effect is more pronounced for a dipole oriented along the anisotropic direction. In contrast, the behavior of the screening length  at constant entropy density depends on the dipole's  orientation, as shown in Figs.~\ref{staticscreening}(right) and \ref{staticscreening1}(right). For dipole's aligned sufficiently close to the anisotropic direction the screening length decreases with the anisotropy, whereas for orientations sufficiently close to the transverse plane the screening length increases with the anisotropy. 
 

\section{Dipole in an anisotropic plasma wind}
\label{screeninganiso}
In this section we will consider a static quark-antiquark pair in an anisotropic plasma that is moving with constant velocity with respect to the dipole --- a dipole in an `anisotropic plasma wind'. We will pay particular attention to the ultra-relativistic limit, which can be understood analytically.\footnote{We recall that we first send the quark mass to infinity and then $v\to 1$ (see Sec.~\ref{intro}).} This limit, together with the static results from Sec.~\ref{static}, will allow us to understand qualitatively the results at any velocity $0 < v< 1$. 

\begin{figure}[t!]
\begin{center}
\includegraphics[scale=0.5]{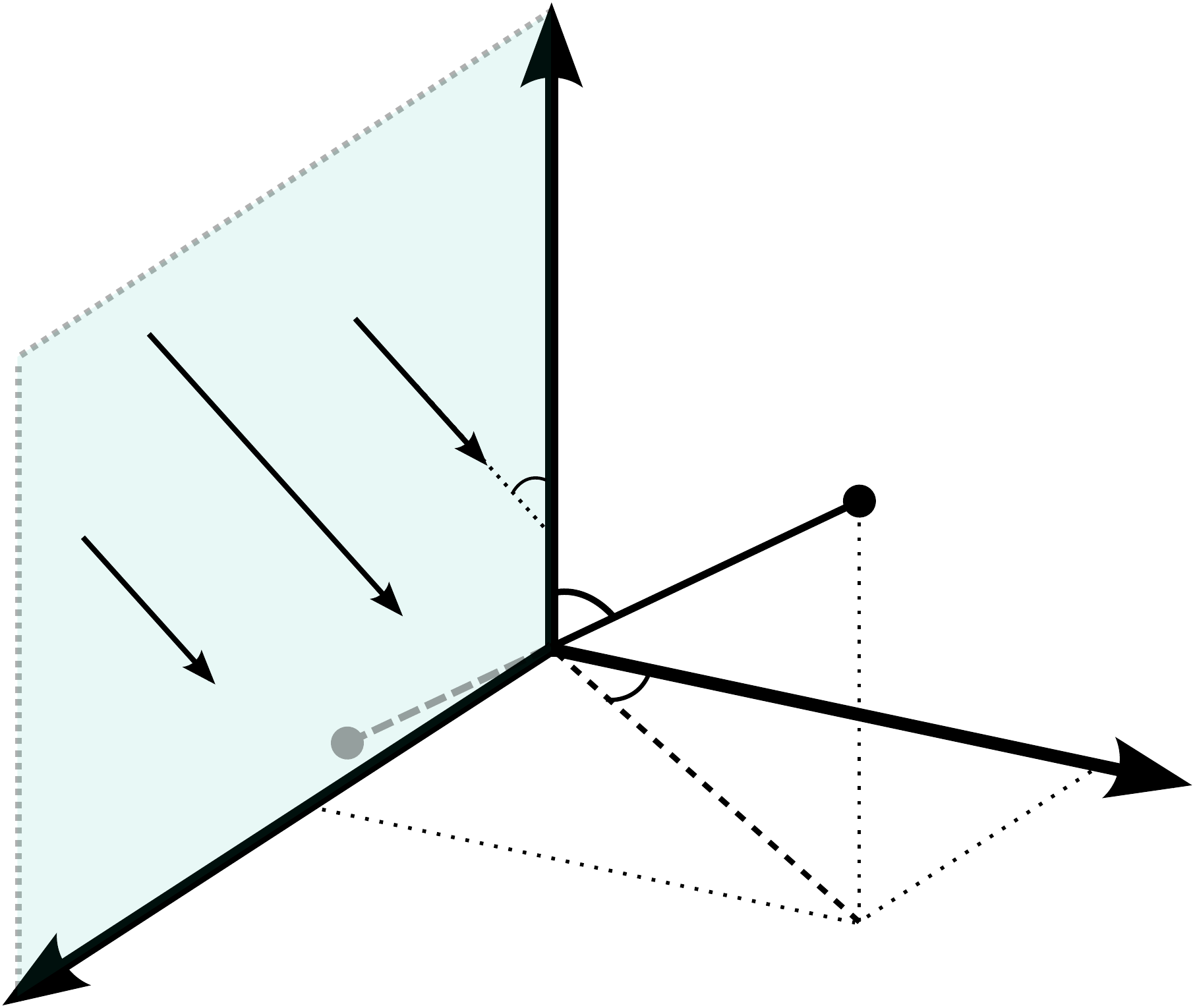}
 \begin{picture}(0,0)
   \put(-280,-5 ){$x$}
\put(0,45 ){$y$}
\put(-150,225 ){$z$}
    \put(-158,125){$\theta_v$}
    \put(-138,95){$\theta$}
    \put(-120,63){$\vp$}
 \end{picture}
\caption{Orientation of the dipole in an anisotropic plasma wind. The wind's velocity lies in the original $xz$-plane (before the boost \eqn{boost}) at an angle $\theta_v$ with respect to the anisotropic direction $z$. The quark lies at angles $\vec{q} = (x,y,z) = \frac{\ell}{2} 
(\sin\theta\sin\varphi,  \sin\theta\cos\varphi , \cos \theta)$ with respect to the relabeled directions (after the boost \eqn{boost}), and the antiquark lies at $-\vec q$.
\label{figure_angles}}
\end{center}
\end{figure}
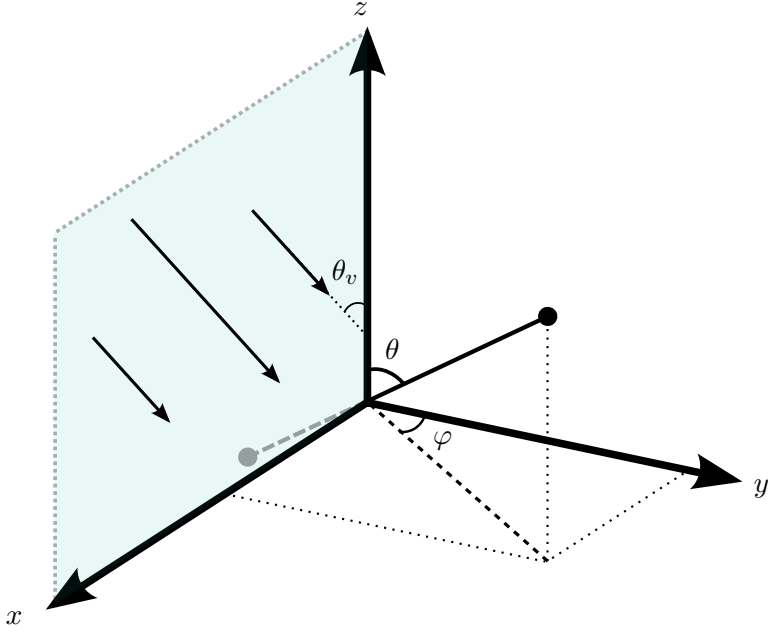

We will first rewrite the solution \eqn{sol2} in a boosted frame, and then place a dipole in it --- see Fig.~\ref{figure_angles}. 
Given the rotational symmetry in the $xy$-plane we assume that the boost velocity is contained in the $xz$-plane, and that it lies at an angle $\theta_v$ with the $z$-axis. Thus we first rotate to a new coordinate system defined through 
\bea
t &=& \tilde t \,, \nonumber \\
x&=&\tilde z\, \sin\theta_v+\tilde x \, \cos\theta_v\,, \nonumber\\
y &=& \tilde y \,, \nonumber \\
z&=&\tilde z \, \cos\theta_v - \tilde x\, \sin\theta_v\,, 
\label{rotation}
\eea
and then perform a boost along the $\tilde z$-direction by setting 
\bea
\tilde t &=& \gamma \left( t'-v\, z' \right) \,, \nonumber \\ 
\tilde x &=& x' \,, \nonumber \\
\tilde y &=& y' \,, \nonumber \\
\tilde z &=& \gamma \left( -v\, t'+z'  \right) \,,
\label{boost}
\eea
where $\gamma=1/\sqrt{1-v^2}$ is the usual Lorentz factor. Below we will consider a dipole with an arbitrary orientation with respect to both the velocity of the plasma and the anisotropic direction $z$  --- see Fig.~\ref{figure_angles}.  We parametrize the orientation of the dipole by two angles $\theta, \vp$ so that the quark lies at
\be
\vec{q} = (x',y',z') = \frac{\ell}{2} 
(\sin\theta\sin\varphi,  \sin\theta\cos\varphi , \cos \theta)
\ee
and the antiquark lies at $-\vec{q}$. 

For notational simplicity, below we will  drop the primes in the final set of coordinates. To avoid confusion, we emphasize that the direction $\theta_v$ of the plasma wind is always measured with respect to the original $(x,y,z)$ axes, i.e.~before the rotation and the boost above. In particular, motion within (outside) the transverse plane refers to a dipole in a plasma wind with $\theta_v=\pi/2$ ($\theta_v \neq \pi/2$). In contrast,  the orientation of the dipole is measured with respect to the final set of coordinates $(x',y',z')$. However, if instead of specifying the dipole's orientation through a pair $(\theta, \vp)$ we specify it by saying that the dipole is aligned with the $x$-, $y$- or $z$-directions, then we are referring to the original directions. Just as an illustration, consider the case of a plasma wind blowing along the original $x$-direction, i.e.~a plasma wind with $\theta_v=\pi/2$.  Then we see from \eqn{rotation} and \eqn{boost} that $(x,z) \sim (z', x')$. Thus in this case by `a dipole oriented along the $x$-direction' we mean a dipole  with $\theta=0$.

After dropping the primes from the final set of coordinates in \eqn{boost} the five-dimensional part of the metric \eqn{sol2} takes the form
\bea
ds^2&=&\frac{L^2}{u^2}\left(-g_{tt}dt^2+g_{xx}dx^2+dy^2+g_{zz}dz^2+g_{tx}dt\,dx+g_{tz}dt\, dz+g_{xz}dx\, dz+\frac{du^2}{\cf}\right)\,,
\cr &&
\label{metricrotatedboosted}
\eea
where
\bea
g_{tt}&=&\frac{\cb\cf-v^2(\sin^2\theta_v+\ch \cos^2\theta_v)}{1-v^2} 
\,, \\[2mm]
g_{xx}&=&\cos^2\theta_v+\ch \sin^2\theta_v\,,\\[2mm]
g_{zz}&=&\frac{\sin^2\theta_v+\ch \cos^2\theta_v-v^2 \cb\cf}{1-v^2}\,,\\[2mm]
g_{tx}&=&\frac{(\ch-1)v}{\sqrt{1-v^2}}\sin(2\theta_v)\,,\\[2mm]
g_{tz}&=&\frac{2v(\cb\cf-\sin^2\theta_v-\ch\cos^2\theta_v)}{1-v^2}\,,\\[2mm]
g_{xz}&=&\frac{1-\ch}{\sqrt{1-v^2}}\sin(2\theta_v)\,.
\label{below}
\eea

In order to determine the screening length for a generic velocity we need to compare the actions of a bound and an unbound quark-antiquark pair, as in the static case of Sec.~\ref{static}. However, in the ultra-relativistic this is not strictly necessary because $L_s=L_\mt{max}$ (see Sec.~\ref{preli}). In other words, in this limit we only need to determine the maximum possible quark-antiquark separation for which a bound state exists. Nevertheless, for completeness we will briefly present  the analysis of the unbound configuration. Each of the strings in the unbound pair is one of the trailing strings studied in \cite{cfmt}, so the reader is referred to this reference for additional details. Note, however, that \cite{cfmt} worked in the plasma rest frame. Here we will work in the dipole's rest and focus on the ultra-relativistic limit. 


\subsection{Unbound quark-antiquark pair}
As in Sec.~\ref{static} we fix the static gauge $t=\tau$, $\sigma=u$, and specify the embedding of the unbound string as 
\be
x \to x(u) \sac z \to z(u) \,.
\ee
The embedding in the $y$-direction is simply $y=0$ because of rotational symmetry in the $xy$-plane and because the string is unbound. As we will see below, in the case of a bound string (dipole) the boundary conditions will generically imply a non-trivial embedding $y(u)$. 

The action for the unbound string reads 
\be
S_\mt{unbound}=
- \frac{L^2}{2\pi \alpha'} \, 2 \int dt \int_0^{\uh} du \, \frac{1}{u^2}\sqrt{\cf^{-1} K_0+K_{xx}x'^2+K_{zz}z'^2+K_{xz}x'z'}\,,
\ee
where
\bea
K_0 &=&
g_{tt}\,,\nonumber \\[2mm]
K_{xx} &=& \frac{\cb \cf(\cos^2\theta_v+\ch\sin^2\theta_v)-\ch v^2}{1-v^2} \,,\nonumber\\[2mm]
K_{zz} &=& \cb\cf \left(\sin^2 \theta_v +\ch \cos^2\theta_v\right)
 \,,\nonumber\\[2mm]
K_{xz} &=& \frac{\cb\cf(1-\ch)}{\sqrt{1-v^2}}  \sin(2 \theta_v)\,.
\label{K}
\eea
Introducing the conjugate momenta
\bea
\Pi_x=\frac{\partial \cl_\mt{unbound}}{\partial x'}\,,\qquad
\Pi_z=\frac{\partial \cl_\mt{unbound}}{\partial z'}
\eea
and solving for $x', z'$ we find
\be
x'=\frac{u^2}{\cf \sqrt{\cb\ch}} \frac{N_x}{\sqrt{D}} \sac
z'=\frac{u^2}{\cf \sqrt{\cb\ch}} \frac{N_z}{\sqrt{D}} \,,
\label{xprimezprime}
\ee
where
\bea
N_x &=& K_{zz} \Pi_x  -\frac{1}{2} K_{xz}\Pi_z \,, \nonumber \\
N_z &=& -\frac{1}{2}K_{xz} \Pi_x  + K_{xx} \Pi_z\,, \nonumber \\[1mm]
D &=& \cb\ch\cf K_0-u^4\left(K_{zz}\Pi_x^2+K_{xx}\Pi_z^2-K_{xz}\Pi_x\Pi_z\right) \,.
\eea
Substituting into the action we arrive at
\be
S_\mt{unbound}=
- \frac{L^2}{2\pi \alpha'} \, 2 \int dt \int_0^{\uh} du \, \frac{\sqrt{\cb\ch} K_0}{u^2\sqrt{D}} \,.
\label{Sdrag}
\ee
The momenta are determined by the condition that \eqn{xprimezprime} remain real for a string that extends all the way from the boundary to the horizon. Following  \cite{cfmt} we analyze this condition by noting that $D$ can be rewritten as 
\be
D = \frac{2 u^4}{K_{xz}} N_x N_z - 
b \Big[ \Pi_x \Pi_z - c \Big]
\Big[ \cb\cf-v^2(\sin^2\theta_v+\ch \cos^2\theta_v) \Big]
\label{Dnice}
\ee
where
\be
b= \frac{\ch u^4}{(1-\ch) \sqrt{1-v^2} \,\sin \theta_v \cos \theta_v}\sac
c= \frac{\cb \cf (1-\ch)\sin \theta_v \cos \theta_v}{u^4 \sqrt{1-v^2}} \,.
\ee
As in \cite{cfmt} we must require that the zeros of the second summand in \eqn{Dnice} coincide with one another and with those of $N_x$ and $N_z$. One of the zeros of the second summand occurs at a critical value $u=u_c$ such that 
\be 
 \cb_c \, \cf_c -d_c  \, v^2 = 0 \sac 
 d_c \equiv  \ch_c \cos^2 \theta_v + \sin^2 \theta_v \,,
 \label{hc}
\ee
where $\cb_c=\cb(u_c)$, etc. At this point we have
\bea
\left. N_x N_z \right|_{u_c}=&& \frac{v^4\cos\theta_v \sin\theta_v}{\sqrt{1-v^2}}(\ch_c-1)\, d_c 
 \Big[ d_c \Pi_x+ 
\frac{(\ch_c-1)\cos\theta_v \sin\theta_v}{\sqrt{1-v^2}}\, \Pi_z \Big]^2 \,.
\eea
Noting that $\ch_c>1$ and that $K_{xz}<0$, we see that $D$ would be negative at $u_c$ unless the momenta are related through
\be
\Pi_x = \frac{(1-\ch_c)\cos\theta_v \sin\theta_v}
{ d_c \sqrt{1-v^2}}\, \Pi_z \,.
\ee
Assuming  this relation and requiring that the other zero in the second summand of \eqn{Dnice} coincide with $u_c$ yields 
\be
\Pi_z^2 = \frac{\cb_c \cf_c \,d_c}{u_c^4} \sac 
\Pi_x^2 = \frac{\cb_c \cf_c (\ch_c -1)^2 \cos^2 \theta_v \sin^2 \theta_v}{u_c^4 (1-v^2) d_c} \,.
\label{result}
\ee
Note that $\Pi_z$ does not vanish for any value of $\theta_v$, whereas $\Pi_x$ vanishes  if $\theta_v=0,\pi/2$. The reason is that for these two particular orientations the plasma wind blows along the original 
$z$- or $x$-directions and the string orients itself with the corresponding axis \cite{cfmt}. As a consequence, the momentum along the orthogonal axis vanishes. However, the changes of coordinates \eqn{rotation} and \eqn{boost} always relabel the direction of motion as $z$, so after these changes the non-vanishing momentum is  labelled $\Pi_z$ irrespectively of whether $\theta_v=0$ or 
$\theta_v=\pi/2$.

We will analyze in detail the ultra-relativistic limit. This is facilitated by explicitly distinguishing the case of motion outside the transverse plane ($\theta_v\neq \pi/2$) and motion within the transverse plane ($\theta_v=\pi/2$).


\subsubsection{Ultra-relativistic motion outside the transverse plane}
In the ultra-relativistic limit $u_c$ approaches the boundary, i.e.~$u_c\to 0$, and we can use the near-boundary expansion (\ref{expansion}) to determine it. The condition \eqn{hc} yields in this limit \cite{cfmt}
\be
u_\mt{c}^2\simeq \frac{4(1-v^2)}{a^2 \cos^2 \theta_v}  \qquad\qquad  [\theta_v\neq \pi/2]\,,
\label{near1}
\ee
which when substituted in \eqn{result} gives the momenta
\be
\Pi_z^2 \simeq \frac{a^4 \cos^4 \theta_v}{16 (1-v^2)^2} \sac 
 \Pi_x^2 \simeq \frac{a^4 \cos^2 \theta_v \sin^2 \theta_v}
{16 (1-v^2)} \,.
\label{near2}
\ee
In these expressions we have ignored subleading terms in an expansion in $1-v^2$, for example we have set $v\simeq 1$, $\ch_c\simeq 1$, etc. Note that in this expansion $\Pi_x$ is subleading with respect to $\Pi_z$.

For later use we must evaluate how $S_\mt{unbound}$ scales with $1-v^2$ in the limit $v\to 1$. For this purpose we split the integration region, and hence the action \eqn{Sdrag}, as
\be
S_\mt{unbound}=S_\mt{unbound}^{(1)}+S_\mt{unbound}^{(2)}\,,
\ee
where $S_\mt{unbound}^{(1)}$ is the action with the integral in $u$ ranging between $0$ and $u_c$, and $S_\mt{unbound}^{(2)}$ is the action with the integral in $u$ ranging between $u_c$ and $u_\mt{H}$. The reason for this separation is that in the first interval $u$ is small and hence we will be able to use the near-boundary expressions (\ref{expansion}), \eqn{near1} and \eqn{near2}. In order to exhibit the dependence on $1-v^2$ of 
$S_\mt{unbound}^{(1)}$ explicitly, it is convenient to work with a rescaled variable $r$ which remains finite in the $v\to 1$ limit, defined though   
\be
u=r\sqrt{1-v^2} \sac u_c=r_c \sqrt{1-v^2}  \,.
\label{uonehalf}
\ee
In terms of this variable we get 
\be
S_\mt{unbound}^{(1)}=
- \frac{L^2}{2\pi \alpha'} \frac{2}{\sqrt{1-v^2}}\int dt \int_0^{r_c} dr \, \frac{1-\frac{1}{4}a^2r^2 \cos^2\theta_v+\ldots}
{r^2\sqrt{1-\frac{1}{4}a^2r^2 \cos^2\theta_v
-\frac{1}{16} a^4 r^4  \cos^4\theta_v +\dots}}\,.
\label{cancels}
\ee
The divergence near $r=0$ will cancel out with that in the action for the bound string. The integrand is smooth across $r=r_c$.  The crucial point is that the result is $O\left[(1-v^2)^{-1/2}\right]$ in the counting in powers of $1-v^2$, and we will find this same scaling in the bound string action (see below). In contrast, $S_\mt{unbound}^{(2)}$ scales as $1-v^2$ in the ultra-relativistic limit. The reason is that $u$ is not small in units of $1-v^2$ in the corresponding region of integration, so all the dependence comes from the fact that the action \eqn{Sdrag} scales as $1/\Pi_z  \sim 1-v^2$ in this region.


\subsubsection{Ultra-relativistic motion within the transverse plane}

In this case $\theta_v=\pi/2$ and hence we see from \eqn{result} that 
$\Pi_x=0$.  The condition \eqn{hc} now gives \cite{cfmt}
\be
u_c^2\simeq \sqrt{\frac{1-v^2}{C}}\,,
\label{uu}
\ee
where 
\be
C= \frac{121}{576}a^4-\cf_4-\cb_4 \,,
\label{CCC}
\ee
and we recall that $\cf_4, \cb_4$ are the coefficients that enter the near-boundary expansion \eqn{expansion}. Substituting \eqn{uu} into \eqn{result} and dropping subleading terms as before we obtain the momentum in the $z$-direction (recall that this corresponds to the original $x$-direction): 
\be
\Pi_z\simeq\frac{1}{u_c^2}=\sqrt{\frac{C}{1-v^2}}\,.
\ee
It is now convenient to work with a rescaled radial coordinate $r$ defined through
\be
u=r(1-v^2)^{1/4}\,.
\ee
Splitting the unbound string action as before, we find
\be
S_\mt{unbound}^{(1)}=
- \frac{L^2}{2\pi \alpha'} \frac{2}{(1-v^2)^{1/4}}\int dt \int_0^{r_c} dr \, \frac{1-C r^4+\ldots}{r^2\sqrt{1-2Cr^4+\dots}}\,.
\label{cancel2}
\ee
Again, the divergence near $r=0$ will cancel out with that in the action for the bound string, which will also be of $O\left[(1-v^2)^{-1/4}\right]$ in the counting in powers of $1-v^2$ (see below). In contrast, $S_\mt{unbound}^{(2)}$  scales as $1/\Pi_x\sim\sqrt{1-v^2}$ in the ultra-relativistic limit, and is therefore subleading.

In summary, we find that in the ultra-relativistic limit
\bea
S_\mt{unbound}=\left\{
\begin{array}{l l}
O\left[ (1-v^2)^{-1/2}\right] \,\, \mbox{if \,} \, \theta_v \neq \pi/2  & ~~~~~~ \mbox{[outside the transverse plane]}\\ & \\
 O\left[(1-v^2)^{-1/4}\right]  \,\, \mbox{if \,} \, \theta_v = \pi/2     & ~~~~~~  \mbox{[within the transverse plane]}\,.
 \end{array}\right.
 \label{result-action}
\eea


\subsection{Bound quark-antiquark pair}
We now consider a dipole with an arbitrary orientation with respect to both the velocity of the plasma and the anisotropic direction $z$  --- see Fig.~\ref{figure_angles}.  As before we fix the static gauge 
$\tau=t, \sigma=u$ and specify the string embedding via three functions $(x(u), y(u), z(u))$ subject to the boundary conditions
\bea
\frac{\ell}{2} \sin\theta\sin\varphi &=& \int_0^{u_\mt{max}} x'du 
\,,\nonumber \\
\frac{\ell}{2} \sin\theta\cos\varphi &=& \int_0^{u_\mt{max}} y'du 
\,,\nonumber \\
\frac{\ell}{2} \cos\theta &=& \int_0^{u_\mt{max}} z'du \,,
\label{bbcc}
\eea
where  $\umax$  is the turning point of the U-shaped string. 
The integral in the action of the bound string  extends only up to this point and now includes a term proportional to $y'^2$:
\bea
S &=& - \frac{L^2}{2\pi \alpha'} \, 2 \int dt \int_0^{\umax} du \, 
\frac{1}{u^2}\sqrt{\cf^{-1}K_0 + K_{xx} x'^2 + K_{yy} y'^2 + 
  K_{zz} z'^2 +  K_{xz}  x'  z'}\,.\cr &&
\label{action-bound}
\eea
All the $K$'s were defined in \eqn{K} except for $K_{yy}$, which is given by
\bea
K_{yy} &=&\frac{\cb \cf - v^2( \sin^2 \theta_v + 
\ch \cos^2 \theta_v )}{1-v^2} \,.
\label{Kyy}
\eea
The momenta are defined as 
\bea
\Pi_x=\frac{\partial \cl}{\partial x'}\,,\qquad
\Pi_y=\frac{\partial \cl}{\partial y'}\,,\qquad
\Pi_z=\frac{\partial \cl}{\partial z'}\,.
\eea
Inverting these equations we get
\bea 
x'&=&
\frac{u^2}{\cf \sqrt{\cb\ch}\sqrt{D}} \left(K_{zz} \Pi_x  -\frac{1}{2} K_{xz}\Pi_z \right)\,, \nonumber \\[2mm]
y'&=&
\frac{u^2\sqrt{\cb \ch}}{\sqrt{D}} \Pi_y \,, \nonumber \\
z'&=& 
\frac{u^2}
{\cf \sqrt{\cb \ch}\sqrt{D}} \left(-\frac{1}{2}K_{xz} \Pi_x  + K_{xx} \Pi_z  \right)\,,
\label{xp}
\eea
where
\be
D =  \cb \ch \cf  K_0 - u^4 \left(K_{zz}\, \Pi_x^2 +\cb \cf \ch\,   \Pi_y^2 +
K_{xx} \,\Pi_z^2 -K_{xz}\, \Pi_x \Pi_z \right) \,.
\label{D}
\ee
Substituting these expressions into the action (\ref{action-bound}) we get
\be
S = - \frac{L^2}{2\pi \alpha'} \, 2 \int dt \int_0^{\umax} du \, 
\frac{\sqrt{\cb\ch}K_0}{u^2\sqrt{D}}\,.
\label{action-bound-2}
\ee
As in the case of the unbound string, we will now distinguish between the cases of motion outside and within the transverse plane, focusing on the ultra-relativistic limit. 

\subsubsection{Ultra-relativistic motion outside the transverse plane}
\label{outside}
 The turn-around point $\umax$ is defined by the condition $D(\umax)=0$. In the ultra-relativistic limit we expect that this point approaches the boundary for the string solution of interest, as in the isotropic case. Thus in this limit $\umax$ can be determined by using the near-boundary expansions of the metric functions (\ref{expansion}).

In the limit $u\to 0$ we find the following expansions:
\bea
K_{zz}&\simeq& 1+\frac{a^2u^2 \cos^2\theta_v}{4}+ \cdots \,,\\[1mm]
K_{xz}&\simeq&0-\frac{a^2u^2 \sin \theta_v \cos \theta_v}
{2\sqrt{1-v^2}} + \cdots 
\,,\\[1mm]
K_{xx}&\simeq&1-\frac{a^2u^2 \cos^2 \theta_v}{4(1-v^2)} + \cdots 
\,,
\eea
from which it follows that 
\be
D\simeq1- \frac{a^2u^2 \cos^2\theta_v}{4(1-v^2)}
-u^4(\Pi_x^2+\Pi_y^2+\Pi_z^2) + \cdots \,.
\label{denomin}
\ee
Similarly, the boundary conditions \eqn{bbcc} take the form
\bea
\frac{\ell}{2}\sin\theta\sin\vp&\simeq&\int_0^{u_\mt{max}}du\, 
\frac{u^2}{\sqrt{D}}\, \Pi_x
+ \cdots \,,  \\[2mm]
\frac{\ell}{2}\sin\theta\cos\vp&\simeq&\int_0^{u_\mt{max}}du\, 
\frac{u^2}{\sqrt{D}}\, 
 \Pi_y + \cdots \,, \nonumber \\[2mm]
\frac{\ell}{2}\cos\theta&\simeq&\int_0^{u_\mt{max}}du\,
\frac{u^2}{\sqrt{D}}
\left(1-\frac{a^2u^2 \cos^2\theta_v }{4(1-v^2)}\right)\Pi_z
+ \cdots \,, \nonumber
\label{bc1}
\eea

\vspace{1mm}
\noindent
In the ultra-relativistic limit, all the terms that we have omitted in the equations above, in particular in \eqn{denomin} and \eqn{bc1}, are subleading  with respect to the terms that we have retained
provided the radial coordinate and the momenta scale as
\bea
u=r\sqrt{1-v^2}\,,\qquad \Pi_i=\frac{p_i}{1-v^2}\,,
\label{scaling1}
\eea
where $r$ and $p_i$ are kept fixed in the limit $v \to 1$. In terms of these rescaled variables \eqn{scaling1} the boundary conditions \eqn{bc1} take the form
\bea
\frac{\ell}{2}\sin\theta\sin\vp&\simeq&
\sqrt{1-v^2}\, p_x\, {\cal I}_2(p,\theta_v)\,,\nonumber \\[1mm]
\frac{\ell}{2}\sin\theta\cos\vp&\simeq&
\sqrt{1-v^2}\, p_y\, {\cal I}_2(p,\theta_v)\,, \nonumber \\[1mm]
\frac{\ell}{2}\cos\theta&\simeq&
\sqrt{1-v^2}\, p_z\left({\cal I}_2(p,\theta_v)-\frac{a^2\cos^2\theta_v}{4}{\cal I}_4(p,\theta_v)\right)\,,
\label{bc1scaled}
\eea
where the integral
\be
{\cal I}_n(p,\theta_v)\equiv \int_0^{r_\mt{max}}dr\frac{r^n}{\sqrt{1-\frac{a^2r^2}{4}\cos^2\theta_v-r^4(p_x^2+p_y^2+p_z^2)}}
\label{integral1}
\ee 
is of $O(1)$ in the counting in powers in $(1-v^2)$, and is finite if $n \geq0$. Further noting that  
\be
K_0 = 1-\frac{a^2 u^2 \cos^2\theta_v}{4(1-v^2)} +O(u^4)
\simeq 1-\frac{a^2 r^2 \cos^2\theta_v}{4}  
\,,
\ee
we see that the bound action scales as
\bea
S &\simeq& - \frac{L^2}{2\pi \alpha'} \frac{2}{\sqrt{1-v^2}}\left({\cal I}_{-2}(p,\theta_v)-\frac{a^2\cos^2\theta_v}{4}{\cal I}_0(p,\theta_v)\right)\int dt\,.
\label{action-bound-exp}
\eea
Since both this bound action and the unbound action \eqn{cancels} scale as $(1-v^2)^{-1/2}$, the divergence at $r=0$ in the bound action coming from the  ${\cal I}_{-2}(p,\theta_v)$ integral would exactly cancel that in the unbound action in the difference \eqn{criterion}. Moreover, by comparing the two actions we would conclude that the momenta $p_i$ introduced in \eqn{scaling1} are indeed of $O(1)$ in the counting in powers of $(1-v^2)$ in the ultra-relativistic limit. It would then follow that the integrals 
${\cal I}_n(p,\theta_v)$ are also of $O(1)$, and therefore that the screening length scales as $L_s \sim (1-v^2)^{1/2}$ in the ultra-relativistic limit. However, as explained below \eqn{below}, in the ultra-relativistic $L_s = L_\mt{max}$ is simply the maximum possible separation between a bound quark-antiquark pair, so it can be determined by maximizing $\ell$ in \eqn{bc1scaled} with respect to the momenta. Since the integrals are bounded from above for any value of the $p_i$, and the maximum is $v$-independent, it follows that 
$L_s=L\mt{max}\sim (1-v^2)^{1/2}$.

\subsubsection{Ultra-relativistic motion within the transverse plane}
\label{within}
In this case $\theta_v=\pi/2$ and the expansions of $D$ and of the boundary conditions \eqn{bbcc} become
\be
D \simeq 1-\frac{C u^4}{1-v^2} -u^4(\Pi_x^2+\Pi_y^2+\Pi_z^2) +
\cdots 
\ee
and
\bea
\frac{\ell}{2}\sin\theta\sin\vp&\simeq&\int_0^{u_\mt{max}}du\, u^2\frac{\Pi_x}
{\sqrt{1-\frac{C u^4}{1-v^2} -u^4(\Pi_x^2+\Pi_y^2+\Pi_z^2)}}
+\cdots \,,\nonumber \\[1mm]
\frac{\ell}{2}\sin\theta\cos\vp&\simeq&\int_0^{u_\mt{max}}du\, u^2\frac{\Pi_y}
{\sqrt{1-\frac{C u^4}{1-v^2} -u^4(\Pi_x^2+\Pi_y^2+\Pi_z^2)}}
+\cdots \,,\nonumber \\[1mm]
\frac{\ell}{2}\cos\theta&\simeq&\int_0^{u_\mt{max}}du\, u^2\frac{
\left(1-\frac{C u^4}{1-v^2} \right)\Pi_z}
{\sqrt{1-\frac{C u^4}{1-v^2} -u^4(\Pi_x^2+\Pi_y^2+\Pi_z^2)}}
+\cdots \,, \nonumber 
\label{bc3}
\eea
where $C$ was defined in \eqn{CCC}. As in the previous section, in the ultra-relativistic limit all the terms that we have omitted in the equations above are subleading  with respect to the terms that we have retained provided the radial coordinate and the momenta scale in this case as
\bea
u=r(1-v^2)^{1/4} \,,\qquad \Pi_i=\frac{p_i}{\sqrt{1-v^2}} \,,
\label{scaling2}
\eea
where $r$ and $p_i$ are kept fixed in the limit $v \to 1$. In terms of the rescaled variables the boundary conditions \eqn{bc3} become
\bea
\frac{\ell}{2}\sin\theta\sin\vp&\simeq&
(1-v^2)^{1/4}\, p_x\, {\cal J}_2(p)
\,,\nonumber \\[1mm]
\frac{\ell}{2}\sin\theta\cos\vp&\simeq&
(1-v^2)^{1/4}\, p_y\, {\cal J}_2(p)
\,,\nonumber \\[1mm]
\frac{\ell}{2}\cos\theta&\simeq&
(1-v^2)^{1/4}\, p_z\,\left( {\cal J}_2(p)-C{\cal J}_6(p)\right)\,,
\label{bc3scaled}
\eea
where the integral
\be
{\cal J}_n(p)=\int_0^{r_\mt{max}}dr\, \frac{r^n}{\sqrt{1-r^4(C+p_x^2+p_y^2+p_z^2)}}
\ee
is of $O(1)$ in the counting in powers in $(1-v^2)$, and is finite if $n \geq 0$. Further noting that 
\be
K_0 = 1-\frac{C}{1-v^2}u^4+O(u^6) \simeq 1-C r^4 \,,
\ee
we see that the bound  action becomes
\be
S \simeq - \frac{L^2}{2\pi \alpha'} \frac{2}{(1-v^2)^{1/4}} \Big(
{\cal J}_{-2}(p)-C{\cal J}_2(p)\Big)\int dt \,.
\label{action-bound-exp-1}
\ee
Since both this bound action and the unbound action \eqn{cancel2} scale as $(1-v^2)^{-1/4}$, the divergence at $r=0$ in the bound action coming from the  ${\cal J}_{-2}(p)$ integral would exactly cancel that in the unbound action in the difference \eqn{criterion}. Moreover, by comparing the two actions we would conclude that the momenta $p_i$ introduced in \eqn{scaling2} are indeed of $O(1)$ in the counting in powers of $(1-v^2)$ in the ultra-relativistic limit. It would then follow that the integrals 
${\cal J}_n(p)$ are also of $O(1)$, and therefore that the screening length scales as $L_s \sim (1-v^2)^{1/4}$ in the ultra-relativistic limit. However, as explained below \eqn{below}, in the ultra-relativistic $L_s = L_\mt{max}$ is simply the maximum possible separation between a bound quark-antiquark pair, so it can be determined by maximizing $\ell$ in \eqn{bc3scaled} with respect to the momenta. Since the integrals are bounded from above for any value of the $p_i$, and the maximum is 
$v$-independent, it follows that $L_s=L\mt{max}\sim (1-v^2)^{1/4}$.
 
In summary, we conclude that in the dipole rest frame the screening length scales in the ultra-relativistic limit as
\vskip -7mm
\bea
L_s\sim\left\{
\begin{array}{l l}
 (1-v^2)^{1/2}  \,\, \mbox{if \,} \, \theta_v \neq \pi/2  & ~~~~~~ \mbox{[motion outside the transverse plane]} \\ & \\
 (1-v^2)^{1/4}   \,\, \mbox{if \,} \, \theta_v = \pi/2  & ~~~~~~ \mbox{[motion within the transverse plane]} 
 \end{array}\right.
 \label{result-scaling}
\eea
irrespectively of the dipole orientation.

\subsection{Isotropic limit}
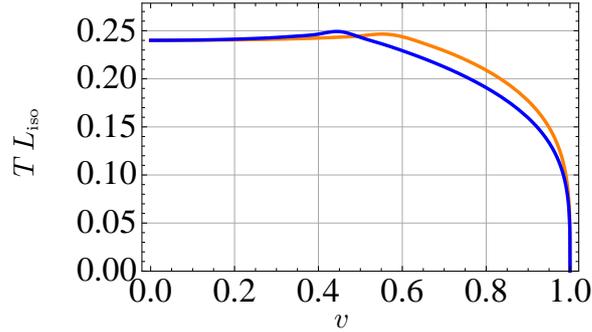
\begin{figure}[tb]
\begin{center}
\includegraphics[scale=0.80]{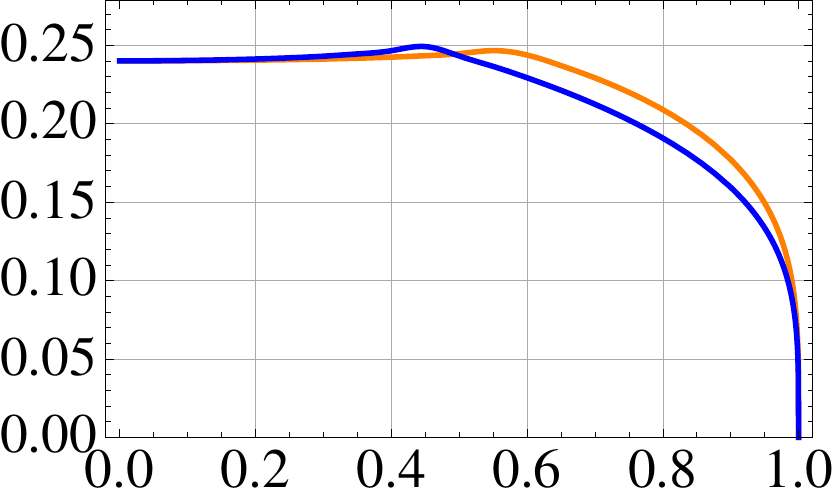}
 \begin{picture}(0,0)
   \put(-100,-7){$v$}
      \put(-220,50){\rotatebox{90}{$T\,L_\mt{iso} $}}
 \end{picture}
\caption{\small  Screening length for a dipole moving through an isotropic plasma in a direction orthogonal (top, blue curve) or parallel (bottom, orange curve) to its orientation. 
\label{lviso}
}
 \end{center} 
 \end{figure}

The results above reduce to the isotropic result of Ref.~\cite{liu1,liu2} in the limit $a\to0$. This limit is most easily recovered from the results for motion within the transverse plane, since some of the terms in the expansions in Section \ref{outside} vanish if $a=0$, thus invalidating the analysis. In contrast, setting $a=0$ in Section \ref{within} boils down to simply setting $C$ to its isotropic value, which from \eqn{CCC} and \eqn{iso} is
\be
C = - \cf_4 = \frac{1}{\uh^4} = \pi^4 T^4 \,.
\ee
Since the value of $C$ does not affect the ultra-relativistic scaling of the screening length, we recover the scaling 
\be
L_\mt{iso} \sim (1-v^2)^{1/4} ~~~~~~ 
\mbox{[isotropic plasma]}
\label{isoscaling}
\ee
found in the isotropic case by the authors of \cite{liu1,liu2}. As in the anisotropic case, the ultra-relativistic scaling of the screening length is independent of the dipole's orientation. In fact, even for $v<1$, the isotropic screening length depends only mildly on the dipole's orientation, as shown in Fig.~\ref{lviso}. 

\subsection{Numerical results for generic velocities}
\label{genericvelocities}
Away from the ultra-relativistic limit the screening length must be obtained numerically. For this reason we have focused on a few representative cases, namely those in which both the direction of the plasma wind and the dipole's orientation are aligned with one of the original $x$, $y$, or $z$ axes. Given the rotational symmetry in the $xy$-plane, there are only five inequivalent cases to consider, because if the wind `blows' in the $z$-direction then orienting the dipole along $x$ or $y$ gives identical physics. In each case, we plot the screening length both as a function of the velocity $v$ for different degrees of anisotropy $a$, and also as a function of the degree of anisotropy for different values of the velocity. In each case the result can be qualitatively understood combining the static results from Sec.~\ref{static} and the ultra-relativistic behavior derived analytically in Section \ref{screeninganiso}. We recall that in all cases below, by `a dipole oriented along $x$, $y$ or $z$' we are referring to the original directions before the rotation \eqn{rotation} and the boost \eqn{boost}.

\begin{figure}[t!!!]
\begin{center}
\begin{tabular}{cc}
\includegraphics[scale=0.75]{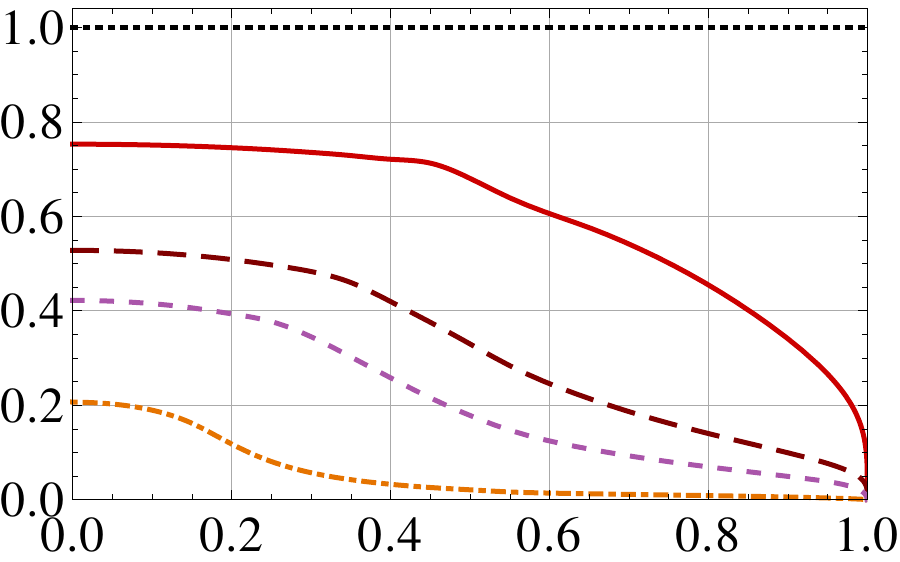}
\put(-95,-10){\small $v$}
\put(-215,40){\rotatebox{90}{$L_s/L_{\text{iso}}(T)$}}
\qquad
\includegraphics[scale=0.75]{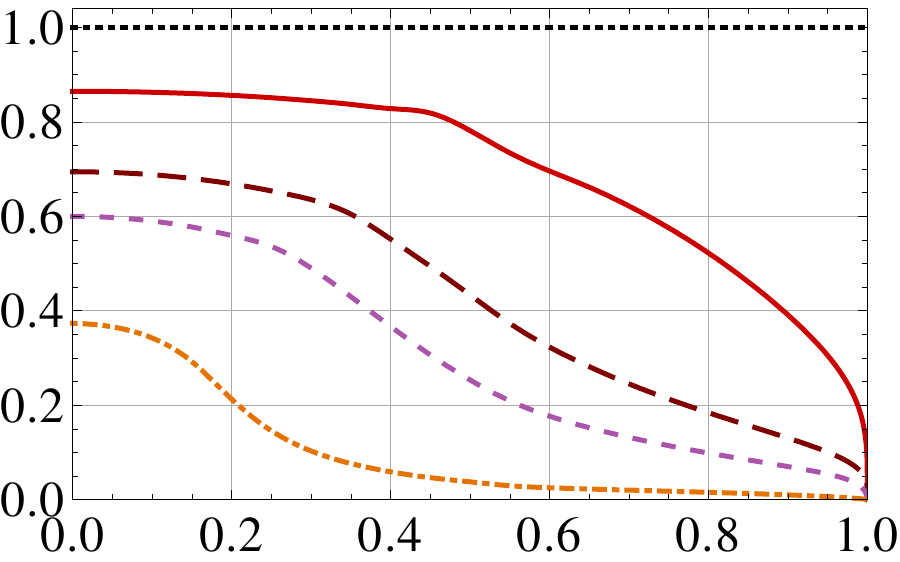}
\put(-95,-10){\small $v$}
\put(-210,40){\rotatebox{90}{$L_s/L_{\text{iso}}(s)$}}
\end{tabular}
\caption{\small Screening length for a plasma wind along the $z$-direction and a dipole oriented along the $z$-direction, for four different values of the anisotropy (from top to bottom) $a/T=12.2,\, 42.6,\, 86, \,744$. The corresponding values in units of the entropy density are (in the same order) $a\nc^{2/3}/s^{1/3}=6.2, 19, 35, 242$.
The screening length is plotted in the appropriate units to facilitate comparison with the isotropic result for a plasma at the same temperature (left), or at the same entropy density (right). The isotropic result is plotted in Fig.~\ref{lviso}, and its ultra-relativistic behavior is given in eq.~(\ref{isoscaling}). At $v=0$ the curves agree with the $\theta=0$ values in  Fig.~\ref{staticscreening1}. As $v\to 1$ they 
 vanish as $(1-v^2)^{1/4}$, in agreement with (\ref{result-scaling})(top line) and \eqn{isoscaling}.
}
\label{extzmovzdiffa}
 \end{center} 
 \end{figure}

\begin{figure}[h!!]
\begin{center}
\begin{tabular}{cc}
\includegraphics[scale=0.75]{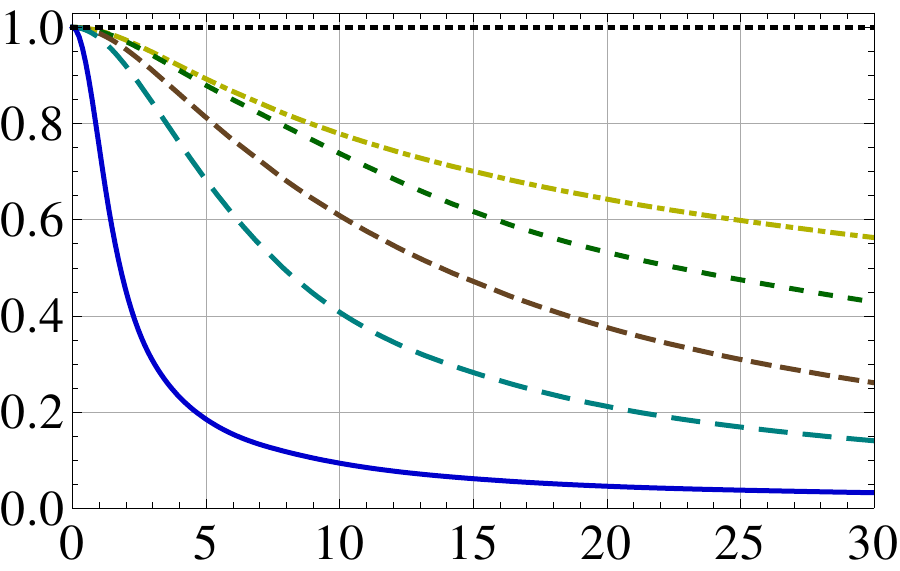}
\put(-109,-10){\small $a/T$}
\put(-215,40){\rotatebox{90}{$L_s/L_{\text{iso}}(T)$}}
\qquad
\includegraphics[scale=0.75]{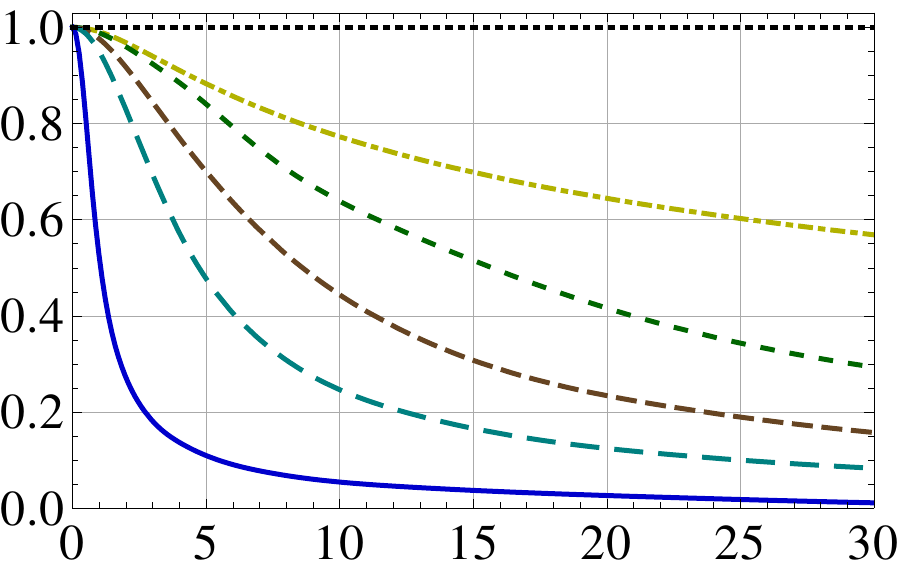}
\put(-109,-10){\small $a\nc^{2/3}/s^{1/3}$}
\put(-210,40){\rotatebox{90}{$L_s/L_{\text{iso}}(s)$}}
\end{tabular} 
\caption{\small Screening length for a plasma wind along the $z$-direction and a dipole oriented along the $z$-direction, at five different velocities (from top to bottom) $v=0.25, 0.5, 0.7, 0.9, 0.9995$. The screening length is plotted in the appropriate units to facilitate comparison with the isotropic result for a plasma at the same temperature (left), or at the same entropy density (right). The isotropic result is plotted in Fig.~\ref{lviso}, and its ultra-relativistic behavior is given in eq.~(\ref{isoscaling}).
\label{extzmovzdiffv}
}
 \end{center} 
 \end{figure}

\paragraph{Wind along $z$ and dipole along $z$.}
The numerical results are shown in Figs.~\ref{extzmovzdiffa} and \ref{extzmovzdiffv}. The curves in Fig.~\ref{extzmovzdiffa} 
start at $v=0$ with the same value as the $\theta=0$ static result shown in  Fig.~\ref{staticscreening1}, and that they vanish as $(1-v^2)^{1/4}$ in the limit $v\to1$, in agreement with (\ref{result-scaling})(top line) and \eqn{isoscaling}. The screening length decreases with the anisotropy, irrespectively of whether $T$ or $s$ are kept fixed.

\begin{figure}[tb]
\begin{center}
\begin{tabular}{cc}
\includegraphics[scale=0.75]{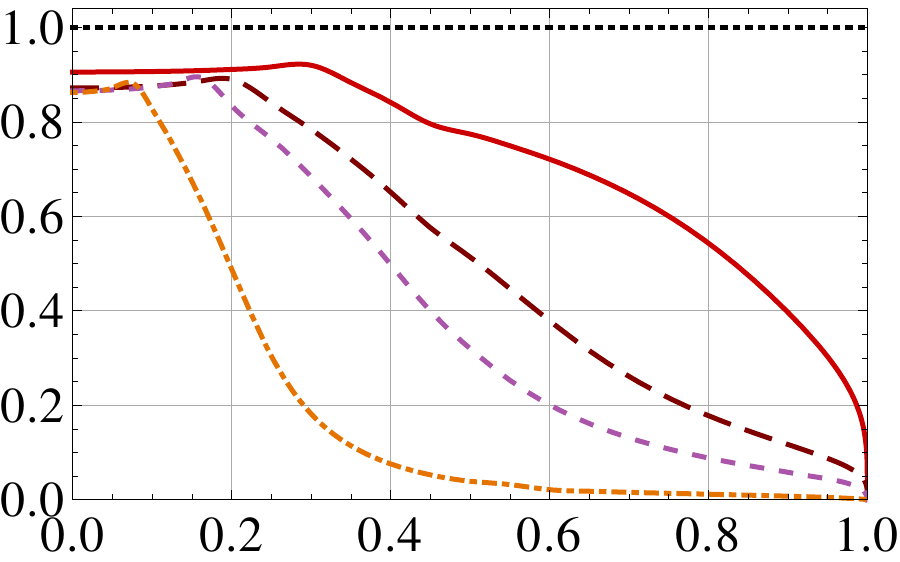}
\put(-109,-10){\small $v$}
\put(-215,40){\rotatebox{90}{$L_{\text{ani}}/L_{\text{iso}}(T)$}}
\qquad
\includegraphics[scale=0.75]{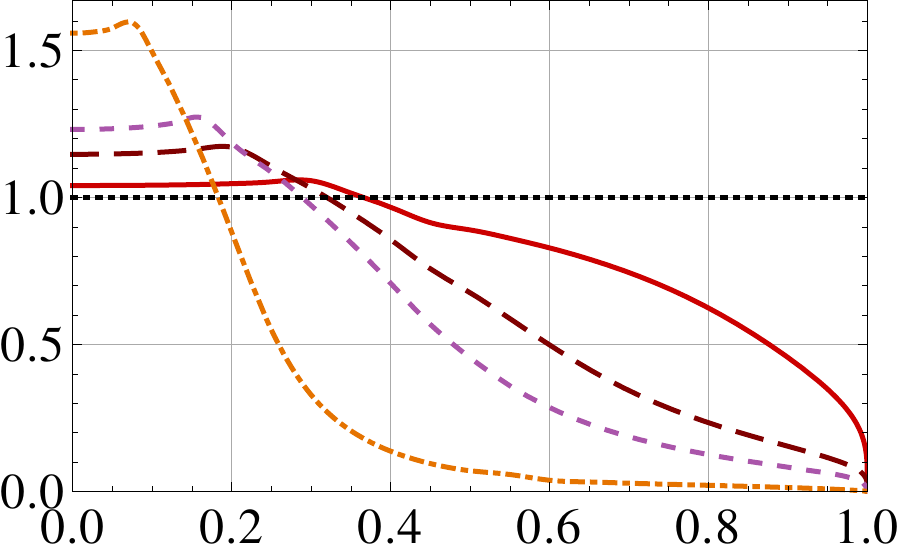}
\put(-109,-10){\small $v$}
\put(-210,40){\rotatebox{90}{$L_{\text{ani}}/L_{\text{iso}}(s)$}}
\end{tabular}
\caption{\small Screening length for a plasma wind along the $z$-direction and a dipole oriented along the $x$-direction, for four different values of the anisotropy $a/T=12.2$ (red, solid), 42.6 (maroon, coarsely dashed), 86 (violet, dashed), 744 (orange, dot-dashed).  The corresponding values in units of the entropy density are (in the same order) $a\nc^{2/3}/s^{1/3}=6.2, 19, 35, 242$.
The screening length is plotted in the appropriate units to facilitate comparison with the isotropic result for a plasma at the same temperature (left), or at the same entropy density (right). The isotropic result is plotted in Fig.~\ref{lviso}, and its ultra-relativistic behavior is given in eq.~(\ref{isoscaling}). At $v=0$ the curves agree with the $\theta=\pi/2$ values in  Fig.~\ref{staticscreening1}.  As $v\to 1$ they 
 vanish as $(1-v^2)^{1/4}$, in agreement with (\ref{result-scaling})(top line) and \eqn{isoscaling}.
\label{extxmovzdiffa}
}
 \end{center} 
 \end{figure}

\begin{figure}[tb]
\begin{center}
\begin{tabular}{cc}
\includegraphics[scale=0.80]{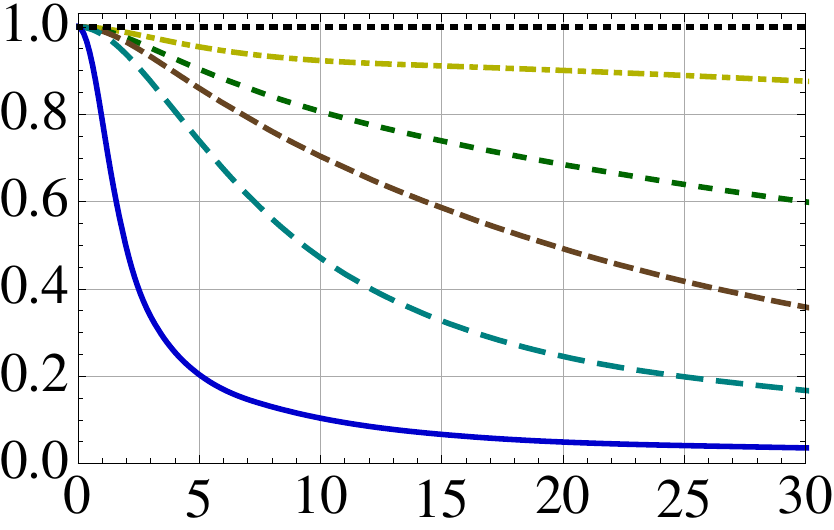}
\put(-109,-10){\small $a/T$}
\put(-210,40){\rotatebox{90}{$L_{\text{ani}}/L_{\text{iso}}(T)$}}
\qquad\,\,\,\,
\includegraphics[scale=0.80]{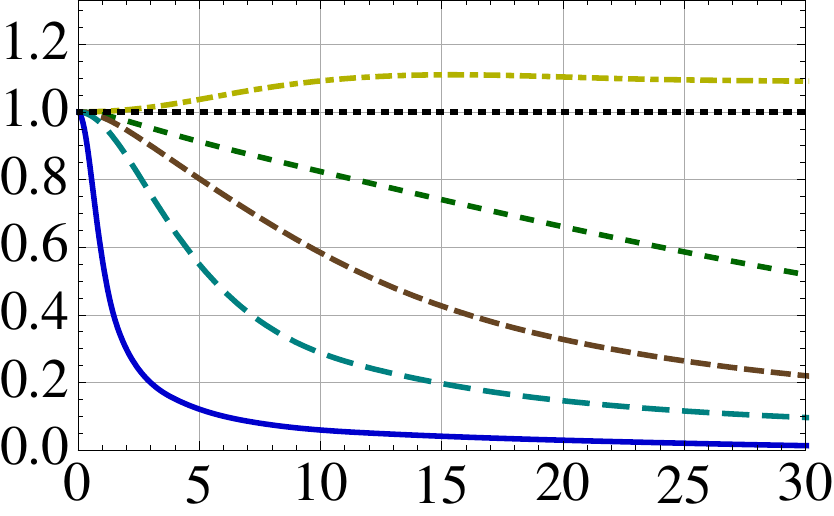}
\put(-109,-10){\small $a/s^{1/3}$}
\put(-210,40){\rotatebox{90}{$L_{\text{ani}}/L_{\text{iso}}(s)$}}
\end{tabular}
\caption{\small  Screening length for a plasma wind along the $z$-direction and a dipole oriented along the $x$-direction, at five different velocities (from top to bottom) $v=0.25, 0.5, 0.7, 0.9, 0.9995$. The screening length is plotted in the appropriate units to facilitate comparison with the isotropic result for a plasma at the same temperature (left), or at the same entropy density (right). The isotropic result is plotted in Fig.~\ref{lviso}, and its ultra-relativistic behavior is given in eq.~(\ref{isoscaling}). 
\label{extxmovzdiffv}
}
 \end{center} 
 \end{figure}

\paragraph{Wind along $z$ and dipole along $x$.}
The numerical results are shown in Figs.~\ref{extxmovzdiffa} and \ref{extxmovzdiffv}. We see that the curves in Fig.~\ref{extxmovzdiffa} 
start at $v=0$ with the same value as the $\theta=\pi/2$ static result shown in  Fig.~\ref{staticscreening1}, and that they vanish as $(1-v^2)^{1/4}$ in the limit $v\to1$, in agreement with 
(\ref{result-scaling})(top line) and \eqn{isoscaling}. In this case the screening length decreases with the anisotropy for any velocity provided the temperature is kept fixed. The same behavior is found at constant entropy density for high enough velocities, whereas for low velocities the screening length at constant $s$ actually increases  with $a$.


\begin{figure}[tb]
\begin{center}
\begin{tabular}{cc}
\includegraphics[scale=0.75]{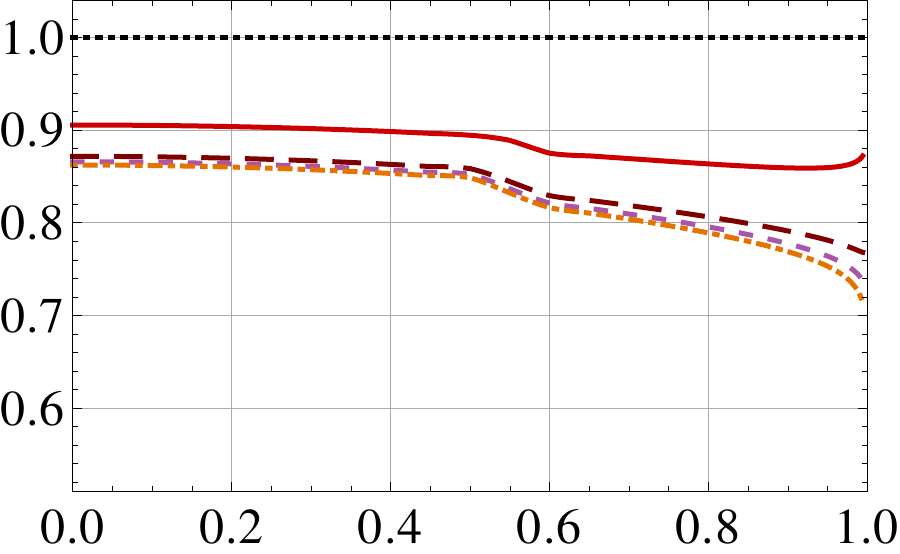}
\put(-95,-10){\small $v$}
\put(-215,40){\rotatebox{90}{$L_s/L_{\text{iso}}(T)$}}
\qquad
\includegraphics[scale=0.75]{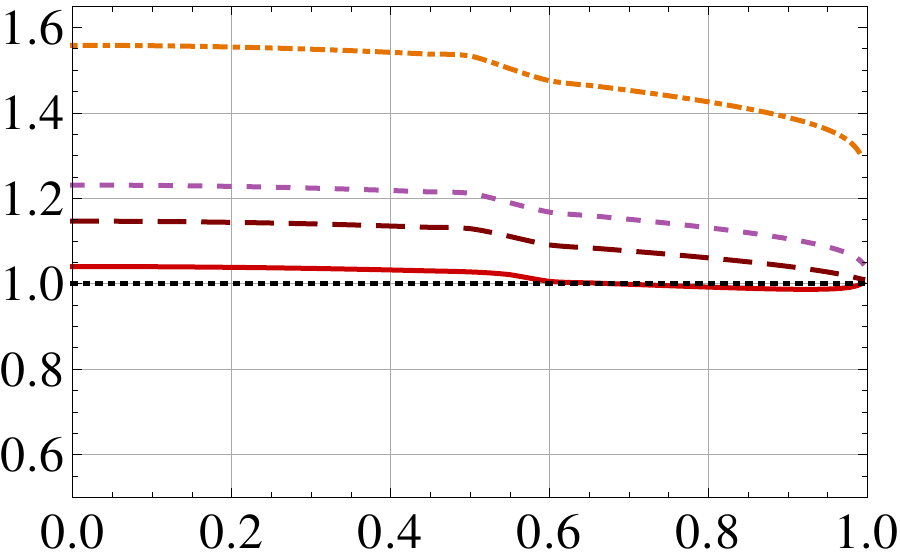}
\put(-95,-10){\small $v$}
\put(-210,40){\rotatebox{90}{$L_s/L_{\text{iso}}(s)$}}
\end{tabular}
\caption{\small Screening length for a plasma wind along the $x$-direction and a dipole oriented along the $x$-direction, for four different values of the anisotropy (from top to bottom) $a/T=12.2,\, 42.6,\, 86, \,744$. The corresponding values in units of the entropy density are (in the same order) $a\nc^{2/3}/s^{1/3}=6.2, 19, 35, 242$.
The screening length is plotted in the appropriate units to facilitate comparison with the isotropic result for a plasma at the same temperature (left), or at the same entropy density (right). The isotropic result is plotted in Fig.~\ref{lviso}, and its ultra-relativistic behavior is given in eq.~(\ref{isoscaling}). At $v=0$ the curves agree with the 
$\theta=\pi/2$ values in  Fig.~\ref{staticscreening1}.  As $v\to 1$ they 
approach a finite, non-zero value, in agreement with (\ref{result-scaling})(bottom line) and \eqn{isoscaling}.}
\label{extxmovxdiffa}
 \end{center} 
 \end{figure}

\begin{figure}[h!!]
\begin{center}
\begin{tabular}{cc}
\includegraphics[scale=0.75]{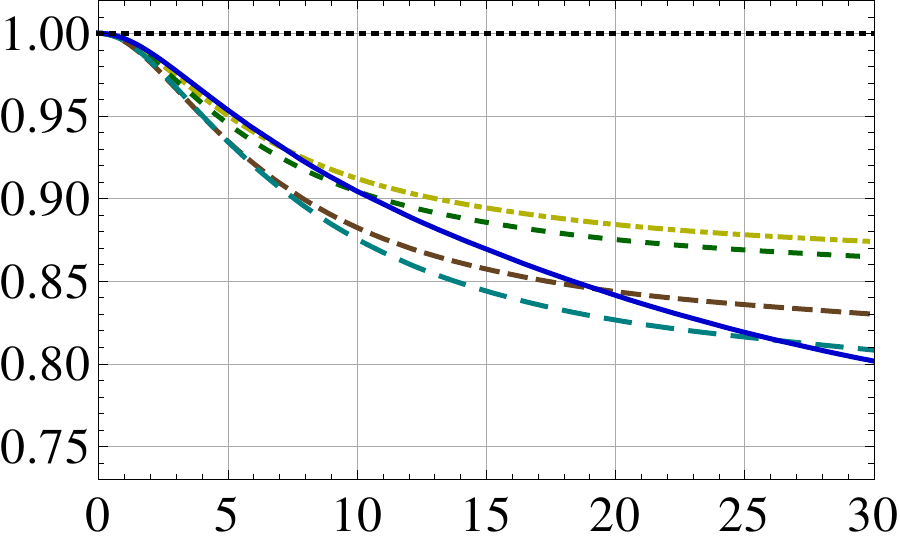}
\put(-109,-10){\small $a/T$}
\put(-215,40){\rotatebox{90}{$L_s/L_{\text{iso}}(T)$}}
\qquad
\includegraphics[scale=0.75]{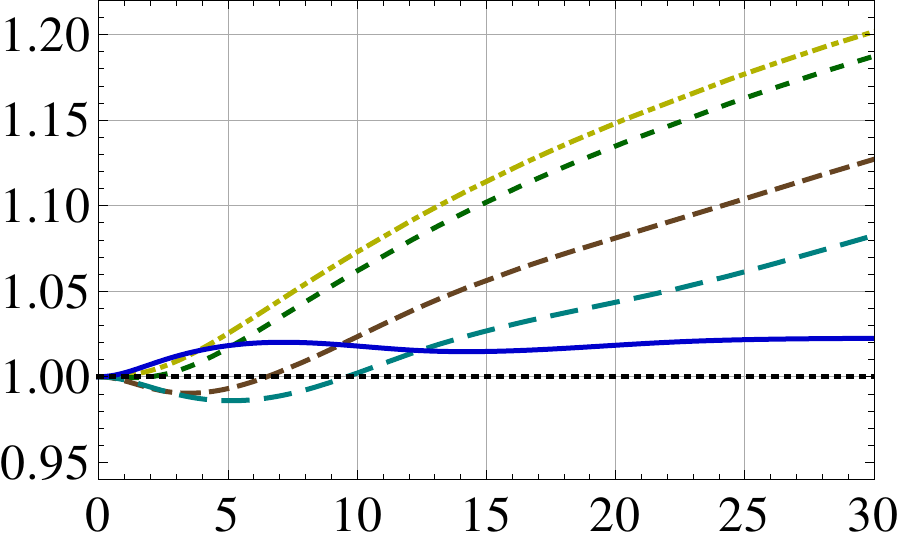}
\put(-109,-10){\small $a\nc^{2/3}/s^{1/3}$}
\put(-210,40){\rotatebox{90}{$L_s/L_{\text{iso}}(s)$}}
\end{tabular}
\caption{\small Screening length for a plasma wind along the $x$-direction and a dipole oriented along the $x$-direction, at five different velocities $v=$0.25 (yellow, dot-dashed), 0.5 (green, short dashed), 0.7 (brown, medium dashed), 0.9 (cyan, long dashed), 0.9995 (blue, solid). The screening length is plotted in the appropriate units to facilitate comparison with the isotropic result for a plasma at the same temperature (left), or at the same entropy density (right). The isotropic result is plotted in Fig.~\ref{lviso}, and its ultra-relativistic behavior is given in eq.~(\ref{isoscaling}).
\label{extxmovxdiffv}
}
 \end{center} 
 \end{figure}

\paragraph{Wind along $x$ and dipole along $x$.}
The numerical results are shown in Figs.~\ref{extxmovxdiffa} and \ref{extxmovxdiffv}. The curves in Fig.~\ref{extxmovxdiffa} 
start at $v=0$ with the same value as the $\theta=\pi/2$ static result shown in  Fig.~\ref{staticscreening1}, and that they approach a finite, non-zero value as $v \to 1$, in agreement with (\ref{result-scaling})(bottom line) and \eqn{isoscaling}. As in previous cases, the screening length decreases with the anisotropy for any velocity provided the temperature is kept fixed. The opposite behavior is found at constant $s$.
 
\begin{figure}[tb]
\begin{center}
\begin{tabular}{cc}
\includegraphics[scale=0.75]{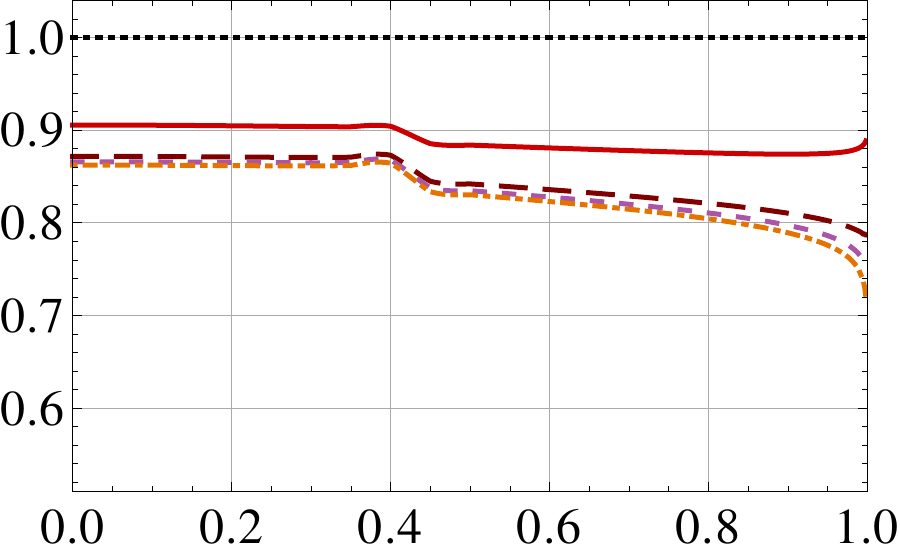}
\put(-109,-10){\small $v$}
\put(-215,40){\rotatebox{90}{$L_{\text{ani}}/L_{\text{iso}}(T)$}}
\qquad
\includegraphics[scale=0.75]{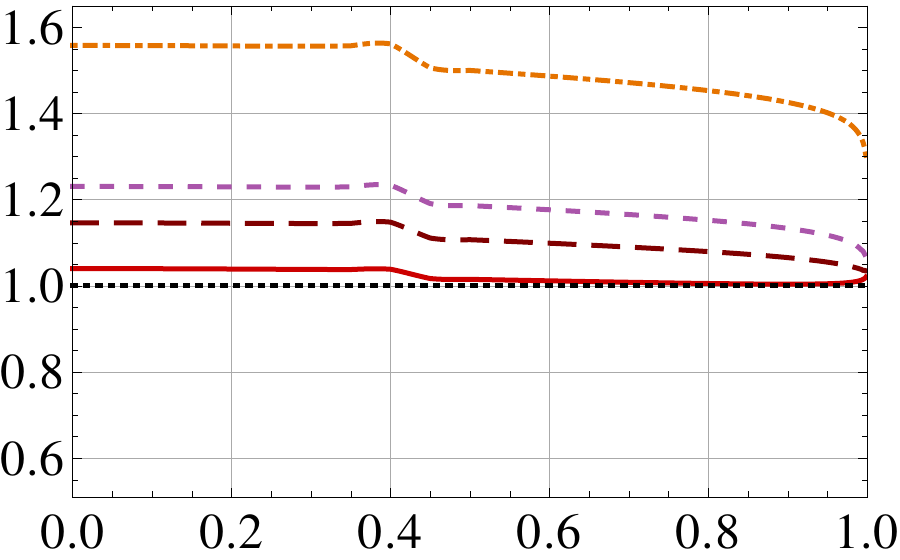}
\put(-109,-10){\small $v$}
\put(-210,40){\rotatebox{90}{$L_{\text{ani}}/L_{\text{iso}}(s)$}}
\end{tabular}
\caption{\small Screening length for a plasma wind along the $x$-direction and a dipole oriented along the $y$-direction, for four different values of the anisotropy $a/T=12.2$ (red, solid), 42.6 (maroon, coarsely dashed), 86 (violet, dashed), 744 (orange, dot-dashed).  
The corresponding values in units of the entropy density are (in the same order) $a\nc^{2/3}/s^{1/3}=6.2, 19, 35, 242$. 
The screening length is plotted in the appropriate units to facilitate comparison with the isotropic result for a plasma at the same temperature (left), or at the same entropy density (right). The isotropic result is plotted in Fig.~\ref{lviso}, and its ultra-relativistic behavior is given in eq.~(\ref{isoscaling}). At $v=0$ the curves agree with the 
$\theta=\pi/2$ values in  Fig.~\ref{staticscreening1}.  As $v\to 1$ they 
approach a finite, non-zero value, in agreement with (\ref{result-scaling})(bottom line) and \eqn{isoscaling}.
\label{extymovxdiffa}
}
 \end{center} 
 \end{figure}

\begin{figure}[tb]
\begin{center}
\begin{tabular}{cc}
\includegraphics[scale=0.75]{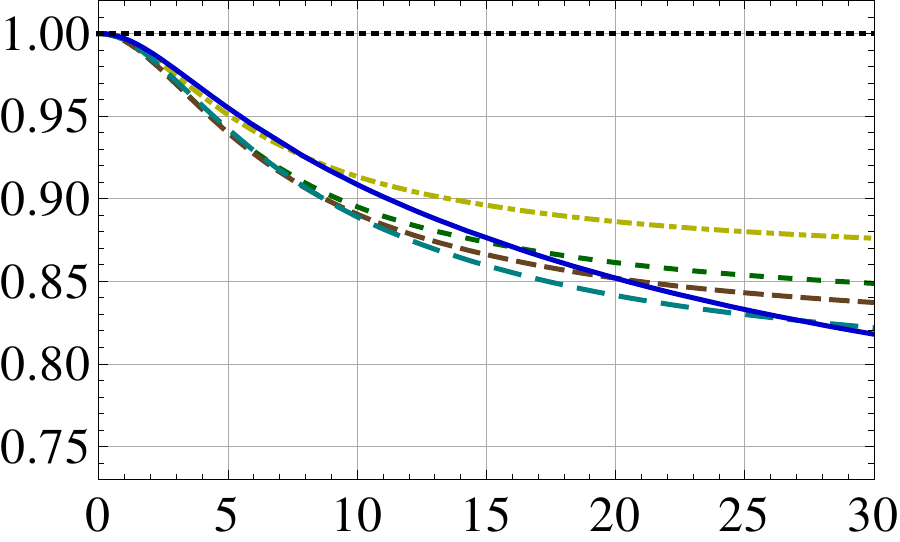}
\put(-109,-10){\small $a/T$}
\put(-215,40){\rotatebox{90}{$L_{\text{ani}}/L_{\text{iso}}(T)$}}
\qquad
\includegraphics[scale=0.80]{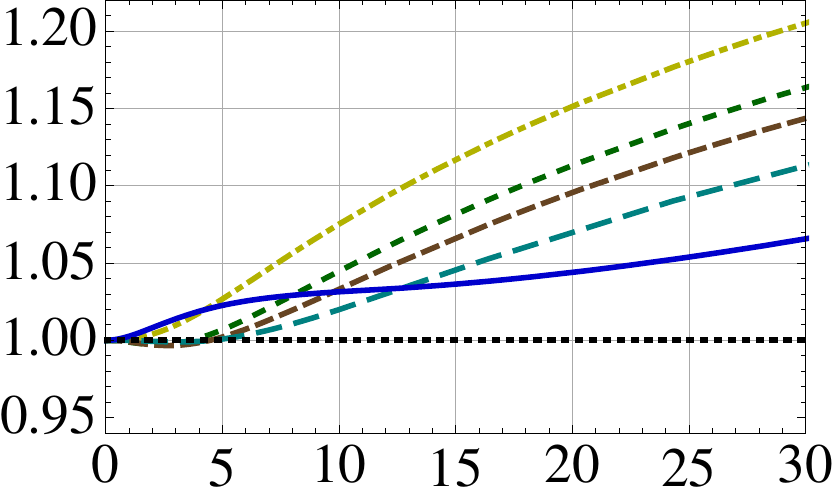}
\put(-109,-10){\small $a \nc^{2/3}/s^{1/3}$}
\put(-210,40){\rotatebox{90}{$L_{\text{ani}}/L_{\text{iso}}(s)$}}
\end{tabular}
\caption{\small  Screening length for a plasma wind along the $x$-direction and a dipole oriented along the $y$-direction, at five different velocities $v=$0.25 (yellow, dot-dashed), 0.5 (green, short dashed), 0.7 (brown, medium dashed), 0.9 (cyan, long dashed), 0.9995 (blue, solid). The screening length is plotted in the appropriate units to facilitate comparison with the isotropic result for a plasma at the same temperature (left), or at the same entropy density (right). The isotropic result is plotted in Fig.~\ref{lviso}, and its ultra-relativistic behavior is given in eq.~(\ref{isoscaling}).
\label{extymovxdiffv}
}
 \end{center} 
 \end{figure}

\paragraph{Wind along $x$ and dipole along $y$.}
The numerical results are shown in Figs.~\ref{extymovxdiffa} and \ref{extymovxdiffv}. We see that the curves in Fig.~\ref{extymovxdiffa} 
start at $v=0$ with the same value as the $\theta=\pi/2$ static result shown in  Fig.~\ref{staticscreening1}, and that they approach a finite, non-zero value as $v \to 1$, in agreement with (\ref{result-scaling})(bottom line)  and \eqn{isoscaling}. The qualitative behavior in as in the case of motion and orientation along $x$. 

\begin{figure}[tb]
\begin{center}
\begin{tabular}{cc}
\includegraphics[scale=0.75]{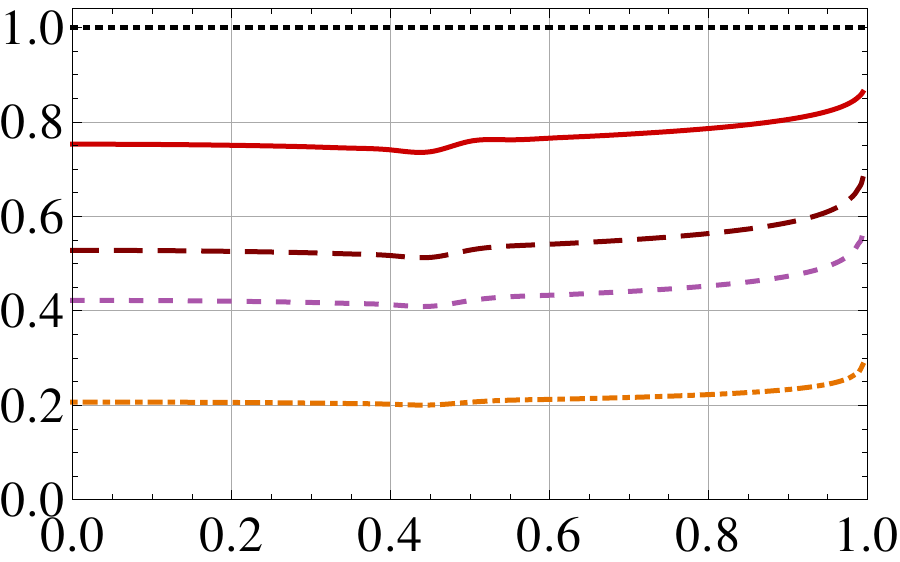}
\put(-109,-10){\small $v$}
\put(-215,40){\rotatebox{90}{$L_{\text{ani}}/L_{\text{iso}}(T)$}}
\qquad
\includegraphics[scale=0.75]{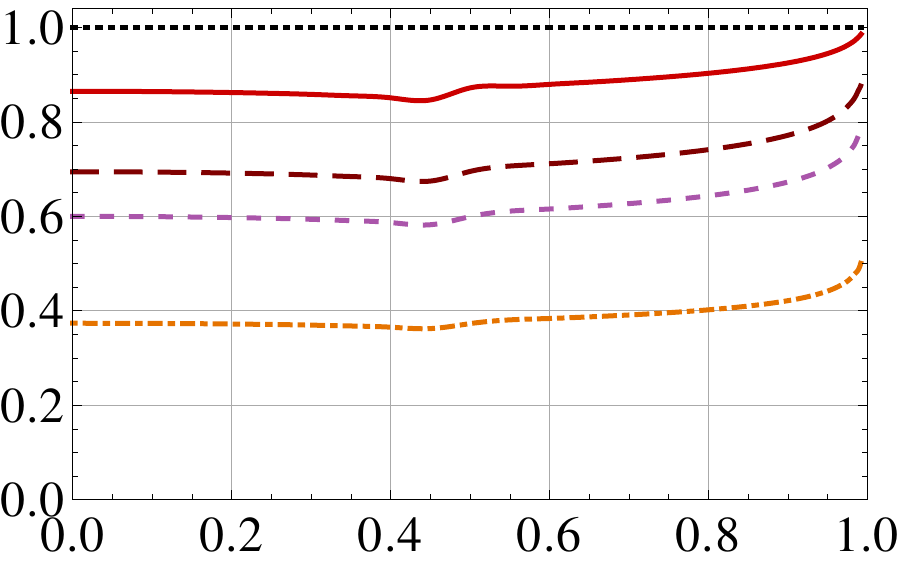}
\put(-109,-10){\small $v$}
\put(-210,40){\rotatebox{90}{$L_{\text{ani}}/L_{\text{iso}}(s)$}}
\end{tabular}
\caption{\small Screening length for a plasma wind along the $x$-direction and a dipole oriented along the $z$-direction, for four different values of the anisotropy (from top to bottom) $a/T=12.2, \, 42.6, \, 86, \, 744$.  The corresponding values in units of the entropy density are (in the same order) $a\nc^{2/3}/s^{1/3}=6.2, 19, 35, 242$.
The screening length is plotted in the appropriate units to facilitate comparison with the isotropic result for a plasma at the same temperature (left), or at the same entropy density (right). The isotropic result is plotted in Fig.~\ref{lviso}, and its ultra-relativistic behavior is given in eq.~(\ref{isoscaling}). At $v=0$ the curves agree with the 
$\theta=0$ values in  Fig.~\ref{staticscreening1}.  As $v\to 1$ they 
approach a finite, non-zero value, in agreement with (\ref{result-scaling})(bottom line) and \eqn{isoscaling}.
\label{extzmovxdiffa}
}
 \end{center} 
 \end{figure}

\begin{figure}[tb]
\begin{center}
\begin{tabular}{cc}
\includegraphics[scale=0.75]{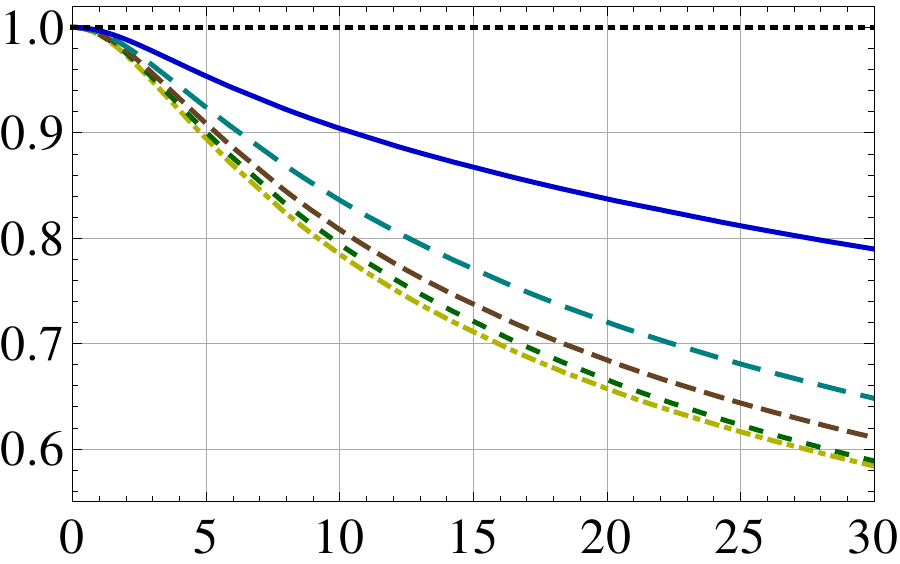}
\put(-109,-10){\small $a/T$}
\put(-215,40){\rotatebox{90}{$L_{\text{ani}}/L_{\text{iso}}(T)$}}
\qquad
\includegraphics[scale=0.75]{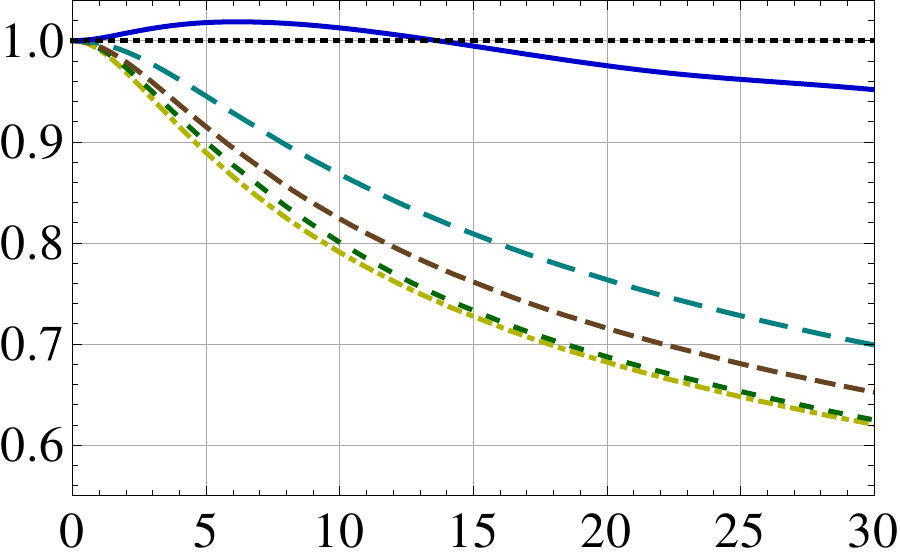}
\put(-109,-10){\small $a/s^{1/3}$}
\put(-210,40){\rotatebox{90}{$L_{\text{ani}}/L_{\text{iso}}(s)$}}
\end{tabular}
\caption{\small  Screening length for a plasma wind along the $x$-direction and a dipole oriented along the $z$-direction, at five different velocities (from bottom to top) $v=0.25, 0.5, 0.7, 0.9, 0.9995$. The screening length is plotted in the appropriate units to facilitate comparison with the isotropic result for a plasma at the same temperature (left), or at the same entropy density (right). The isotropic result is plotted in Fig.~\ref{lviso}, and its ultra-relativistic behavior is given in eq.~(\ref{isoscaling}).
\label{extzmovxdiffv}
} 
 \end{center} 
 \end{figure}

\paragraph{Wind along $x$ and dipole along $z$.}
The numerical results are shown in Figs.~\ref{extzmovxdiffa} and \ref{extzmovxdiffv}. We see that the curves in Fig.~\ref{extzmovxdiffa} 
start at $v=0$ with the same value as the $\theta=0$ static result shown in  Fig.~\ref{staticscreening1}, and that they approach a finite, non-zero value as $v \to 1$, in agreement with (\ref{result-scaling})(bottom line)  and \eqn{isoscaling}. The screening length decreases with the anisotropy for any velocity provided the temperature is kept fixed. The same is true at large anisotropies if the entropy density is kept fixed.


\section{Dissociation temperature and dissociation anisotropy}
\label{diss_sec}
In  previous sections we have focused on computing the screening length in an anisotropic plasma, $L_s(T,a)$, and on comparing it to its isotropic counterpart $L_\mt{iso} = L_s (T,0)$. The screening length characterizes the dissociation of a quark-antiquark pair for fixed $T$ and $a$: A pair separated a distance $\ell < L_s$ forms a bound state, but if $\ell$ is increased above $L_s$ then the bound state dissociates.  Similarly, one may define a dissociation temperature $T_\mt{diss}(a,\ell)$ that characterizes the dissociation of a quark-antiquark pair of fixed size $\ell$ in a plasma with a given degree of anisotropy $a$: for $T< T_\mt{diss}$ the pair forms a bound state, but if $T$ is increased above $T_\mt{diss}$ then the bound state dissociates. Analogously, one may define a dissociation anisotropy $a_\mt{diss}(T,\ell)$ such that a bound state forms for $a<a_\mt{diss}$ but not for 
$a>a_\mt{diss}$. It is useful to think of the three-dimensional space parametrized by $(T,a,\ell)$ as divided  in two disconnected regions by a two-dimensional surface: in one region quark-antiquark pairs bind together, while in the other one they do not. The functions $L_s(T,a)$,  $T_\mt{diss}(a,\ell)$ and $a_\mt{diss}(T,\ell)$ are then simply different parametrizations of the dividing surface. It is therefore clear that if a triplet $(T,a,\ell)$ lies on the dividing surface then 
\be
T L_s(a,T) = \Tdiss(a,\ell) \ell \sac a L_s(T,a) = \adiss (T,\ell) \ell 
\sac \mbox{etc.}
\label{interpret}
\ee

In this section we will focus on the qualitative form of  $T_\mt{diss}$ and $a_\mt{diss}$. As we will see, most of the analysis follows from the asymptotic behavior of the screening length for $a\gg T$. This means that, at the qualitative level, most of the results that we will obtain would also apply if we were to replace the temperature by the entropy density as one of our variables. The reason is that, by virtue of \eqn{larges},  the limit $a \gg T$ corresponds to the limit $a \gg s^{1/3}$ and vice versa. In addition, we will see that for generic dipole's orientations and velocities, the large-anisotropy limit is entirely controlled by the near-boundary behavior of the metric at $O(u^2)$, which depends solely on $a$ and is therefore completely insensitive to the values of the temperature or of the entropy density. 

\begin{figure}[t]
\begin{center}
\includegraphics[scale=0.85]{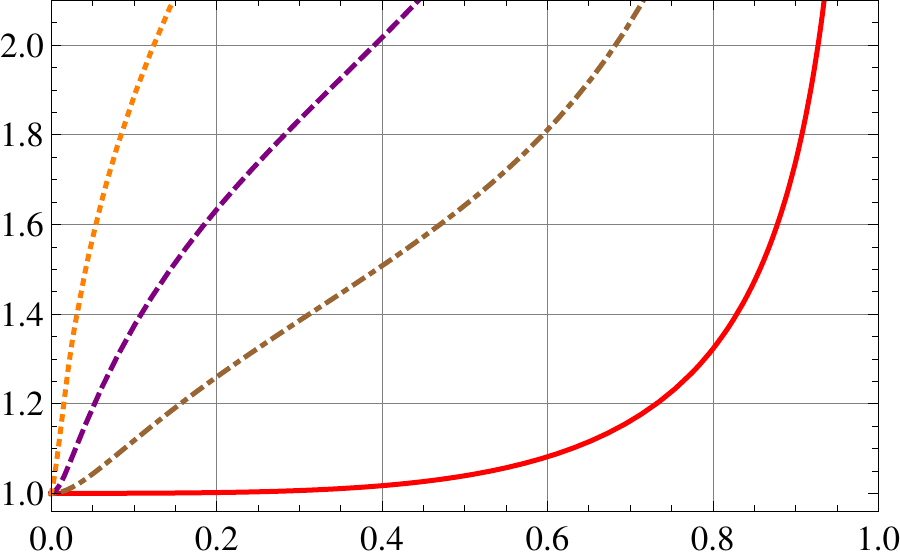}
 \begin{picture}(0,0)
\put(-120,-10){\small $u/\uh$}
\put(-245,55){\rotatebox{90}{$v_\mt{proper}/v_z$}}
 \end{picture}
\caption{\small Proper velocity in the $z$-direction at a position $u$ away from the boundary, as defined in \eqn{proper}, for different values of $a/T$. From right to left, $a/T=1.38, 33, 86, 249$.
\label{propervz}}
\end{center}
\end{figure}
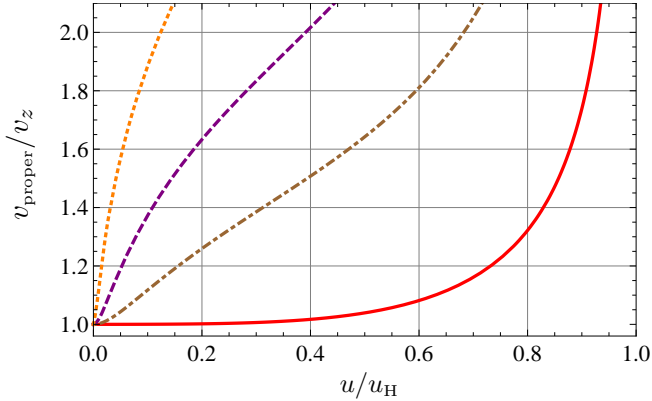

The key point in the large-$a$ analysis is the requirement that no point on the string can move faster than the local speed of light in the bulk. Consider a meson moving with a velocity $v$ that has a non-zero component $v_z$ along the $z$-direction. Then we see from  \eqn{sol2} that the proper velocity along this direction of a point on the string sitting at a value $u$ of the radial coordinate is 
\be
v_\mt{proper}(u) = v_z \sqrt{-\frac{g_{zz} (u)}{g_{tt} (u)}} =
 v_z \sqrt{\frac{\ch (u)}{\cf(u) \cb (u)}} \,.
\label{proper}
\ee
The function $\ch(u)$ increases monotonically from the boundary to the horizon, and is does so more steeply as $a/T$ increases, as illustrated in Fig.~\ref{plots}. The combination 
$\cf(u) \cb (u)$ has the opposite behavior, as expected from the fact that gravity is attractive: it decreases monotonically from the boundary to the horizon. In the isotropic case $\ch=1$ and $\cf \cb$ decreases more steeply as $T$ increases. This is thus the first hint that increasing the anisotropy has an effect similar to increasing the temperature: both make $v_\mt{proper}(u)$ a more steeply increasing function away from the boundary. We have illustrated the effect of the anisotropy in Fig.~\ref{propervz}, where we see that $v_\mt{proper}/v_z$ becomes a steeper function of $u$ as $a/T$ increases. 

It follows that, for fixed $v_z \neq 0$, there is a maximum value of $\umax$ beyond which $v_\mt{proper}$ becomes superluminal, so no string solution can penetrate to $u>\umax$. As we will corroborate numerically, this upper bound on $\umax$ translates into an upper bound on $L_s$. Moreover, $\umax$ decreases as $a/T$ increases. This means that for sufficiently large anisotropies we can use the near-boundary expansions \eqn{expansion} in order to determine  $L_s$, in analogy to what we did in the ultra-relativistic limit. As in that case, for $v_z \neq 0$ the analysis is controlled by the $O(u^2)$ terms in \eqn{expansion}. The key point is that these terms depend on $a$ but not on $T$, so by dimensional analysis it follows that $\umax \sim a^{-1}$ and $L_s \sim a^{-1}$ in the limit $a/T \gg 1$. This limit can be understood as $a\to \infty$ at fixed $T$, or as $T\to 0$ at fixed $a$. We thus conclude that, even at $T=0$, a generic meson will dissociate for a sufficiently large anisotropy $\adiss$. 

Mesons at rest and mesons whose velocity is exactly aligned with the transverse plane constitute an exception to the  argument above, since in this case $v_z=0$ and their physics is mostly insensitive to the function $\ch(u)$ which characterizes the anisotropic direction. Therefore in this case we expect that $\umax$ and $L_s$ will remain finite as we send $a\to \infty$ at fixed $T$, and hence that dimensional analysis will imply $L_s \sim T^{-1}$. 

In summary, the heuristic argument above suggests that in the limit $a/T \gg 1$ we should have
\vskip -5mm
\bea
L_s(T,a)  \sim\left\{
\begin{array}{l l}
\mbox{const.} \times T^{-1}  \,\,\,\,\,\,\,
\mbox{if the meson is static or in motion within the transverse plane, }  & ~~~~~~ 
\\ & \\
  \mbox{const.}  \times a^{-1}\,\,\,\,\,\, 
   \mbox{otherwise.}  & ~~~~~~
 \end{array}\right.
 \label{asym}
\eea
\vskip 1mm
\noindent
The constants may depend on all the dimensionless parameters such as the velocity and the dipole's orientation. 
We will refer to the behavior in the second line as `generic' and to that in the first line as `non-generic', since the latter only applies if the velocity is exactly zero or if the motion is exactly aligned with the transverse plane. The generic behavior  is of course consistent with the analysis of Sec.~\ref{outside}. Indeed, we saw in that section that for motion outside the transverse plane the ultra-relativistic behavior of $L_s$ is entirely controlled by the $O(u^2)$ terms in the metric, which depend on $a$ but not on $T$. 

\begin{figure}[tb]
\begin{center}
\begin{tabular}{cc}
\includegraphics[scale=\size]{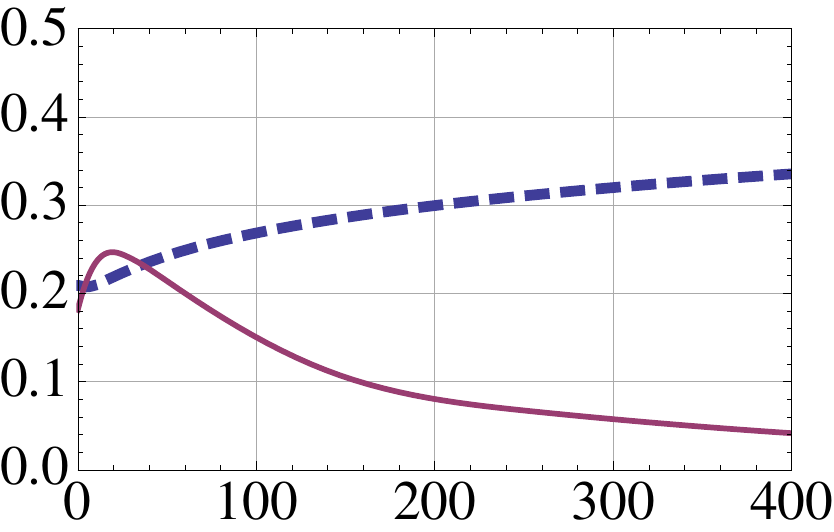}
\put(-80,-10){\small $a/T$}
\put(-160,35){\rotatebox{90}{$T\, \umax$}}
\put(-35,80){\small $z-z$}
 & \qquad
\includegraphics[scale=\size]{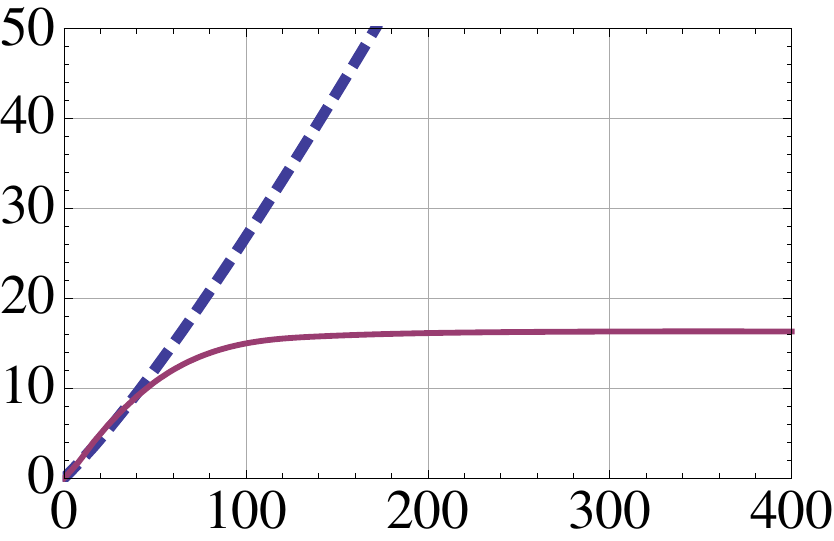}
\put(-80,-10){\small $a/T$}
\put(-157,35){\rotatebox{90}{$a\, \umax$}}
\put(-35,80){\small $z-z$}
\\[\dist]
\includegraphics[scale=\size]{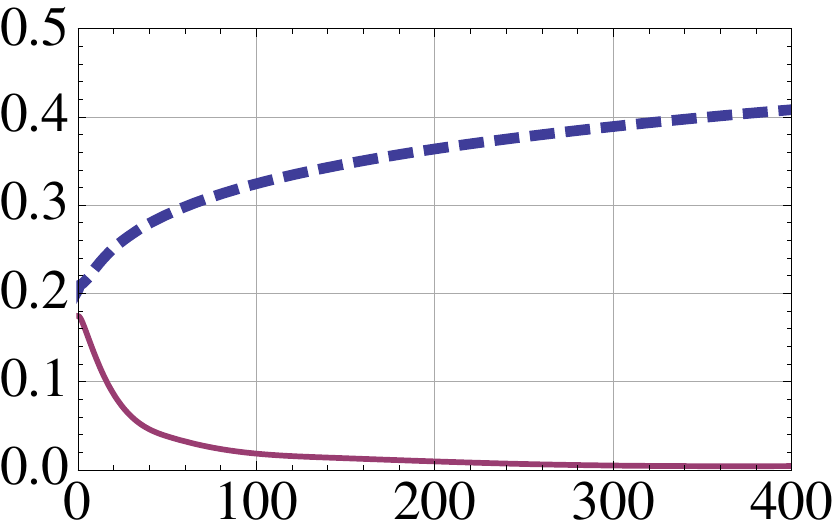}
\put(-80,-10){\small $a/T$}
\put(-160,35){\rotatebox{90}{$T\, \umax$}}
\put(-35,80){\small $z-x$}
 & \qquad
\includegraphics[scale=0.63]{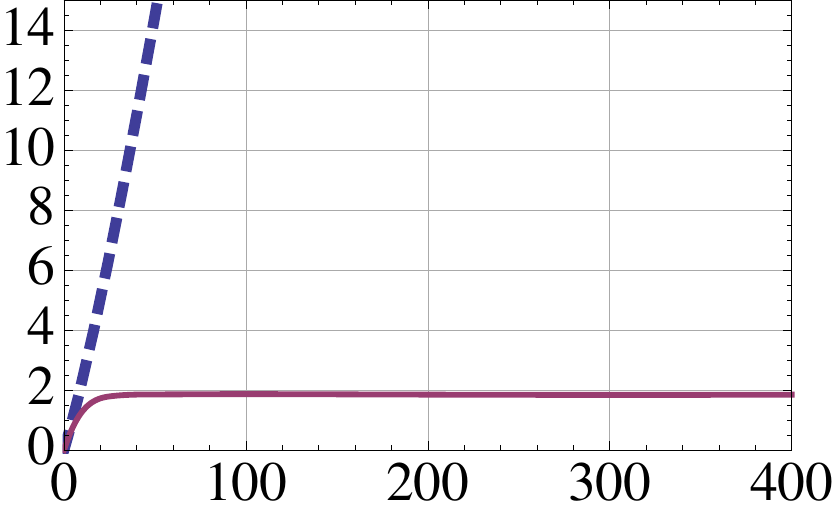}
\put(-80,-10){\small $a/T$}
\put(-157,35){\rotatebox{90}{$a\, \umax$}}
\put(-35,78){\small $z-x$}
\\[\dist]
\includegraphics[scale=\size]{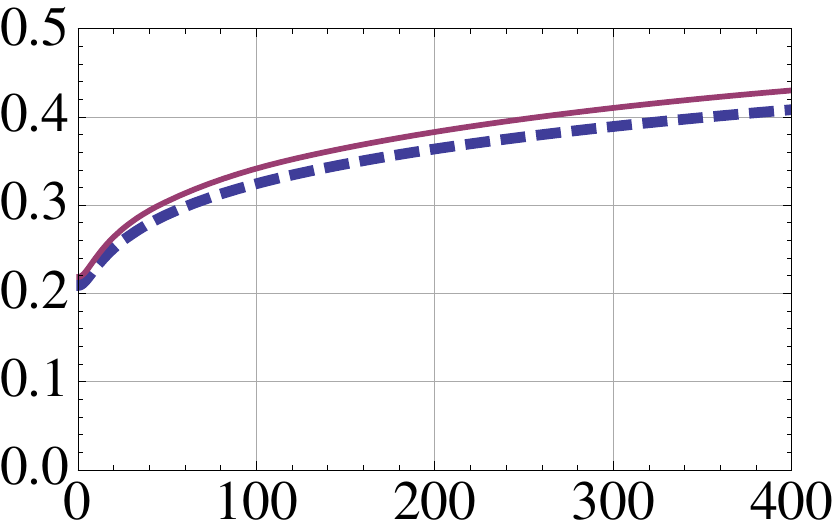}
\put(-80,-10){\small $a/T$}
\put(-160,35){\rotatebox{90}{$T\, \umax$}}
\put(-35,80){\small $x-x$}
 & \qquad
\includegraphics[scale=0.63]{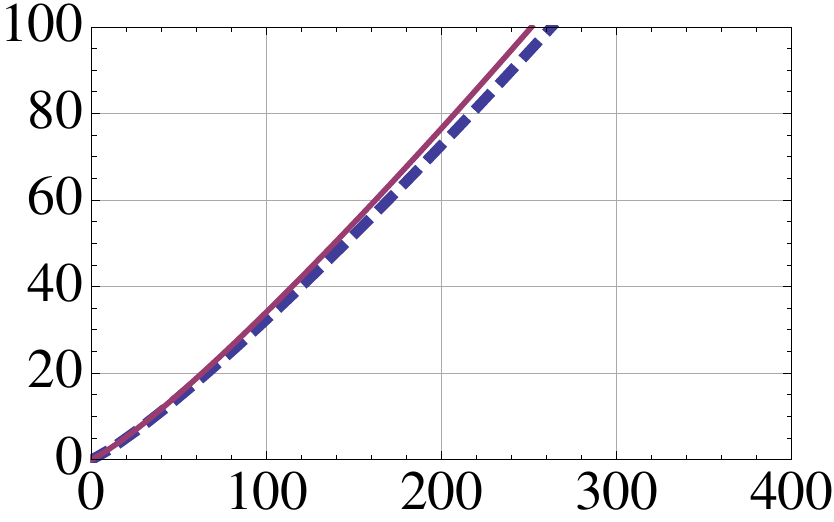}
\put(-80,-10){\small $a/T$}
\put(-157,35){\rotatebox{90}{$a\, \umax$}}
\put(-35,78){\small $x-x$}
\\[\dist]
\includegraphics[scale=\size]{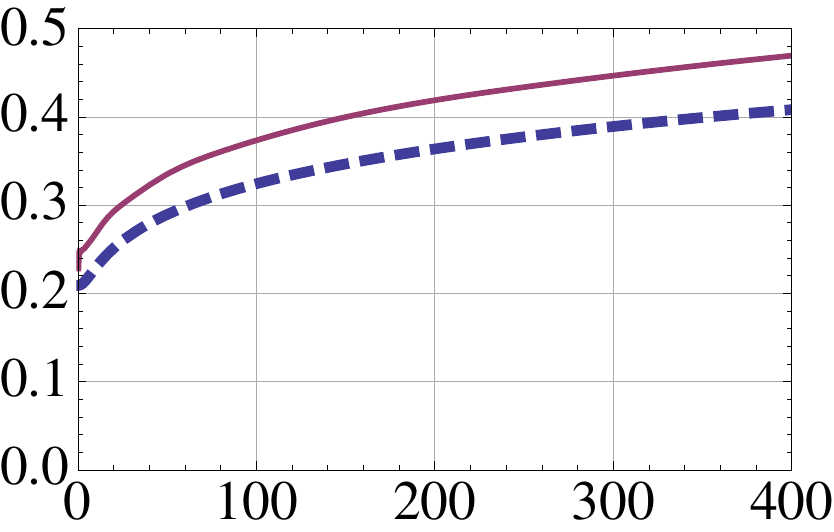}
\put(-80,-10){\small $a/T$}
\put(-160,35){\rotatebox{90}{$T\, \umax$}}
\put(-35,78){\small $x-y$}
 & \qquad
\includegraphics[scale=0.62]{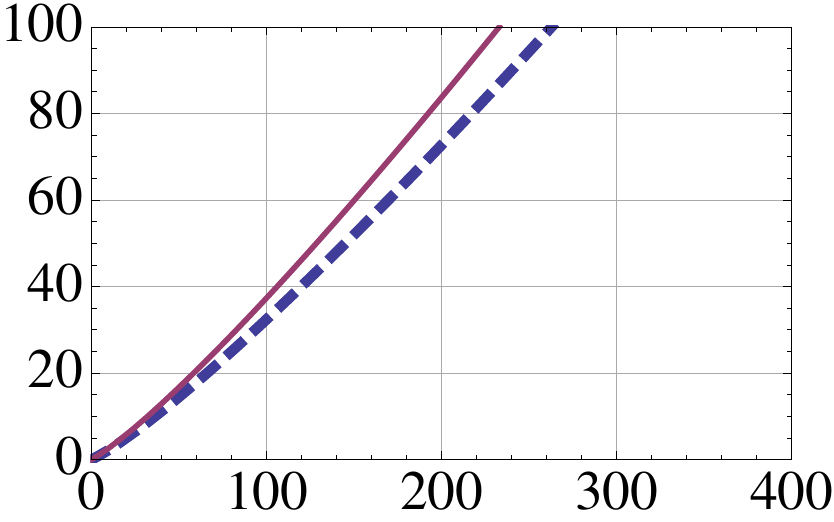}
\put(-80,-10){\small $a/T$}
\put(-157,35){\rotatebox{90}{$a\, \umax$}}
\put(-35,78){\small $x-y$}
\\[\dist]
\includegraphics[scale=\size]{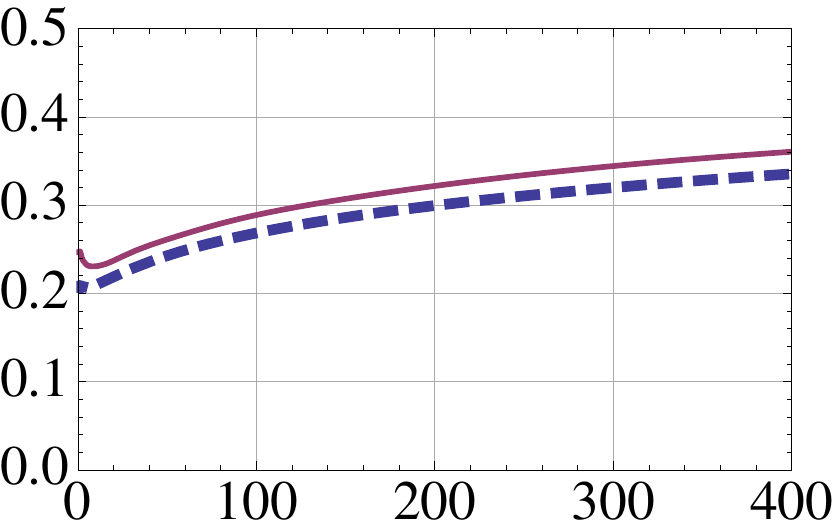}
\put(-80,-10){\small $a/T$}
\put(-160,35){\rotatebox{90}{$T\, \umax$}}
\put(-35,78){\small $x-z$}
 & \qquad
\includegraphics[scale=\size]{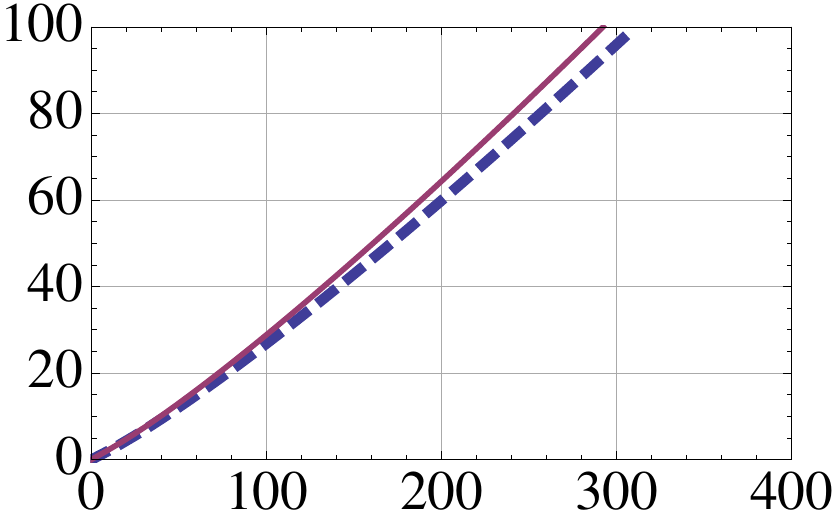}
\put(-80,-10){\small $a/T$}
\put(-157,35){\rotatebox{90}{$a\, \umax$}}
\put(-35,78){\small $x-z$}
\end{tabular}
\caption{\small Value of $\umax$ in units of $1/T$ (left) or $1/a$ (right), as a function of the ratio $a/T$, for a dipole at rest (dashed, blue curve) and for a dipole moving with $v=0.45$ (continuous, magenta curve). The first letter on the top right corner of each plot indicates the direction of motion, and the second one indicates the orientation of the dipole. 
\label{umaxplots}
}
 \end{center}  
 \end{figure}

Fig.~\ref{umaxplots} shows our numerical results for $\umax$, in units of $T^{-1}$ and $a^{-1}$, as a function of $a/T$, for the five physically distinct cases discussed in Sec.~\ref{genericvelocities}.
From the continuous, magenta curves in the first two rows we see that $\umax$ goes to zero at large $a/T$ in the cases of motion along $z$, irrespectively  of  the dipole's orientation. In contrast, we see that $\umax$ does not go zero for a static meson (dashed, blue curves) or for a meson moving along the $x$-direction (continuous, magenta curves in the last three rows).

\begin{figure}[tb]
\begin{center}
\begin{tabular}{cc}
\includegraphics[scale=0.64]{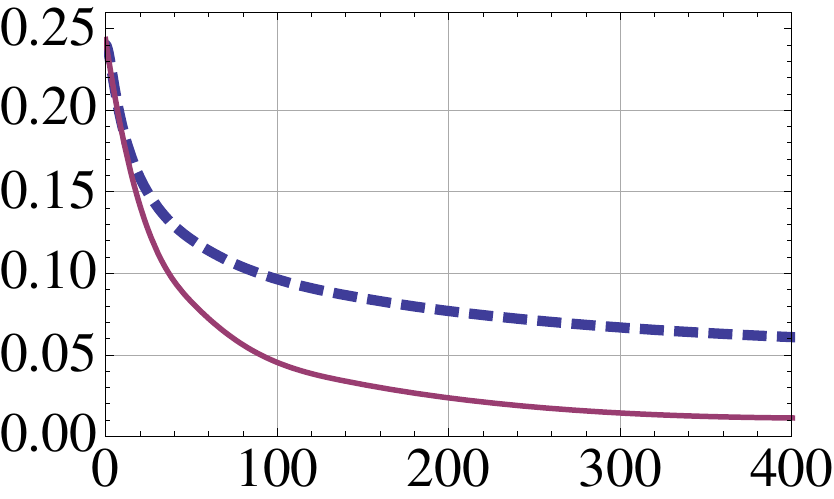}
\put(-80,-10){\small $a/T$}
\put(-167,35){\rotatebox{90}{$T\, L_s$}}
\put(-35,80){\small $z-z$}
 & \qquad
\includegraphics[scale=\sizetwo]{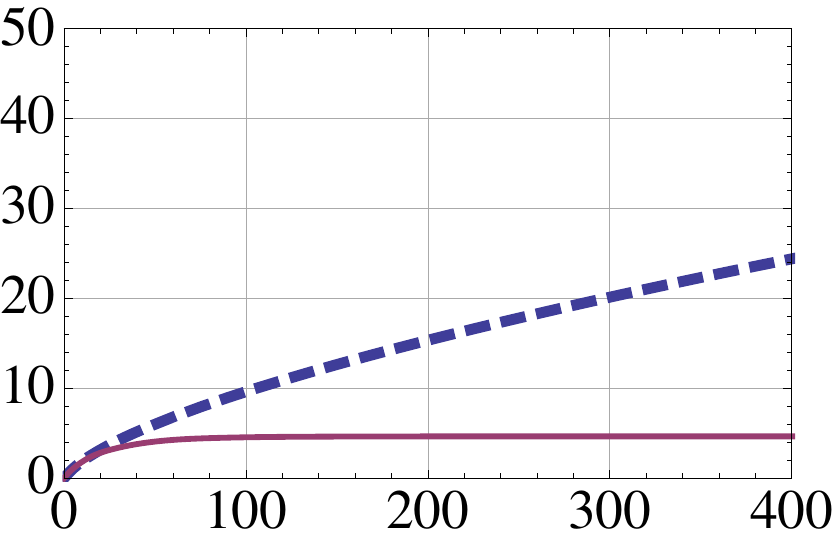}
\put(-80,-10){\small $a/T$}
\put(-157,35){\rotatebox{90}{$a\, L_s$}}
\put(-35,80){\small $z-z$}
\\[\dist]
\includegraphics[scale=0.64]{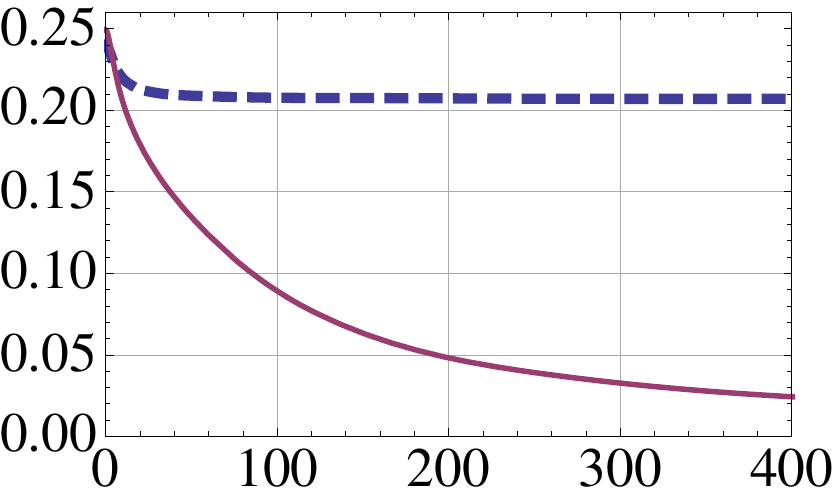}
\put(-80,-10){\small $a/T$}
\put(-167,35){\rotatebox{90}{$T\, L_s$}}
\put(-35,80){\small $z-x$}
 & \qquad
\includegraphics[scale=\sizetwo]{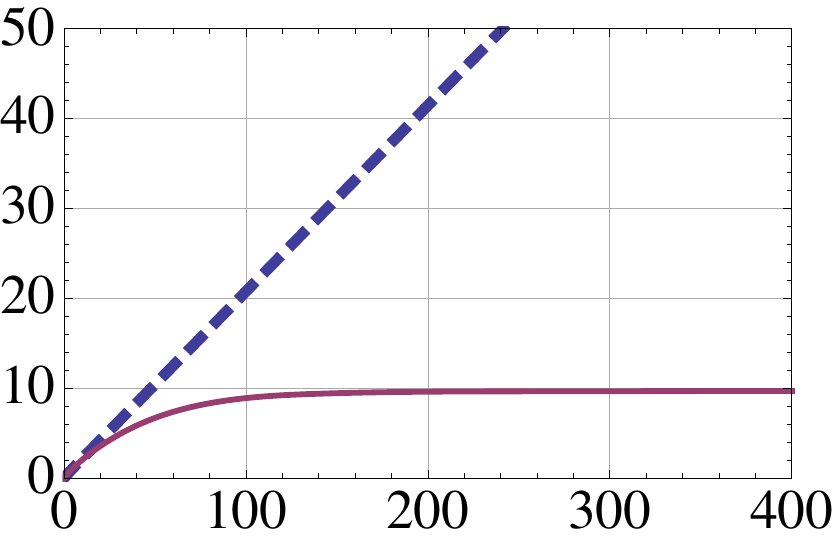}
\put(-80,-10){\small $a/T$}
\put(-157,35){\rotatebox{90}{$a\, L_s$}}
\put(-35,78){\small $z-x$}
\\[\dist]
\includegraphics[scale=0.64]{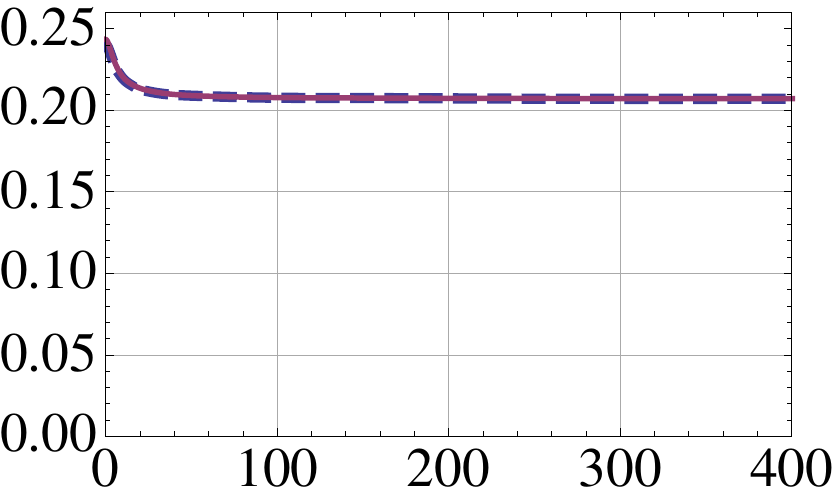}
\put(-80,-10){\small $a/T$}
\put(-167,35){\rotatebox{90}{$T\, L_s$}}
\put(-35,80){\small $x-x$}
 & \qquad
\includegraphics[scale=\sizetwo]{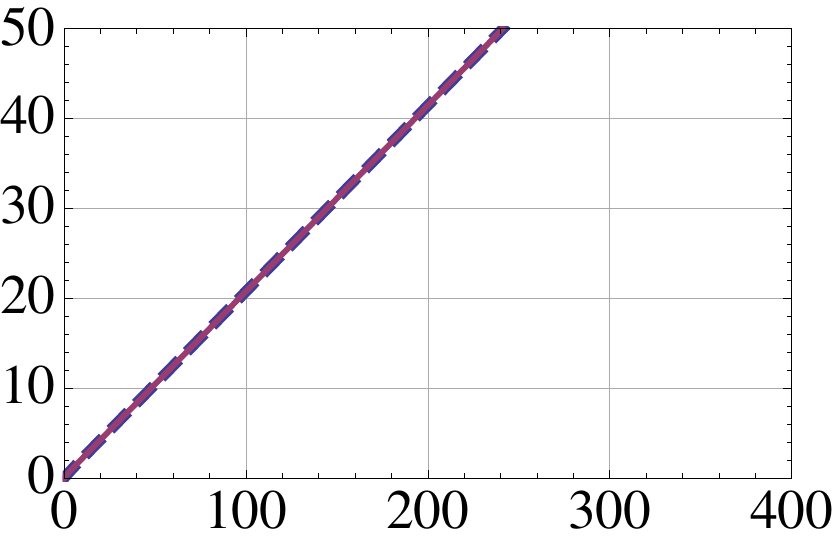}
\put(-80,-10){\small $a/T$}
\put(-157,35){\rotatebox{90}{$a\, L_s$}}
\put(-35,78){\small $x-x$}
\\[\dist]
\includegraphics[scale=0.64]{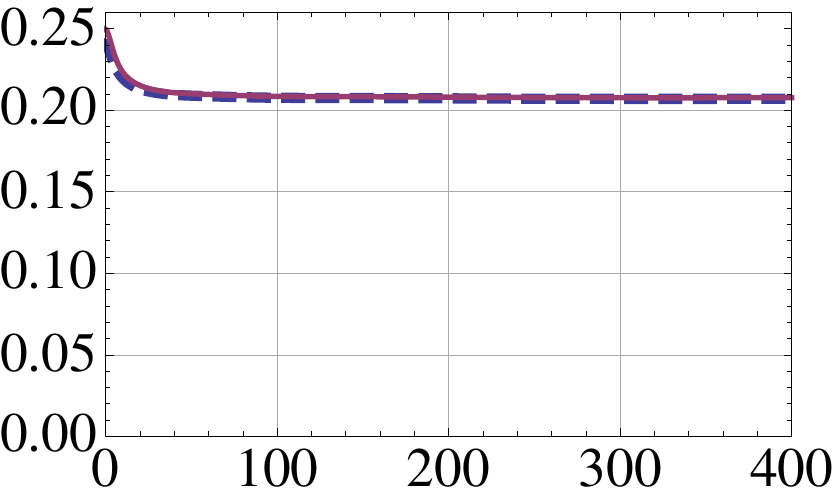}
\put(-80,-10){\small $a/T$}
\put(-167,35){\rotatebox{90}{$T\, L_s$}}
\put(-35,78){\small $x-y$}
 & \qquad
\includegraphics[scale=\sizetwo]{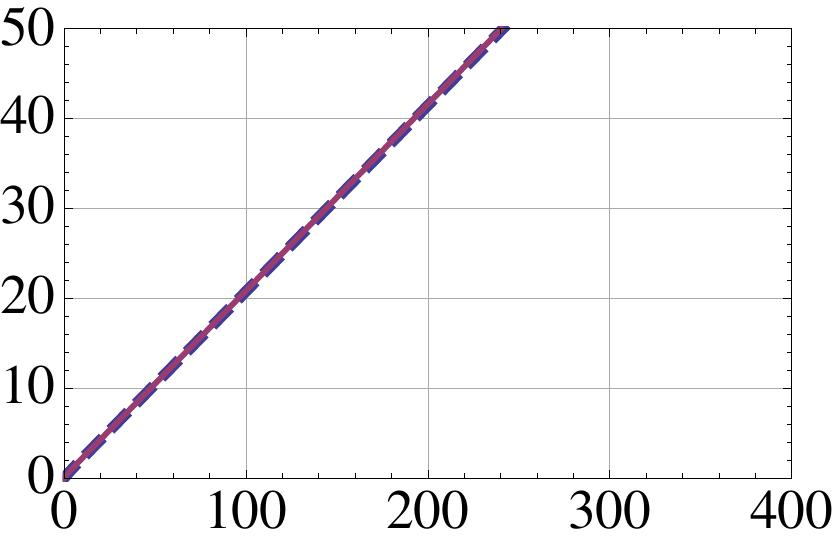}
\put(-80,-10){\small $a/T$}
\put(-157,35){\rotatebox{90}{$a\, L_s$}}
\put(-35,78){\small $x-y$}
\\[\dist]
\includegraphics[scale=0.64]{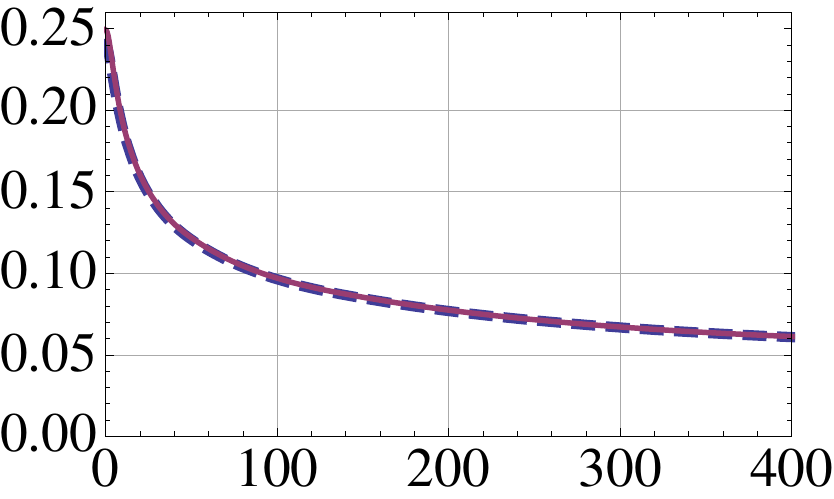}
\put(-80,-10){\small $a/T$}
\put(-167,35){\rotatebox{90}{$T\, L_s$}}
\put(-35,78){\small $x-z$}
 & \qquad
\includegraphics[scale=\sizetwo]{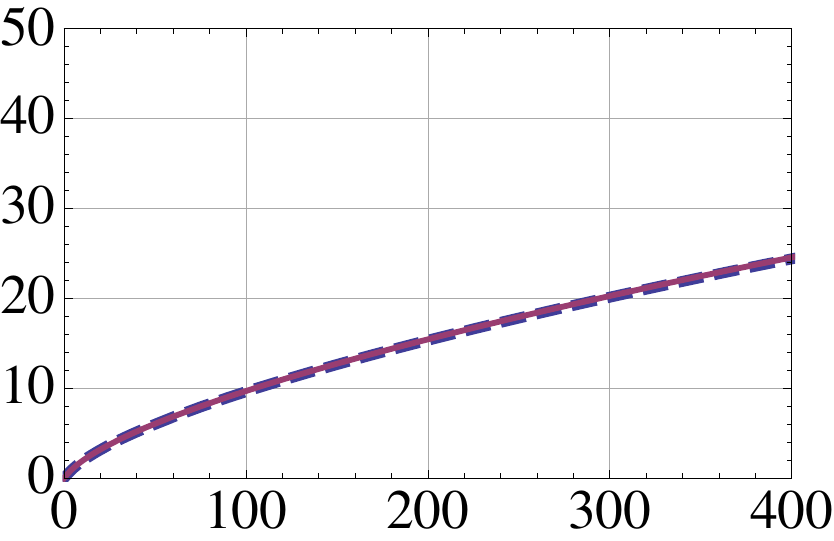}
\put(-80,-10){\small $a/T$}
\put(-157,35){\rotatebox{90}{$a\, L_s$}}
\put(-35,78){\small $x-z$}
\end{tabular}
\caption{\small Screening length in units of $1/T$ (left) or $1/a$ (right), as a function of the ratio $a/T$,  for a dipole at rest (dashed, blue curve) and for a dipole moving with $v=0.45$ (continuous, magenta curve). The first letter on the top right corner of each plot indicates the direction of motion, and the second one indicates the orientation of the dipole.
\label{Lsplots}
}
 \end{center}  
 \end{figure}
Recalling that the isotropic screening length is of the form 
$L_\mt{iso} \propto 1/T$, we see that the quantity plotted on the vertical axes in Figs.~\ref{staticscreening}, \ref{extzmovzdiffv}, \ref{extxmovzdiffv}, \ref{extxmovxdiffv}, \ref{extymovxdiffv} and \ref{extzmovxdiffv} is precisely proportional to $T L_s(T,a)$. However, the asymptotic behavior \eqn{asym} is not apparent in these plots because in most cases the horizontal axes do not extend to high enough values of $a/T$. For this reason we have illustrated the two possible asymptotic behaviors of $L_s$ in  Fig.~\ref{Lsplots}, where we have extended the horizontal axes to  larger values of $a/T$. We see from the continuous, magenta curves in the first two rows that $L_s  \sim 1/a$ for motion along the $z$-direction. For motion within the transverse plane we see from the same curves in the last three rows that $L_s \sim 1/T$. This approximate scaling relation seems to hold quite precisely for a dipole oriented within the transverse plane (3rd and 4th rows), whereas for a dipole oriented in the $z$-direction the product $T L_s$ seems to retain a slight (perhaps logarithmic) dependence on $a/T$ at large $a/T$. We can draw similar conclusions from the  dashed, blue curves in the figure, which correspond to static mesons. We see that for mesons oriented within the transverse plane (2nd, 3rd and 4th rows) the relation $T L_s \sim \mbox{constant}$ holds quite precisely, whereas for mesons oriented in the $z$-direction (1st and 5th rows) there seems to be some slight residual dependence on $a/T$ at large $a/T$.

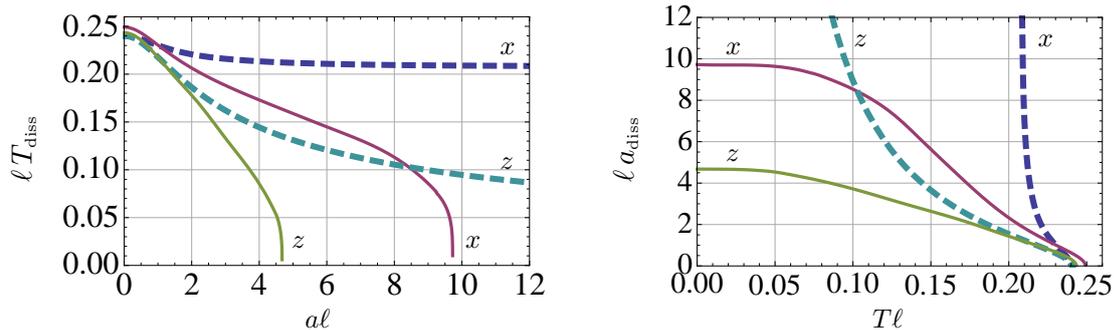
\begin{figure}[tb]
\begin{center}
\begin{tabular}{cc}
\includegraphics[scale=0.75]{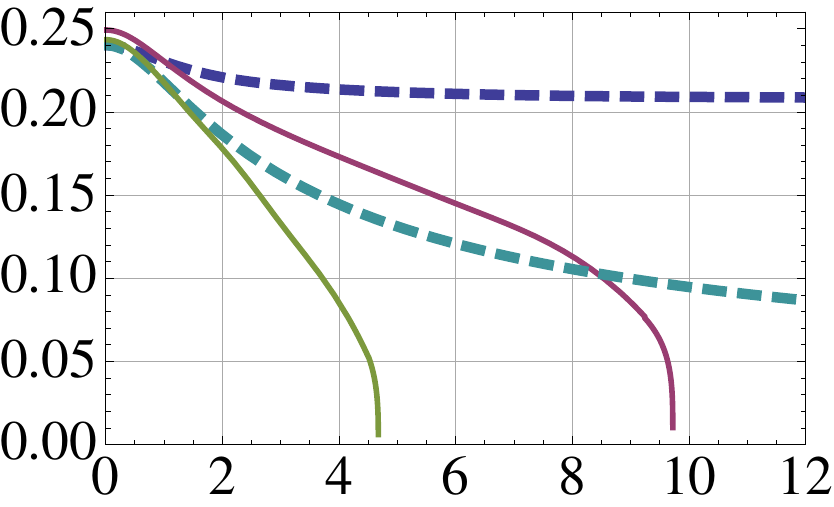}
\put(-90,-10){\small $a\ell$}
\put(-17,93){\small $x$}
\put(-17,50){\small $z$}
\put(-30,20){\small $x$}
\put(-95,20){\small $z$}
\put(-200,45){\rotatebox{90}{$\ell\, \Tdiss$}}
\qquad \qquad
\includegraphics[scale=0.70]{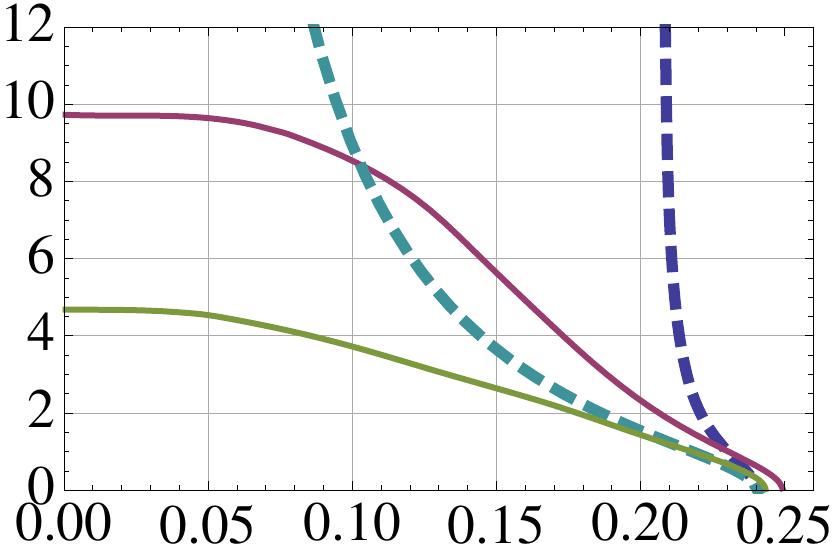}
\put(-90,-10){\small $T\ell$}
\put(-28,96){\small $x$}
\put(-97,96){\small $z$}
\put(-145,93){\small $x$}
\put(-145,53){\small $z$}
\put(-187,45){\rotatebox{90}{$\ell\, \adiss$}}
\end{tabular}
\caption{\small Dissociation temperature (left) $\Tdiss(a,\ell)=\ell^{-1} f(a\ell)$ and dissociation anisotropy (right) $\adiss(T,\ell) = \ell^{-1} g(T\ell)$ for a dipole at rest (dashed curves) and for a dipole moving along the $z$-direction with $v=0.45$ (continuous curves). The orientation of the dipole is indicated by a letter next to each curve.
\label{genericTdiss}}
 \end{center}   
 \end{figure}

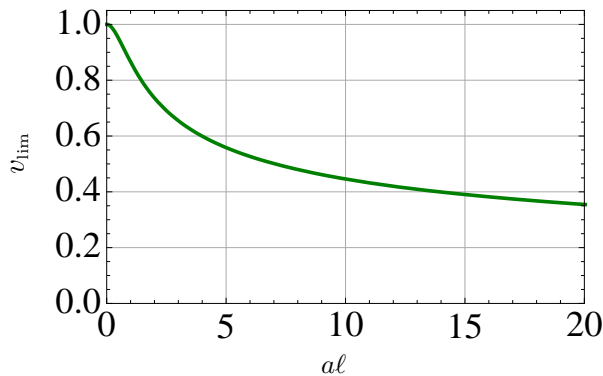
\begin{figure}[t]
\begin{center}
\includegraphics[scale=0.85]{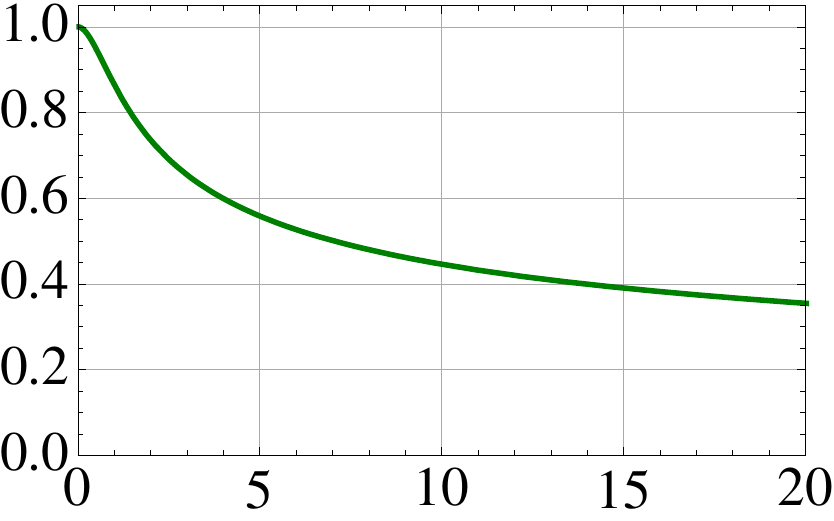}
 \begin{picture}(0,0)
\put(-109,-10){\small $a\ell$}
\put(-225,65){\rotatebox{90}{$v_{\mt{lim}}$}}
 \end{picture}
\caption{\small Limiting velocity, for fixed anisotropy and $T=0$, beyond which a meson oriented along the $x$-direction and moving along the $z$-direction will dissociate.
\label{genericvmax}}
\end{center}
\end{figure}

Combining the two plots on the left and the right columns of Fig.~\ref{Lsplots} we can eliminate $a/T$ and obtain $T L_s$ as a function of $a L_s$ and vice versa. Recalling \eqn{interpret} we see that we can interpret the result in the first case as $\Tdiss(a,\ell) = \ell^{-1} f(a\ell)$, whereas in the second case we get $\adiss(T,\ell) = \ell^{-1} g(T \ell)$. The functions $f$ and $g$ are the curves shown in Fig.~\ref{genericTdiss}(left) and Fig.~\ref{genericTdiss}(right), respectively. The right plot is of course the mirror image along a 45 degree line of the left plot. 
We see in Fig.~\ref{genericTdiss}(left) that the dissociation temperature decreases monotonically with increasing anisotropy and vanishes at $a \ell \simeq 9.75$ (for the chosen velocity and orientation). On the right plot this corresponds to the dissociation anisotropy at zero temperature. As anticipated above, even at zero temperature, a generic meson of size $\ell$ will dissociate if the anisotropy is increased above  $\adiss(T=0,\ell) \propto 1/\ell$. The proportionality constant in this relation is a decreasing function of the meson velocity in the plasma. This implies that for a fixed anisotropy there is a limiting velocity $\vlim$ above which a meson will dissociate, even at zero temperature. The form of $v_\mt{lim}(a\ell)$ for $T=0$ is plotted in Fig.~\ref{genericvmax}.

The existence of a limiting velocity for quarkonium mesons is well known in a strongly coupled isotropic plasma \cite{limiting1,limiting2}, in which case the dissociation at $v=v_\mt{lim}$ is caused by the temperature. What we see here is that in our anisotropic plasma this behavior persists as $T\to 0$ for generic motion. In this limit it is the anisotropy that is responsible for the dissociation. In the case of ultra-relativistic motion the relation between $\adiss$ or $\Tdiss$ and $v_\mt{lim}$ can be obtained by combining the scalings \eqn{result-scaling} and \eqn{asym}. For generic motion these relations yield
\be
\adiss(T,\ell) \sim \frac{1}{\ell} (1-\vlim^2)^{1/2} \,, 
\qquad \qquad [a\gg T \,, \vlim \lesssim 1]
\ee
whereas for motion within the transverse plane we obtain
\be
\Tdiss(a,\ell) \sim \frac{1}{\ell} (1-\vlim^2)^{1/4} \,.
\qquad \qquad [a\gg T \,, \vlim \lesssim 1]
\label{TT}
\ee
 
\begin{figure}[tb]
\begin{center}
\begin{tabular}{cc}
\includegraphics[scale=0.75]{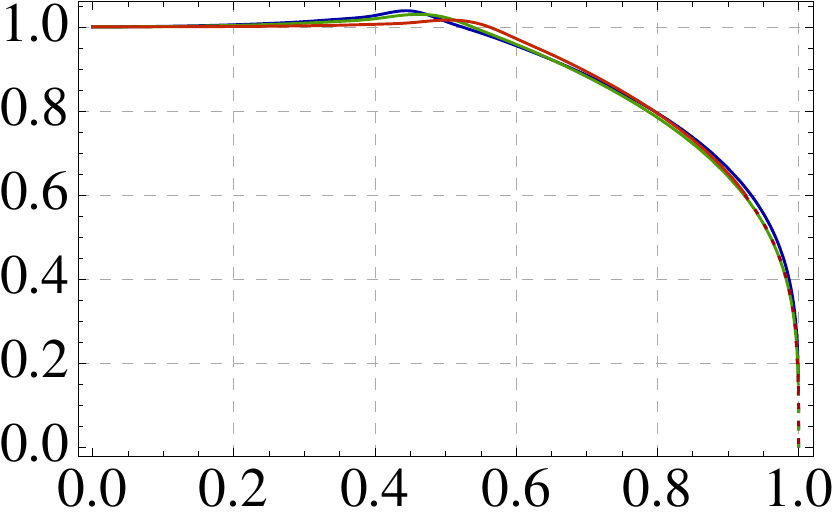}
\put(-80,-10){\small $v$}
\put(-200,30){\rotatebox{90}{$\Tdiss(v)/\Tdiss(0)$}}
\qquad \qquad
\includegraphics[scale=0.75]{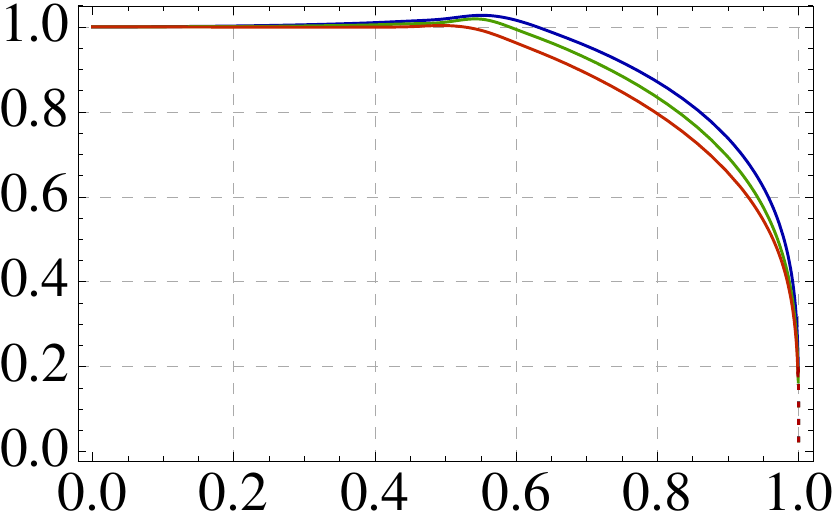}
\put(-80,-10){\small $v$}
\put(-200,30){\rotatebox{90}{$\Tdiss(v)/\Tdiss(0)$}}
\end{tabular}
\caption{\small Dissociation temperature for a meson moving along the $x$-direction and oriented along the $z$-direction (left) or along the $x$-direction (right).  Each curve corresponds to a fixed value of the product  $a\ell = 0 \, \mbox{(blue curve)}, 1.4 \, \mbox{(green curve)}, 25 \, \mbox{(red curve)}$.  
\label{Tdissalongx}
}
 \end{center} 
 \end{figure}

The scaling \eqn{TT} agrees with the isotropic result \cite{liu1,liu2} and illustrates the fact that, for motion within the transverse plane, the limiting velocity in our anisotropic plasma approaches unity as $T\to 0$. This behavior is the same for a meson at rest, as illustrated in Fig.~\ref{genericTdiss}, where we see that a sufficiently small meson will remain bound in the plasma for any value of the anisotropy provided the plasma is cold enough. In fact, the form of the dissociation temperature for all anisotropies and all velocities within the transverse plane is qualitatively analogous to that of the isotropic case, as shown in Fig.~\ref{Tdissalongx}. 
The fact that the curves in this figure approximately overlap one another signals that the dependence of the dissociation temperature on $v$ and $a \ell$ can be approximately factorized over the entire range $0\leq v \leq 1$.

\begin{figure}[tb]
\begin{center}
\begin{tabular}{cc}
\includegraphics[scale=0.75]{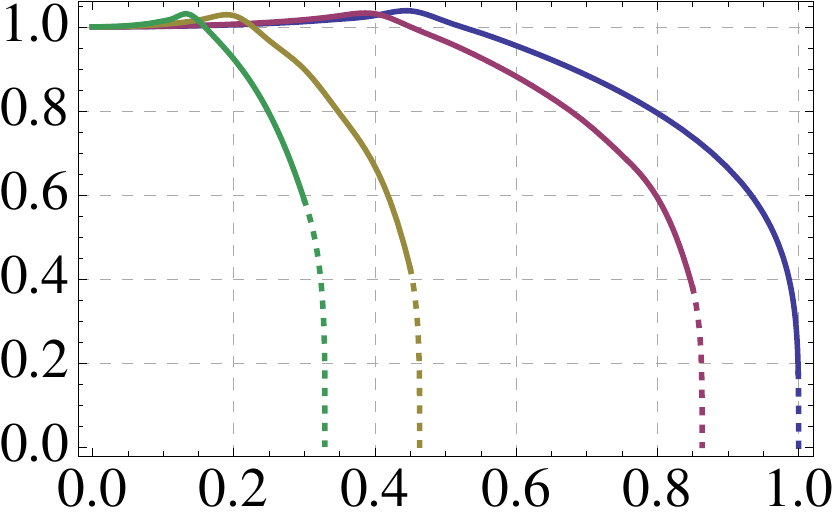}
\put(-80,-10){\small $v$}
\put(-200,30){\rotatebox{90}{$\Tdiss(v)/\Tdiss(0)$}}
\qquad\qquad
\includegraphics[scale=0.75]{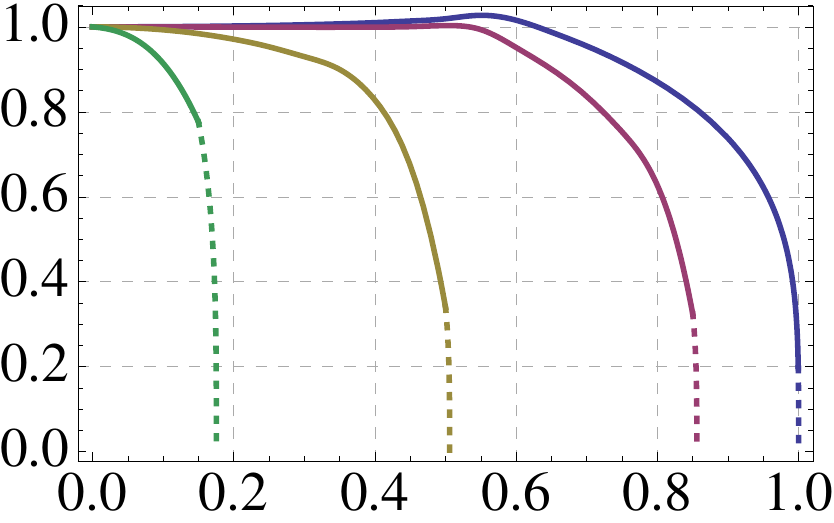}
\put(-80,-10){\small $v$}
\put(-200,30){\rotatebox{90}{$\Tdiss(v)/\Tdiss(0)$}}
\end{tabular}
\caption{\small Dissociation temperature for a meson moving along the $z$-direction and oriented along the $x$-direction (left) or along the $z$-direction (right).  Each curve corresponds to a fixed value of the product  $a\ell$. From right to left, $a\ell=0,1,5.4,25$.
\label{Tdissalongz}
}
 \end{center} 
 \end{figure}

In contrast, for generic motion we saw above that the limiting velocity is subluminal even at $T=0$, 
$v_\mt{lim} (T=0,a\ell)<1$. Increasing the temperature simply decreases the value of the limiting velocity, 
\mbox{$v_\mt{lim}(T\ell,a\ell) < v_\mt{lim} (T=0,a\ell)$}. Turning these statements around we see that, at a fixed anisotropy, the dissociation temperature is a decreasing function of the velocity that vanishes at 
$v=v_\mt{lim} (T=0,a\ell)$. This is illustrated in Fig.~\ref{Tdissalongz}, where we see that $v_\mt{lim} (T=0,a\ell)$ decreases as the anisotropy increases, in agreement with  Fig.~\ref{genericvmax}.
In order to facilitate comparison with the isotropic results of \cite{liu1,liu2,chernicoff1}, in Fig.~\ref{Tdissalongz} we have chosen to normalize the dissociation temperature by its value at $v=0$ instead of by the dipole's size $\ell$. Our numerical results suggest that as $v$ approaches $v_\mt{lim}$ the dissociation temperature may vanish as
\be
\frac{\Tdiss (v,a\ell)}{\Tdiss (0, a\ell)} \sim 
\left( v_\mt{lim}^2 - v^2 \right)^{\varepsilon} \,. 
\label{diss1}
\ee
In this equation $\vlim = \vlim(T=0,a\ell)$ and $\varepsilon=\varepsilon(a\ell)>0$ is an anisotropy-dependent exponent. Unfortunately, the limit $v \to \vlim$ is difficult to analyze numerically, so our results are not precise enough to allow us to establish \eqn{diss1} unambiguously. To emphasize this point, in Fig.~\ref{Tdissalongz} we have plotted as discontinuous the part of the curves between the last two data points. The last point lies on the horizontal axis at $(v,T)=(\vlim, 0)$, and the penultimate point lies at a certain height at $(v \lesssim \vlim, T>0)$. Since this last bit of the curves is an interpolation between these data points, it is difficult to establish whether the slopes of the curves diverge as they meet the horizontal axis, as would be implied by the scaling \eqn{diss1}. Presumably, this scaling could be verified or falsified analytically by including the first correction in $T/a$ to the  scaling in the second line of \eqn{asym}.


\section{Discussion}
\label{discussion}
We have considered an anisotropic ${\cal N}=4$ SYM plasma in which the $x,y$ directions are rotationally symmetric, but the $z$-direction is not. In the context of heavy ion collisions the latter would correspond to the beam direction, and the former to the transverse plane. The screening length of a quarkonium meson in motion in the plasma depends on the relative orientation between these directions, on the one hand, and the direction of motion of the meson and its orientation, on the other. This dependence can be parametrized by three angles  $(\theta_v, \theta,\vp)$, as shown in Fig.~\ref{figure_angles}. We have determined the screening length for the most general geometric parameters and for any anisotropy. Our results are valid in the strong-coupling, large-$\nc$ limit, since we have obtained them by means of the gravity dual \cite{prl,jhep} of the anisotropic ${\cal N}=4$  plasma. The anisotropy is induced by a position-dependent theta term in the gauge theory, or equivalently by a position-dependent  axion on the gravity side. One may therefore wonder how sensitive the conclusions may be to the specific source of the anisotropy. In this respect it is useful to note that the gravity calculation involves only the coupling of the string to the background metric. This means that any anisotropy that gives rise to a qualitatively similar metric (and no Neveu-Schwarz $B$-field) will yield qualitatively similar results for the screening length, irrespectively of the form of the rest of supergravity fields. 

An example of a rather robust conclusion is the ultra-relativistic behavior\footnote{We recall that we first send the quark mass to infinity and then $v\to 1$ (see Sec.~\ref{intro}).} of the screening length \eqn{result-scaling}, which for motion not exactly aligned with the transverse plane is $L_s \sim (1-v^2)^{1/2}$. The 1/2 exponent contrasts with the 1/4 isotropic result \cite{liu1,liu2}, and  follows from the fact that the near-boundary fall-off of the metric \eqn{sol2} takes the schematic form
\be
g_{\mu\nu} = \frac{L^2}{u^2} \left( \eta_{\mu\nu} + 
u^2 g^{(2)}_{\mu\nu} + u^4 g^{(4)}_{\mu\nu} + \cdots
\right) \,.
\label{near}
\ee
As $v$ grows closer and closer to 1 the point of maximum penetration of the string into the bulk, $\umax$, moves closer and closer to the AdS boundary at $u=0$. As a consequence, the physics in this limit is solely controlled by the near-boundary behavior of the metric. For generic motion the behavior is in fact governed by the $O(u^2)$ terms alone, and a simple scaling argument then leads to the 1/2 exponent above. In the isotropic case the $O(u^2)$ terms are absent and the same scaling argument leads to the 1/4 exponent. 

In fact, a similar reasoning allowed us to determine the large-anisotropy limit. Since the metric component $g_{zz} \propto \ch(u)$ grows as one moves from the boundary to the horizon, a subluminal velocity of the meson at the boundary would eventually translate into a superluminal proper velocity \eqn{proper} at a sufficiently large value of   $u$.\footnote{Note that the overall conformal factor $1/u^2$ in \eqn{sol2} plays no role in this argument, since it cancels out in the ratio \eqn{proper}.} This sets an upper limit on the maximum penetration length $\umax$ of the string into the bulk and hence on $L_s$. Moreover, $g_{zz}$ becomes steeper as $a/T$ increases, so in the limit $a/T \gg 1$ the point $\umax$ approaches the AdS boundary (unless the motion is aligned with the transverse plane), just as in the ultra-relativistic limit. In this limit the physics is again controlled by the $O(u^2)$ terms in the metric, which depend on $a$ but not on $T$. Therefore dimensional analysis implies that $L_s = \mbox{const.}\times a^{-1}$, were the proportionality `constant' is a decreasing function of the velocity. This led us to one of our main conclusions: even in the limit $T\to 0$, a generic meson of size $\ell$ will dissociate at some high enough anisotropy $\adiss \sim \ell^{-1}$. Similarly, for fixed $a$ and $T$, even if $T=0$, a generic meson will dissociate if its velocity exceeds a limiting velocity $\vlim (a,T)<1$, as shown in Fig.~\ref{genericvmax} for $T=0$. As explained in Sec.~\ref{diss_sec}, the conclusions in this paragraph would remain unchanged if we worked at constant entropy density instead of at constant temperature, since in the limit  $a\gg s^{1/3}$ the physics would again be controlled only by the $O(u^2)$ terms in the metric.

The above discussion makes it clear that, at the qualitative level, much of the physics depends only on a few features of the solution: The presence of the $g^{(2)}_{\mu\nu}$ term in the near-boundary expansion of the metric, the fact that the metric \eqn{near} be non-boost-invariant at order $u^2$ (i.e.~that $g^{(2)}_{\mu\nu}$ not be proportional to $\eta_{\mu\nu}$), and the fact that $g_{zz}$ increases  as a function of both $u$ and $a/T$.\footnote{Again, up to possible overall conformal factors.} The second condition is necessary because otherwise the physics of a meson in motion would be equivalent to that of a meson at rest, and we have seen that the latter is very similar to that of a meson in an isotropic plasma. The third condition ensures that $\umax$ moves close to the boundary as $a/T$ increases. Note that adding temperature to an otherwise boost-invariant metric  will only affect $g^{(4)}_{\mu\nu}$, and thus this is not enough to make $g^{(2)}_{\mu\nu}$ non-boost-invariant. This conclusion is consistent with the fact that $g^{(2)}_{\mu\nu}$ is only a function of the external sources which the theory is coupled to.

\begin{figure}[tb]
\begin{center}
\includegraphics[scale=0.73]{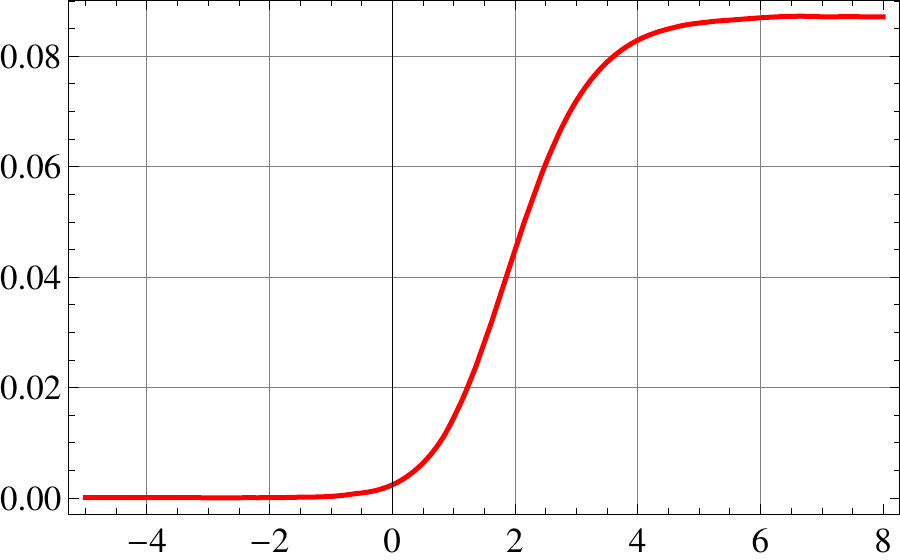}
\put(-109,-10){\small $\log (a/T)$}
\put(-210,40){\rotatebox{90}
{$\log (s^\mt{2D}/ s_\mt{iso}^\mt{2D})$}}
\caption{\small  Log-log plot of the entropy density per unit 2-area in the $xy$-directions on a constant-$z$ slice as a function of $a/T$, normalized to the isotropic result $s_\mt{iso}^\mt{2D} = \frac{\pi}{2} \nc^2 T^2$. 
\label{scalings2D}
}
 \end{center} 
 \end{figure}

From the gauge theory viewpoint, some heuristic intuition  can be gained  by recalling that the anisotropy is induced by dissolving along the $z$-direction objects that extend along the $xy$-directions \cite{ALT,prl,jhep}. The number density of such objects along the $z$-direction, $dn/dz$, is proportional to $a$. On the gravity side these are D7-branes that wrap the five-sphere in the metric \eqn{sol2}, extend along the $xy$-directions, and are homogeneously distributed in the $z$-direction.  Increasing $a$ has a large effect on the entropy density per unit 3-volume in the $xyz$-directions, in the sense that $s/T^3 \to \infty$ as $a/T \to \infty$, as shown in Fig.~\ref{scalings}. In contrast, the entropy density per unit 2-area in the $xy$-directions on a constant-$z$ slice, $s^\mt{2D}/T^2$,  approaches a constant in the  limit $a/T \to \infty$. This is illustrated in Fig.~\ref{scalings2D}, which is based on our numerical calculations, but it can also be proven analytically following the argument in Sec.~2.5 of Ref.~\cite{ALT}. In view of these differences, it is perhaps not surprising that the anisotropy has the largest effect on the physics of mesons moving along the $z$-direction, and the smallest effect on the physics of mesons moving within the transverse plane. Mesons at rest are also more sensitive to the anisotropy if they extend along the $z$-direction than if they are contained within the transverse plane. Presumably, the correct intuition behind this physics is that moving  against the D7-branes is  harder than moving along them.

We close with a few comments on existing weak-coupling results on the physics of quarkonium dissociation in the real-world QGP. In the isotropic case the velocity dependence of the heavy quark potential has been studied using perturbative and effective field theory methods, see e.g.~\cite{Song:2007gm,Dominguez:2008be,Escobedo:2011ie,talk}. These analyses include  modifications of both the real and imaginary parts of the potential, which are related to screening and to the thermal width of the states, respectively. They find that meson dissociation at non-zero velocity results form a complex interplay between the real and the imaginary parts of the potential.  However, the  general trend that seems to emerge is that screening effects increase with the velocity, while the width of the states decreases.  The behavior of the real part is thus in qualitative agreement with the isotropic limit of our results. However, the extraction of a screening length from these analyses is not immediate due to the fact that the real part of the potential is not  approximately Yukawa-like  \cite{Escobedo:2011ie,talk}, in contrast with the holographic result. In any case, an interesting consequence of the dominance of the real part of the potential is that, at sufficiently high velocities, dissociation is caused by screening rather than by Landau damping \cite{Escobedo:2011ie,talk}. In the holographic framework, the thermal widths of our mesons could presumably be computed along the lines of \cite{Faulkner:2008qk}. 
  
To the best of our knowledge no results at non-zero velocity exist in the presence of anisotropies, so in this case we will limit ourselves to the static situation. We emphasize though that any comparison between these results and ours should be interpreted with caution, because the sources of anisotropy in the QGP created in a heavy ion collision and in our system are different. In the QGP the anisotropy is dynamical in the sense that it is due to the initial distribution of particles in momentum space, which will evolve in time and eventually become isotropic. In contrast, in our case the anisotropy is due to an external source that keeps the system in an equilibrium anisotropic state that will not evolve in time.  We hope that, nevertheless, our system might provide a good toy model for processes whose characteristic time scale is sufficiently shorter than the time scale controlling the time evolution of the QGP.  
 
A general conclusion of Refs.~\cite{dumitru1,dumitru2,burnier} is that, if the comparison between the anisotropic plasma and its isotropic counterpart is made at equal temperatures, then the screening length increases with the anisotropy. This effect occurs for dipoles oriented both along and orthogonally to the anisotropic direction, but it is more pronounced for dipoles along the anisotropic direction. The dependence on the anisotropy in these weak-coupling results is the opposite of what we find in our strongly coupled plasma. In our case the screening length in the anisotropic plasma is smaller than in its isotropic counterpart if both plasmas are taken to have the same temperature, as shown in Fig.~\ref{staticscreening}(left). We also find that the effect is more pronounced for dipoles extending along the anisotropic direction, as illustrated in Fig.~\ref{staticscreening1}(left).

Refs.~\cite{burnier,philipsen} argued that if the comparison between the anisotropic and the isotropic plasmas is made at equal entropy densities, then the physics of quarkonium dissociation exhibits little or no sensitivity to the value of the anisotropy. This is again in contrast to our results since, as shown in Fig.~\ref{staticscreening}(right) and Fig.~\ref{staticscreening1}(right), the screening length in this case is just as sensitive to the anisotropy as in the equal-temperature comparison. The difference in the equal-entropy case is simply that the screening length may increase or decrease with the anisotropy depending on the dipole's orientation.


\begin{acknowledgments}
It is a pleasure to thank M.~Strickland for helpful discussions. MC is supported by a postdoctoral fellowship from Mexico's National Council of Science and Technology (CONACyT). We acknowledge financial support from 2009-SGR-168, MEC FPA2010-20807-C02-01, MEC FPA2010-20807-C02-02 and CPAN CSD2007-00042 Consolider-Ingenio 2010 (MC, DF and DM), and from DE-FG02-95ER40896 and CNPq (DT).
\end{acknowledgments}


\end{document}